European Parliament

# DIRECTORATE-GENERAL FOR INTERNAL POLICIES

## POLICY DEPARTMENT C
### CITIZENS' RIGHTS AND CONSTITUTIONAL AFFAIRS

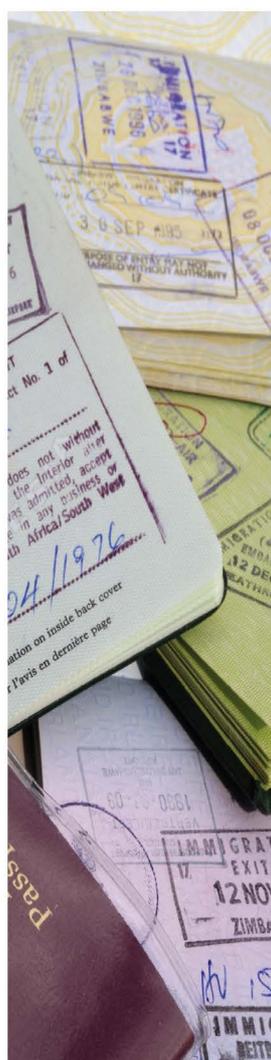

Constitutional Affairs

**Justice, Freedom and Security**

Gender Equality

Legal and Parliamentary Affairs

Petitions

# An Assessment of the Commission's Proposal on Privacy and Electronic Communications

## STUDY FOR THE LIBE COMMITTEE

EN

2017

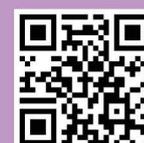

# An assessment of the Commission's Proposal on Privacy and Electronic Communications

## STUDY

**Abstract**

This study, commissioned by the European Parliament's Policy Department for Citizens' Rights and Constitutional Affairs at the request of the LIBE Committee, appraises the European Commission's proposal for an ePrivacy Regulation. The study assesses whether the proposal would ensure that the right to the protection of personal data, the right to respect for private life and communications, and related rights enjoy a high standard of protection. The study also highlights the proposal's potential benefits and drawbacks more generally.

**ABOUT THE PUBLICATION**

This research paper was requested by the European Parliament's Committee on Civil Liberties, Justice and Home Affairs, and was commissioned, overseen and published by the Policy Department for Citizens' Rights and Constitutional Affairs.

Policy Departments provide independent expertise, both in-house and externally, to support European Parliament committees and other parliamentary bodies in shaping legislation and exercising democratic scrutiny over EU external and internal policies.

To contact the Policy Department for Citizens' Rights and Constitutional Affairs or to subscribe to its newsletter please write to:
Poldep-citizens@ep.europa.eu

**Research Administrator Responsible**

Kristiina MILT
Policy Department C: Citizens' Rights and Constitutional Affairs
B-1047 Brussels
E-mail: poldep-citizens@ep.europa.eu

**AUTHORS**

Dr. Frederik ZUIDERVEEN BORGESIUS (project leader and editor)
Dr. Joris VAN HOBOKEN
Ronan FAHY
Dr. Kristina IRION
Max ROZENDAAL

All authors work at the IViR Institute for Information Law, University of Amsterdam.

# CONTENTS

# LIST OF ABBREVIATIONS

| | |
|---|---|
| **CJEU** | Court of Justice of the European Union |
| **EDPS** | European Data Protection Supervisor |
| **ePrivacy proposal** | Commission's proposal on Privacy and Electronic Communications |
| **EU Charter** | Charter of Fundamental Rights of the European Union |
| **GDPR** | General Data Protection Regulation |
| **OTT** | Over The Top service |

# EXECUTIVE SUMMARY

## Background

One of the major challenges in a modern society is protecting privacy (private life), confidentiality of communications, and related rights. On 11 January 2017, the European Commission presented a Proposal for a Regulation of the European Parliament and of the Council, concerning the respect for private life and the protection of personal data in electronic communications and repealing Directive 2002/58/EC (Regulation on Privacy and Electronic Communications). This ePrivacy proposal aims to replace the current ePrivacy Directive (last amended in 2009[1]).

The ePrivacy proposal lays down rules regarding the protection of fundamental rights and freedoms of natural and legal persons in the provision and use of electronic communications services, and in particular, the rights to respect for private life and communications and the protection of natural persons with regard to the processing of personal data. In addition, the proposal aims to ensure free movement of electronic communications data and electronic communications services within the EU. The proposal's provisions aim to particularise and complement the General Data Protection Regulation (GDPR) by laying down specific rules for the purposes mentioned above.

## Main points

We compliment the Commission on the ambitious proposal, which delivers important improvements compared to the current framework for electronic communications privacy. However, the ePrivacy proposal also has its weaknesses.

**In this study we discuss weaknesses of the proposed provisions, and ways to improve these provisions. We recommend that the EU lawmaker pays extra attention to four points; (i) location tracking; (ii) browsers and default settings; (iii) tracking walls; (iv) the confidentiality of communications. Regarding those topics, the ePrivacy proposal does not ensure sufficient protection of the right to privacy and confidentiality of communications. Some provisions in the ePrivacy proposal offer less protection than the GDPR.**

## Location tracking

Article 8(2) concerns the collection of information emitted by user devices. The provision does not sufficiently protect people against secretive or unwanted location tracking. The provision could be interpreted as follows. If an organisation wants to follow people's movements (based on Wi-Fi- or Bluetooth tracking for instance), it merely has to put up posters that say: '*In this city we track your location based on the Wi-Fi and Bluetooth signals of your devices. Turn off your phone or other device, or your Wi-Fi and Bluetooth, if you don't want to be tracked*'. Hence, Article 8(2) allows location tracking without consent and without an opt-out option.

Under that proposed rule, people might never feel free from surveillance when they walk or drive around. People would always have to look around whether they see a sign or poster that informs them of location tracking. Moreover, people could only escape location tracking by limiting the functionalities of their phones and other devices. Article 8(2) would reduce the protection that people enjoy under the GDPR, and would violate privacy.

**The proposed Article 8(2) should be significantly amended to protect privacy and related rights. We recommend that collecting Wi-Fi or Bluetooth signals should only**

---

[1] Directive 2009/136/EC of the European Parliament and of the Council of 25 November 2009 amending (…) Directive 2002/58/EC concerning the processing of personal data and the protection of privacy in the electronic communications sector (…) (OJ L 337, 18.12.2009, p. 11–36).

**be allowed after the individual's informed consent.** There should be exceptions to that consent requirement, but only as far as strictly necessary to enable the device to connect to another device. **The lawmaker could consider introducing an exception for anonymous people counting. Data collection for people counting should only be allowed if there are sufficient safeguards, which should include immediate anonymisation.**

## Browsers, default settings, and Do Not Track

Article 10 does not offer sufficient privacy protection. The provision states, in short, that browsers and similar software should offer people the option to allow or reject third party tracking (internet-wide tracking). In an earlier version of the ePrivacy proposal, the provision provided that browsers should have privacy-friendly settings by default. Article 10 is hard to reconcile with the GDPR, which prescribes data protection by design and by default.

**We recommend that the EU lawmaker reinserts the privacy by design approach.** Browsers and similar software should, by default, be set to privacy friendly settings that limit online tracking. **The lawmaker should also consider requiring compliance with Do Not Track (or a similar standard).** Do Not Track should enable people to signal with their browser that they do not want to be tracked. **Do Not Track should apply to all tracking technologies, including cookies and device fingerprinting.** The standard should be user-friendly, and should be backed by law and proper enforcement.

## Tracking walls and other take-it-or-leave-it choices

On the internet, many companies offer people take-it-or-leave-it-choices regarding privacy. For instance, some websites install tracking walls (or 'cookie walls'), barriers that visitors can only pass if they agree to being tracked. If people encounter such take-it-or-leave-it choices, they are likely to consent, even if they do not want to disclose data in exchange for using a website.

**We recommend that the EU lawmaker bans tracking walls, at least in certain circumstances. A complete ban would provide most legal clarity. Under a partial ban, tracking walls are prohibited under certain circumstances (a black list).** For instance, state-funded websites, sites regarding health or other sensitive information, and sites with a monopoly-like position should not be allowed to install tracking walls. **The black list should be complemented with a grey list, with circumstances in which a tracking wall is presumed to be illegal. The lawmaker should also adopt rules for take-it-or-leave-it choices regarding privacy in other contexts.**

## The right to confidentiality of communications and exceptions

The rules regarding confidentiality of communications, and especially regarding the exceptions to that right, need further attention of the EU lawmaker (Article 5 and 6). **We recommend that the EU lawmaker only allows the analysis of communications content and metadata in limited circumstances, and only as far is *strictly* necessary. If no exception applies, the law should ensure that all end-users (for instance the sender and receiver of an email) should give meaningful consent before companies can analyse their communications content or metadata. Furthermore, a phone provider or an internet provider should not be allowed to offer a take-it-or-leave it choice, where people can only subscribe if they allow the provider to analyse their communications content or metadata for marketing purposes. In addition, the definition of metadata should be amended to ensure that metadata generated by 'over the top' service providers are within the scope of the definition.**

## Other provisions

Many other proposed provisions also require clarification or amendments. Below we list some of the most important points.

Article 1 describes the subject matter of the ePrivacy Regulation. **We recommend that the EU lawmaker clarifies that the ePrivacy Regulation does not only protect privacy and communications confidentiality, but also protects the right to impart and receive information, and related rights.** We also recommend stating, in an article rather than in the preamble, that the ePrivacy Regulation does not lower the level of protection enjoyed by natural persons under the GDPR. Furthermore, the ePrivacy Regulation should state that it aims for a 'high' level of protection of privacy, communications confidentiality, and related rights.

Article 2 and 3 concern the regulation's material and territorial scope. **We recommend that the EU lawmaker clarifies the scope of Article 2 and 3; the scope of those provisions seems to be too narrow.**

**Article 4 provides definitions, and refers to the draft European Electronic Communications Code for other definitions. The lawmaker could consider defining the main concepts in the ePrivacy Regulation itself, rather than in the European Electronic Communications Code. If the lawmaker chooses to define relevant concepts in the European Electronic Communications Code, we recommend that, while working on the ePrivacy proposal, the lawmaker pays close attention to the development of that Code.**

The scope of 'electronic communications service' is widened significantly in the ePrivacy proposal, and also encompasses many 'over the top' services that enable communication. **It makes sense to broaden the scope of the ePrivacy rules, especially the rules that protect communications confidentiality.**

**Several definitions in Article 4 should be clarified.** For instance, we recommend that the lawmaker clarifies or amends the definitions of 'end-user', 'direct marketing', and 'electronic communications metadata'.

Article 5 says, in short, that electronic communications are confidential. **The lawmaker should clarify that injecting ads or other content into communications violates the right to communications confidentiality (Article 5). We recommend that the lawmaker considers ensuring that communications data are also protected when the data are stored in the cloud.**

**Electronic communications content and metadata both deserve a high level of protection. Article 6(1) provides exceptions to the principle of communications confidentiality for electronic communications *data* (metadata and content).** Article 6(1) provides exceptions regarding *metadata* and *content*; Article 6(2) regarding *metadata*; Article 6(3) regarding *content*. As noted, the lawmaker should use the phrase 'strictly necessary' (instead of 'necessary') to emphasise that the exceptions (in Article 6 and 8) should be interpreted narrowly.

The ePrivacy proposal broadens the possibilities for telecom providers to process electronic communications data, based on end-users consent. **We recommend that the lawmaker carefully considers whether it accepts that the ePrivacy proposal lowers the protection of privacy and communications confidentiality in the context of telecom providers. As noted, the stricter requirements for telecom providers and consent should be included in an article, rather than in Recital 18.**

**In addition, we urge the lawmaker to carefully consider whether it wants to allow the use of anonymised metadata, without people's consent, for heatmaps or other purposes.** From a privacy perspective, it would probably be better not to allow such practices.

**The lawmaker should clarify that merely contacting another end-user does not signify consent. We also recommend that the lawmaker considers adding a provision that ensures that the rights of non-end-users (mentioned in an email for example) are respected.** As mentioned previously, we recommend that the lawmaker clarifies that the consent of all end-users is required for processing of electronic communications data (if the processing does not fall under the other exceptions).

**We recommend that the lawmaker clarifies the meaning of references to anonymisation in Article 6. More generally, we recommend that the lawmaker keeps in mind that anonymising data does not take away all threats to fundamental rights.**

**The lawmaker should consider introducing a type of household exception (applicable to Article 6).** For such a household exception, inspiration could be drawn from the household exception in the GDPR. Such an exception should only apply to processing specifically requested by the end-user, and the requested processing should not disproportionally affect the fundamental rights of other end-users. **The lawmaker should ensure that such an exception does not create a loophole that enables further processing for other purposes. We recommend that the lawmaker does not add a 'legitimate interests provision' to the ePrivacy Regulation.**

**Article 7 needs clarification. The lawmaker should consider describing specifically for which purposes such storage of anonymised communications content should be allowed – if at all. Moreover, the lawmaker should keep in mind that it is rarely possible to anonymise communications content such as email messages or phone conversations. We also recommend that the lawmaker clarifies the goal and the meaning of the phrase regarding third parties.**

Article 8(1) concerns the protection of information stored in and related to end-users' terminal equipment. Article 8(1) applies, for instance, to cookies. **We recommend that the lawmaker aims for a future-proof scoping of Article 8(1). For instance, Article 8(1) should apply not only to cookies and similar technologies, but also to device fingerprinting.**

**The lawmaker should amend the proposed exception for analytics (web audience measurement), to ensure that privacy and related rights are respected. Furthermore, an exception for necessary security updates is needed. We also recommend that the lawmaker considers adding an exception for employment relationships. As noted previously, Article 8(2), regarding location tracking, should be significantly amended.**

**Regarding Article 9, we recommend that the lawmaker clarifies that the end-user's consent can never legitimise a disproportionate interference with privacy, communications confidentiality, or related rights.** And as noted above, we recommend making compliance with Do Not Track and similar standards obligatory. Even if an obligation to comply with Do Not Track were included in the Regulation, the lawmaker should require privacy-friendly defaults (Article 10). We also recommend a complete or partial ban of tracking walls and other take-it-or-leave-it choices regarding privacy (Article 8, 9, and 10).

Article 11 concerns restrictions on some of the ePrivacy proposal's rights and obligations, including the right to communications confidentiality, for instance for law enforcement. **We recommend that the EU lawmaker adds a duty for providers of electronic communications services to publish statistics about the number of requests they received from authorities.**

**Article 16 concerns unsolicited communications, such as spam. Article 16 should be amended and clarified.** For instance, we recommend that the lawmaker considers extending the scope of Article 16, to protect legal persons against direct marketing communications without consent. **Article 16 should make explicit that withdrawing consent should be at least as easy as giving consent.**

Under Article 17, providers of electronic communications services must alert end-users in case of a particular risk that may compromise the security of networks and services. **We recommend that the EU lawmaker examines whether EU legislation on security of devices should be improved. However, the ePrivacy Regulation may not be the right instrument for such rules. Furthermore, the lawmaker should recognise the value of encryption for the protection of privacy and confidentiality of communications.**

**Article 21, regarding remedies, needs amendments. For instance, collective redress mechanisms (such as in the GDPR) should be made possible.** Article 22 concerns the right to compensation and liability. **Article 22 should be carefully reviewed to ensure that the right to compensation and liability encompasses all foreseeable situations under the ePrivacy Regulation.**

**Article 23 and 24 concern fines and penalties; both provisions require the EU lawmaker's attention.** The maximum fine for violating Article 8 is lower than the maximum fine for violating Articles 5, 6, and 7 of the ePrivacy Regulation. We recommend that the EU lawmaker gives careful consideration to the relative weight of these provisions. We recommend that the lawmaker considers fully harmonising fines and penalties.

**Table 1 below provides an overview of the ePrivacy proposal's provisions. For each provision we indicate the importance of amending the provision.** We recommend major amendments for provisions marked as 'priority 1'. For 'priority 2' and 'priority 3' provisions, amending provisions is slightly less urgent. But we emphasise that the table gives merely a rough indication of priorities: amendments are important for each category. Where no mark has been set in the table, we do not recommend any changes to the provision. **For details, we refer to the comments to each provision in the main text of the study.**

**In conclusion, amendments are especially needed regarding (i) location tracking, (ii) browsers and default settings, (iii) tracking walls, and (iv) the confidentiality of communications.**

**Table 1 : Overview of recommended amendments of the provisions in the ePrivacy proposal**

The table gives a rough indication of the priorities when considering amendments.

|  | Priority 1 | Priority 2 | Priority 3 |
|---|---|---|---|
| Article 1 |  | X |  |
| Article 2 |  | X |  |
| Article 3 |  | X |  |
| Article 4 |  | X |  |
| Article 5 |  | X |  |
| Article 6 | X |  |  |
| Article 7 |  | X |  |
| Article 8 | X |  |  |
| Article 9 | X |  |  |
| Article 10 | X |  |  |
| Article 11 |  | X |  |
| Article 12 |  |  | X |
| Article 13 |  |  | X |
| Article 14 |  |  | X |
| Article 15 |  |  | X |
| Article 16 |  | X |  |
| Article 17 |  | X |  |
| Article 18 |  |  | X |
| Article 19 |  |  | X |
| Article 20 |  |  | X |
| Article 21 |  | X |  |
| Article 22 |  |  | X |
| Article 23 |  |  | X |
| Article 24 |  |  | X |
| Article 25 |  |  | X |
| Article 26 |  |  | X |
| Article 27 |  |  |  |
| Article 28 |  |  |  |
| Article 29 |  |  |  |

**KEY FINDINGS AND RECOMMENDATIONS**

- The ePrivacy proposal has good elements, but should be significantly amended to protect the right to privacy and confidentiality of communications.

- Location tracking, such as Wi-Fi tracking, should only be allowed after people give their consent (with possibly a limited exception for anonymous people counting, if thee are sufficient safeguards for privacy).

- Browsers and similar software should be set to privacy by default. It should be made easier for people to give or refuse consent to online tracking, for instance by requiring companies to comply with the Do Not Track standard.

- Tracking walls and similar take-it-or-leave-it choices regarding privacy should be banned, or banned in certain circumstances.

- Companies should only be allowed to analyse people's communications, such as emails, phone conversation, or chats, or the related metadata, when all end-users give meaningful informed consent, subject to limited, narrow, and specific exceptions. The definition of metadata should be amended.

- Other provisions should also be clarified and amended.

# 1. INTRODUCTION

## 1.1. Privacy, confidentiality of communications, and related rights

### 1.1.1. Privacy and confidentiality of communications

*Everyone has the right to respect for his or her private and family life, home and communications.*

That is the text of Article 7 of the Charter of Fundamental Rights of the European Union, which copies the right almost verbatim from the European Convention on Human Rights.[2] Article 7 of the EU Charter protects, in short, privacy and confidentiality of communications.[3] It follows from the EU Charter that its article 7 offers at least the same protection as Article 8 of the European Convention on Human Rights.[4]

Confidentiality of communications is a right with a long history. King Louis XI of France nationalised the postal service in 1464. Soon the state had organised mail delivery in many European countries. Therefore, many states could read their citizens' letters. In some countries, such as France, the state did systematically read its citizens' letters. In response to such practices, many states in Europe included a right to the confidentiality of correspondence in their constitutions during the nineteenth century. Hence, it was the introduction of a new communication channel (the postal service) that eventually led to the introduction of a new fundamental right.[5]

In the twentieth century, the right to confidentiality of correspondence was extended to a general right to confidentiality of communications in Europe. Article 8 of the European Convention on Human Rights grants everyone 'the right to respect for his private and family life, his home and his *correspondence*'.[6] However, in 1978, the European Court of Human Rights brought telephone calls under the scope of article 8, although the Convention speaks of 'correspondence'.[7]

In 2007, the Court brought internet use under the protection of article 8. After reiterating that phone calls are protected, the Court said that '[i]t follows logically that emails sent from work should be similarly protected under Article 8.'[8] Moreover, the European Court of Human Rights states that Article 8 protects 'information derived from the monitoring of personal internet usage.'[9] The Court added that people have reasonable expectations of privacy regarding their internet use.[10]

All communications are protected under the European Convention on Human Rights – not only private communications. The Court notes that Article 8 'does not use, as it does for the

---

[2] Article 8(1) of the European Convention on Human Rights. The EU Charter of Fundamental Rights uses the more modern and technology neutral term 'communications' instead of 'correspondence', the word used in the European Convention on Human Rights.
[3] In this study, we use 'privacy' and 'private life' interchangeably. Article 7 of the EU Charter of Fundamental Rights and article 8 of the European Convention on Human Rights use the phrase 'respect for private and family life'. See in detail on the (slight) difference between "private life" and "privacy" González Fuster 2014, p. 82- 84; p. 255.
[4] Article 52(3) of the EU Charter of Fundamental Rights.
[5] See on the history of the legal protection of confidentiality of communications Hofman 1995, p. 23 and further; Ruiz 1997, p. 64-70; Steenbruggen 2009, p. 40-44; Arnbak 2016, p.15-54, p. 71-108. See also: EDPS 2016/5, footnote 11 (p. 23) and EDPS 2017/6, p. 36 (footnote 11) for an overview of national constitutions protecting communications confidentiality. See also: Koops et al 2017, section III.D.3, p. 34-37.
[6] Emphasis added.
[7] ECtHR, Klass and others v. Germany, No. 5029/71, 6 September 1978, par. 41.
[8] ECtHR, Copland v. United Kingdom, No. 62617/00, 3 April 2007, par. 41 (capitalisation adapted, internal citations and numbering deleted). See also ECtHR, M.N. and others v. San Marino, No. 28005/12, 7 July 2015, par. 52.
[9] ECtHR, Copland v. United Kingdom, No. 62617/00, 3 April 2007, par. 41 (capitalisation adapted, internal citations and numbering deleted). See also ECtHR, Bărbulescu v Romania, No. 61496/08, 12 January 2016 (referred to Grand Chamber), par. 37; Steenbruggen 2009, p. 136-137.
[10] ECtHR, Copland v. United Kingdom, No. 62617/00, 3 April 2007, par. 42.

word "life", any adjective to qualify the word "correspondence"'.[11] And the protection of Article 8 is not limited to certain communication channels. The European Court of Human Rights states that *all* forms of communications deserve protection:

'In establishing the right of "everyone" to respect for his "correspondence", Article 8 of the Convention protects the confidentiality of "private communications", *whatever the content of the correspondence concerned, and whatever form it may take*. This means that what Article 8 protects is the confidentiality of all the exchanges in which individuals may engage for the purposes of communication.'[12]

The European Court of Human Rights also derives positive duties for states from the Convention.[13] Hence, sometimes the state has to take action to protect people from interferences by other private actors. The Court summarises this as follows.

'The Court reiterates that although the purpose of Article 8 is essentially to protect an individual against arbitrary interference by the public authorities, it does not merely compel the State to abstain from such interference: in addition to this primarily negative undertaking, there may be positive obligations inherent in an effective respect for private life. These obligations may involve the adoption of measures designed to secure respect for private life even in the sphere of the relations of individuals between themselves (…).'[14]

The Charter of Fundamental Rights of the European Union takes a more modern and technology-neutral approach than the European Convention on Human Rights, and protects 'communications' rather than 'correspondence'.[15] The CJEU has confirmed the importance of communications confidentiality, for instance in its *Tele 2 and Watson* judgment.[16] Privacy and communications confidentiality are also protected in, for example, the United Nations Declaration of Human Rights,[17] and the International Covenant on Civil and Political Rights.[18]

The right to communications confidentiality protects the interests of the individual and of society as a whole: (i) the right protects individual (mostly privacy-related) interests of parties who communicate, and (ii) the right protects the trust that society as a whole has in a communication channel.[19]

The current ePrivacy Directive continues the tradition of extending the scope of the right to communications confidentiality to new communication channels.[20] The ePrivacy Directive requires, in short, Member States to ensure the confidentiality of communications and the related traffic data by means of publicly available electronic communications services. In particular, Member States must prohibit tapping, storage or other types of communications surveillance, without the consent of the users.[21] Hence, the provision emphasises Member States' positive obligations regarding confidentiality of communications.[22] In practice, this communications confidentiality in the current ePrivacy Directive applies mostly to internet access providers and phone companies.

---

[11] ECtHR Niemietz v. Germany, No. 13710/88, 16 December 1992, par 32 (internal citations omitted).
[12] ECtHR, Michaud v. France, No. 12323/11, 6 December 2012, par. 90 (internal citations omitted; emphasis added).
[13] See generally on positive obligations: Arnbak 2016, p. 77; Akandji-Kombe 2007; De Hert 2011.
[14] ECtHR, Bărbulescu v Romania, No. 61496/08, 12 January 2016, referred to Grand Chamber on 6 June 2016. See also: ECtHR, Z v. Finland, No. 22009/93, 25 February 1997, par. 36. See also ECtHR, Mosley v. United Kingdom, No. 48009/08, 10 May 2011, par 106.
[15] Article 7 of the EU Charter of Fundamental Rights.
[16] CJEU (Grand Chamber), Judgment of 21 December 2016, cases C-203/15 (Tele2 Sverige AB) and C-698/15 (Watson), ECLI:EU:C:2016:970.
[17] Article 12 of the Universal Declaration of Human Rights, G.A. Res. 217A (III), U.N. Doc. A/810 at 71 (1948).
[18] Article 17 of the International Covenant on Civil and Political Rights, Dec. 16, 1966, S. Treaty Doc. No. 95-20, 6 I.L.M. 368 (1967), 999 U.N.T.S. 171.
[19] Steenbruggen 2009, p. 44-49; p. 354. Asscher 2002, p. 18. Many scholars argue that privacy is also an important value for society, rather than merely an individual interest. See e.g. Regan 1995.
[20] The current ePrivacy Directive was last amended in 2009, by the Directive 2009/136/EC of the European Parliament and of the Council of 25 November 2009 amending (…) Directive 2002/58/EC concerning the processing of personal data and the protection of privacy in the electronic communications sector (…) (OJ L 337, 18.12.2009, p. 11–36).
[21] Article 5(1) of the 2009 ePrivacy Directive.
[22] Steenbruggen 2009, p. 176; p. 356.

The 2017 ePrivacy proposal clarifies that communications confidentiality must also be respected in the context of new communication services such as webmail and WhatsApp.[23] This extension of the protection afforded to the right to communications confidentiality by EU primary legislation is valuable and resolves a certain degree of ambiguity present in the existing rules in the ePrivacy Directive.

While the current ePrivacy Directive is not, or is only partly, applicable for many new communications services, case law of the European Court of Human Rights confirms that any type of communications provider must respect communications confidentiality.[24] Empirical research confirms the importance of privacy and communications confidentiality. In a recent Eurobarometer survey, '72% state that it is very important that the confidentiality of their e-mails and online instant messaging is guaranteed.'[25]

### 1.1.2. Related rights

The right to privacy and to communications confidentiality is essential for the exercise of other fundamental rights such as freedom of expression and freedom of assembly.[26] The early history of the right to communications confidentiality illustrates the connection between that right and the right to freedom of expression. Nowadays the right to confidentiality of communications is often regarded as a privacy-related right.[27] But when it was developed in the late eighteenth century, confidentiality of correspondence was seen as an auxiliary right to safeguard freedom of expression.[28]

Privacy and communications confidentiality are essential for enjoyment of the right to receive and impart information.[29] As Frank La Rue, the former Special Rapporteur on the promotion and protection of the right to freedom of opinion and expression for the United Nations, notes:

'States cannot ensure that individuals are able to freely seek and receive information or express themselves without respecting, protecting and promoting their right to privacy. Privacy and freedom of expression are interlinked and mutually dependent; an infringement upon one can be both the cause and consequence of an infringement upon the other.'[30]

The CJEU notes in its 2016 *Tele 2 and Watson* judgment regarding national data retention laws: 'the retention of traffic and location data could (…) have an effect on the use of means of electronic communication and, consequently, on the exercise by the users thereof of their freedom of expression.'[31] Along similar lines, the European Court of Human Rights states that the 'menace of surveillance can be claimed in itself to restrict free communication.'[32] These cases concern data collection for law enforcement, but similar conclusions can be drawn about commercial data collection.

---

[23] See our comment on Article 4(1)(b), on 'electronic communications service'.
[24] ECtHR, Michaud v. France, No. 12323/11, 6 December 2012, par. 90.
[25] Explanatory memorandum to the ePrivacy proposal, p. 6 (section 3.2). See Eurobarometer 2016, p. 33.
[26] See Article 10, 11, and 12 of the EU Charter of Fundamental Rights. See also Explanatory memorandum to the ePrivacy proposal, p. 9.
[27] See for instance article 7 of the EU Charter of Fundamental Rights.
[28] Ruiz 1997, p. 67. See also Steenbruggen 2009, p. 48-49; p. 75. See also ECtHR, Autronic AG v. Switzerland, No. 12726/87, 22 May 1990, par. 47. See also Arnbak 2016, p. 95: 'The free flow of information is one of the classic rationales of protecting the confidentiality of communications.'
[29] See Steenbruggen 2009, p. 46-49. See also (in the US:) Cohen 1995; Richards 2014.
[30] La Rue 2013, p. 20.
[31] CJEU (Grand Chamber), Judgment of 21 December 2016, cases C-203/15 (Tele2 Sverige AB) and C-698/15 (Watson), ECLI:EU:C:2016:970, par. 101. See also CJEU 8 April 2014, C-293/12 (Digital Rights Ireland), ECLI:EU:C:2014:238, par. 28.
[32] ECtHR, Klass and others v. Germany, No. 5029/71, 6 September 1978, par. 37. See also ECtHR, Liberty and others v. United Kingdom, No. 58243/00, 1 July 2008, par. 56. See also par. 104-105.

Privacy and communications confidentiality are essential for freedom of thought, conscience, and religion, and for freedom of assembly. Those rights would be threatened if people's fear off eaves-dropping or surveillance would lead to people not feeling free to discuss their thoughts or their assemblies.[33] Moreover, as the EU lawmaker notes, privacy anxieties hinder online business.[34]

## 1.2.    Scope of the study

The current ePrivacy Directive aims to protect values such as privacy and communications confidentiality in the electronic communications sector. The ePrivacy Directive's objectives and principles remain sound. But unfortunately, the ePrivacy Directive leaves considerable gaps in user protection.[35] Therefore, after extensive evaluations and deliberations,[36] the European Commission presented a proposal for an ePrivacy Regulation in January 2017. That proposal is the topic of this study.

This is a legal study. The request by the Policy Department for Citizens' Rights and Constitutional Affairs provided the following research objective:

'The study should assess and evaluate the Commission's ePrivacy proposal. The research should assess whether the proposal would ensure that the right to the protection of personal data, the right to respect for private life and communications, and related rights enjoy a high standard of protection, and should assess the proposal's potential benefits and drawbacks more generally. It should also assess whether the proposal ensures consistency with the General Data Protection Regulation (GDPR) and does not lower the level of protection set for by the GDPR.'

While interesting and important, certain questions are outside the scope of this study. We do not discuss questions such as: would it have been better to include the rules on confidentiality of communications in the GDPR? Would it be better to adopt one regulation that includes all the rules on direct marketing, including unsolicited communications, behavioural targeting, and other types of targeted marketing? Is there a need for additional EU legislation on security of devices and communications? Such questions could be examined in other studies.

For several key definitions, the ePrivacy proposal refers to the 'Directive establishing a European Electronic Communications Code', which is being drafted at the same time as the ePrivacy Regulation. For this study, we rely on the draft text of the 'Directive establishing the European Electronic Communications Code, dated 12 October 2016.[37] We recommend that the lawmaker considers defining the main concepts in the ePrivacy Regulation itself, rather than in the European Electronic Communications Code. If the lawmaker chooses to define main concepts in the European Electronic Communications Code, we recommend that, while working on the ePrivacy proposal, the EU lawmaker pays close attention to the development of that Code.[38]

---

[33] See Arnbak 2016, p. 96.
[34] See e.g. Recital 5 of the 2009 ePrivacy Directive; ePrivacy proposal 2017, p. 1. See also ePrivacy Impact Assessment 2017 Pt. 1, p. 16: 'Tracking of surfing behaviour is expected to grow more pervasive in the coming years. (...) The consequence could be to reduce trust in the digital economy and reinforce citizens' feeling of being powerless, i.e. not protected by the law.' See also EDPS 2017/6, p. 6.
[35] See Recital 6 of the ePrivacy proposal, and section 2.3 of its explanatory memorandum.
[36] See for an overview of the evaluation of the ePrivacy Directive, stakeholder consultations, and expertise sought by the Commission: Explanatory memorandum to the ePrivacy proposal, p. 5-8, chapter 3.
[37] Proposal for a Directive of the European Parliament and of the Council establishing the European Electronic Communications Code (Recast) COM/2016/0590 final/2 - 2016/0288 (COD), Corrigendum, <http://eur-lex.europa.eu/legal-content/EN/ALL/?uri=comnat:COM_2016_0590_FIN>.
[38] See our comment on Article 4(1)(b).

The Article 29 Working Party and the European Data Protection Supervisor (EDPS) have published thorough and detailed opinions on the ePrivacy proposal.[39] The Article 29 Working Party is an advisory body in which national Data Protection Authorities cooperate. National authorities have much practical experience with ePrivacy and data protection rules.[40] The EDPS is the EU's independent Data Protection Authority, and advises EU institutions and bodies on all matters concerning personal data processing.[41] The opinions of the Working Party and the EDPS give valuable input for this study. However, we do not necessarily agree with those opinions on all points. In December 2016, a draft of the ePrivacy Regulation was published by and discussed in the media.[42] On occasion, we refer to differences between the ePrivacy Proposal and that draft text.

We compliment the Commission on its extensive preparatory work, and for the ambitious and important proposal, which delivers important improvements with respect to the current framework for electronic communications privacy.

However, the ePrivacy proposal also has its weaknesses. In this study, we highlight some important weaknesses of the ePrivacy proposal, and recommend how they could be addressed. Because of time constraints, we do not discuss every possible aspect of the proposal.

The study builds on, and includes some sentences from, earlier work of the authors.[43] We would like to thank Prof. Nico Van Eijk and Prof. Natali Helberger for comments on the draft text. We would also like to thank Dr. H. Asghari, Dr. Sophie Boerman, Dr. Sanne Kruikemeijer, and Prof. Tal Zarsky.

In the next section, we address the question of whether ePrivacy rules are necessary, in addition to the GDPR.

## 1.3. The need for ePrivacy rules in addition to the GDPR

It is sometimes suggested that a separate ePrivacy Regulation is not needed alongside to the GDPR.[44] However, from a legal perspective, separate ePrivacy rules are not only useful – but also necessary. Since the 1990s, the EU lawmaker has realised that it is necessary to provide harmonised privacy and communications confidentiality rules for the electronic communications sector. In addition to the 1995 Data Protection Directive, the EU adopted a telecommunications privacy Directive in 1997.[45] In 2002, that Directive was replaced by the ePrivacy Directive, which was intended to be more in line with new technologies.[46] In 2009, the ePrivacy Directive was updated by the Citizens' Rights Directive.[47] It makes sense to update the ePrivacy regime again, now that the GDPR will replace the Data Protection Directive.[48] Moreover, developments in the electronic communications landscape require an update of certain parts of the existing ePrivacy Directive.

---

[39] Article 29 Working Part 2017 (WP247); EDPS 2017/6.
[40] See on the Article 29 Working Party: Gutwirth and 7. 2008; Hijmans 2016, chapter 7.
[41] See on the EDPS: Hijmans 2006; Hijmans 2016, chapter 8(4).
[42] December 2016 draft – European Commission, Proposal for a Regulation of the European Parliament and of the Council concerning the respect for private life and the protection of personal data in electronic communications and repealing Directive 2002/58/EC ('Privacy and Electronic Communications Regulation') […](2016) XXX draft (published 14 December 2016) <https://www.politico.eu/wp-content/uploads/2016/12/POLITICO-e-privacy-directive-review-draft-december.pdf> accessed 5 May 2017.
[43] See Van Hoboken and Zuiderveen Borgesius 2015; Zuiderveen Borgesius 2015; Zuiderveen Borgesius FJ 2015a; Zuiderveen Borgesius FJ 2015c, Zuiderveen Borgesius et al 2017b.
[44] For instance, 63.4% of industry respondents to the consultation by the European Commission see no need for special rules for the electronic communications sector on confidentiality of electronic communications (Explanatory memorandum ePrivacy proposal, p. 6, section 3.2).
[45] Council Directive 97/66/EC of 15 December 1997 concerning the processing of personal data and the protection of privacy in the telecommunications sector (ISDN Directive) (OJ L 024, 30.1.1998, p. 1-8). See about the history of the ePrivacy rules: De Hert and Papakonstantinou 2011.
[46] See recital 4 of the ePrivacy Directive.
[47] Council Directive 2009/136/EC of 25 November 2009 amending Directive 2002/22/EC on universal service and users' rights relating to electronic communications networks and services, Directive 2002/58/EC concerning the processing of personal data and the protection of privacy in the electronic communications sector and Regulation (EC) No 2006/2004 on cooperation between national authorities responsible for the enforcement of consumer protection laws (Citizen's Rights Directive) (OJ L 337, 18.12.2009, p. 11–36).

Separate ePrivacy rules are useful and necessary for at least four reasons. First, EU law should protect the fundamental right to privacy and communications confidentiality.[49] Confidentiality and privacy of communications should also be protected when no personal data are involved.[50]

Therefore, the GDPR (which only regulates personal data) does not suffice to protect communications confidentiality. The Charter of Fundamental Rights of the European Union contains a right to the protection of personal data (Article 8): 'Everyone has the right to the protection of personal data concerning him or her.'[51] The GDPR implements Article 8 of the EU Charter, and provides rules that aim to ensure that personal data are only processed lawfully, fairly, and transparently.[52] The GDPR does not aim to protect the right to communications confidentiality, or to protect the right to privacy in general. As the Impact Assessment to the ePrivacy proposal notes, 'data protection rules do not protect, as a rule, the confidentiality of information relating only to legal persons, for instance information such as business secrets or trade negotiations.'[53]

For instance, companies can email each other confidential information, such as trade secrets, bookkeeping figures, or an instruction to buy stocks at a stock exchange. Companies are legal persons (not protected by the GDPR[54]) and not all their communications include personal data. Suppose that that info@large-company.com sends an email with trade secrets (without personal data) to info@big-company.com. In that scenario, the GDPR may not apply. But many companies want such communications to remain confidential.

Moreover, the Court of Justice of the European Union[55] and the European Court of Human Rights[56] confirm 'that professional activities of legal persons may not be excluded from the protection of the right guaranteed by Article 7 of the EU Charter and Article 8 of the ECHR.'[57]

A second reason why separate ePrivacy rules are necessary concerns the internal market. The ePrivacy proposal aims to harmonise rules on the right to privacy and electronic communications confidentiality in the interest of the Digital Single Market.[58] If the EU does not harmonise the regulations protecting privacy and the right to communications confidentiality in the area of electronic communications, Member States are likely to adopt rules to protect those interests. Each Member State might adopt different rules. Such a situation could lead to a patchwork of different national rules, which could hinder the internal market.[59]

Third, separate ePrivacy rules can improve legal clarity. The GDPR contains many general provisions with open norms, because the GDPR lays down an omnibus regime and aims to cover many different situations. Such open norms are too vague for the data and situations regulated in the ePrivacy proposal. For instance, the ePrivacy proposal applies to the content of communications (emails, phone calls, WhatsApp messages, etc.), and the data stored on people's phones. To protect people in such sensitive situations, and to protect such sensitive information, the open norms in the GDPR do not suffice.

---

[48] See recital 173 of the GDPR: the ePrivacy 'Directive 2002/58/EC should be reviewed in particular in order to ensure consistency with this Regulation'.
[49] Article 7 of the EU Charter of Fundamental Rights.
[50] See Arnbak 2016, p. 225.
[51] See generally on the right to protection of personal data: González Fuster 2014.
[52] Article 5(1)(a) of the GDPR.
[53] ePrivacy Impact Assessment 2017 Pt. 1, p. 8.
[54] Recital 14 of the GDPR.
[55] Original footnote: 'See C-450/06 Varec SA, ECLI:EU:C:2008:91, §48.'
[56] See for instance ECtHR, Niemietz v. Germany, No. 13710/88, 16 December 1992, par. 32.
[57] Explanatory memorandum ePrivacy proposal, p. 4, section 2.1 ('confirm' changed to 'confirms' by the authors). See also EDPS 5/2016, p. 7.
[58] Article 1 of the ePrivacy proposal.
[59] See EDPS 2016/5, p. 7.

Fourth, the ePrivacy proposal contains (like the current ePrivacy Directive) useful rules that regulate specific situations in the electronic communications context that are not covered by the GDPR. For instance, the GDPR only partly protects user's devices (such as phones, computers, or connected 'smart' fridges),[60] and only partly protects people against unsolicited communications.[61]

The necessity of ePrivacy rules was broadly confirmed during the stakeholder consultation organised by the Commission: '83.4% of the responding citizens, consumer and civil society organisations and 88.9% of public authorities agree [that there is a need] for special rules for the electronic communications sector on confidentiality of electronic communications'.[62] However, 63.4% of industry respondents disagree.[63] In conclusion, rules such as those included in the ePrivacy proposal are both useful and necessary, alongside the GDPR.

## 1.4.   Outline of the study

In the next chapters, we comment on provisions in the ePrivacy proposal, in roughly the order they appear in the proposal.

We have copied each provision of the ePrivacy proposal into this study. These quotations lead to a rather long study, however this approach enables people to read the study without having to consult the ePrivacy proposal. Most provisions are followed by a comments and recommendations. In the executive summary we highlight the main weaknesses of the proposal, and suggest priorities when amending the proposal.

---

[60] See Article 8 of the ePrivacy proposal. The GDPR does apply to 'personal data' that are stored on users' devices, and could thus provide some protection in certain situations.
[61] See Article 16 of the ePrivacy proposal.
[62] Explanatory memorandum ePrivacy proposal, p. 6, section 3.2. See also European Digital Rights 2017.
[63] Explanatory memorandum ePrivacy proposal, p. 6, section 3.2.

# 2. GENERAL PROVISIONS (PROPOSAL CHAPTER I)

## 2.1. Article 1, subject matter

### 2.1.1. Article 1(1), protecting fundamental rights

Article 1(1) of the ePrivacy proposal states:

'This Regulation lays down rules regarding the protection of fundamental rights and freedoms of natural and legal persons in the provision and use of electronic communications services, and in particular, the rights to respect for private life and communications and the protection of natural persons with regard to the processing of personal data.'

See also Recitals 1-13.

**Comment**
The European Commission has proposed a Regulation, rather than a Directive. The Commission chose a Regulation 'to ensure consistency with the GDPR and legal certainty for users and businesses alike by avoiding divergent interpretation in the Member States. A Regulation can ensure an equal level of protection throughout the Union for users and lower compliance costs for businesses operating across borders.'[64] Most stakeholders, including citizens, consumer and civil society organisations, industry, and public authorities, also prefer a Regulation to a Directive.[65] **Indeed, we recommend adopting a Regulation, rather than a Directive.[66]**

Article 1(1) explicitly mentions privacy and communication confidentiality (Article 7 of the EU Charter) and personal data (Article 8 of the EU Charter). As the Commission notes in the explanatory memorandum to the ePrivacy proposal: 'Effective protection of the confidentiality of communications is essential for exercising the freedom of expression and information and other related rights, such as the right to personal data protection or the freedom of thought, conscience and religion.'[67]

**We recommend that the EU lawmaker clarifies in Article 1(1) and in the preamble that the ePrivacy Regulation also protects the right to impart and receive information (Article 11 of the EU Charter). The EU lawmaker should consider adding references to freedom of thought, conscience and religion, and to freedom of assembly and association as well (Article 10 and 12 of the EU Charter).**

The explanatory memorandum to the ePrivacy proposal notes that the Digital Single Market Strategy and the review of the ePrivacy rules aim 'to provide a high level of privacy protection for users of electronic communications services and a level playing field for all market players.'[68] However, neither the ePrivacy proposal's preamble nor its provisions state that the ePrivacy rules aim for a 'high level' of protection. In contrast, the GDPR[69] and the current Data Protection Directive[70] state in their preambles that they aim for a 'high level of the protection' of personal data.[71] Such phrases can assist courts and stakeholders when interpreting legal provisions. **The EU lawmaker should therefore consider including a**

---

[64] Explanatory memorandum ePrivacy proposal, p. 4, section 2.4.
[65] European Commission 2016, p. 9. See also EDPS 2017/6, p. 8.
[66] See also EDPS 2017/6, p. 7.
[67] Explanatory memorandum to the ePrivacy proposal, p. 9.
[68] Explanatory memorandum to the ePrivacy proposal, p. 2 (section 1.1).
[69] Recital 6 of the GDPR.
[70] Recital 10 of the Data Protection Directive.
[71] The ePrivacy Directive does not contain a similar "high protection" phrase in its preamble. However, the explanatory memorandum of the ePrivacy Directive does contain that phrase. See CJEU (Grand Chamber), Judgment of 21 December 2016, cases C-203/15 (Tele2 Sverige AB) and C-698/15 (Watson), ECLI:EU:C:2016:970, par. 82.

**similar 'high level of protection' phrase in the ePrivacy Regulation's preamble.** The sentence from the explanatory memorandum (quoted above) could provide inspiration.

### 2.1.2.    Article 1(2) and 1(3), internal market, and relationship to GDPR

Articles 1(2) and 1(3) of the ePrivacy proposal state:

'(2) This Regulation ensures free movement of electronic communications data and electronic communications services within the Union, which shall be neither restricted nor prohibited for reasons related to the respect for the private life and communications of natural and legal persons and the protection of natural persons with regard to the processing of personal data. (3) The provisions of this Regulation particularise and complement [the GDPR] by laying down specific rules for the purposes mentioned in paragraphs 1 and 2.'

See also Recital 5.

### Comment

Like the GDPR[72] and the current ePrivacy Directive,[73] the ePrivacy proposal has a dual aim: (i) protecting fundamental rights, and (ii) fostering the internal market.

The provisions of the ePrivacy Regulation 'particularise and complement' the GDPR. The relationship between the ePrivacy proposal and the GDPR is roughly as follows. When both instruments would apply, the ePrivacy Regulation *particularises* the GDPR (as a *lex specialis*). If the ePrivacy Regulation applies to a situation outside the scope of the GDPR, the ePrivacy Regulation *complements* the GDPR.

The phrase 'Lex specialis derogat legi generali' implies: 'whenever two or more norms deal with the same subject matter, priority should be given to the norm that is more specific.'[74] The current ePrivacy Directive often, but not always, functions as a *lex specialis* to general data protection law (the Data Protection Directive[75]).[76]

Like the current ePrivacy Directive, the ePrivacy proposal often, but not always, functions as a *lex specialis* to the GDPR. The GDPR applies, in short, as soon as personal data are processed.[77] The ePrivacy proposal would function as a *lex specialis* to the GDPR under certain circumstances. First, suppose that the GDPR applies to a situation in which electronic communications metadata are processed, which are also personal data. Second, suppose that the ePrivacy Regulation also applies, because electronic communications metadata are processed and Article 6(2) of the ePrivacy Regulation applies. In such a situation, the rule in the ePrivacy Regulation functions as a *lex specialis* to a rule in the GDPR, the requirement for a legal basis for processing.[78] After all, the ePrivacy Regulation and the GDPR deal with the same subject matter (the legal basis for processing) in such a situation. One could also say: in such a situation the ePrivacy Regulation *particularises* the GDPR. Hence, in such a situation, priority should be given to the ePrivacy rule, as that is more specific than the GDPR rule.

But suppose that in a certain situation the ePrivacy Regulation applies, and the GDPR does not apply. For instance, companies might exchange trade secrets over email. If no personal data are included in the emails or otherwise processed, the GDPR would not apply.[79] But the

---

[72] Article 1 of the GDPR.
[73] Article 1(1) of the ePrivacy Directive 2009.
[74] Koskenniemi 2006, p. 8.
[75] Directive 95/46/EC of the European Parliament and of the Council of 24 October 1995 on the protection of individuals with regard to the processing of personal data and on the free movement of such data (OJ L 281, 23.11.1995 p. 31–50).
[76] See: Kotschy 2014; Zuiderveen Borgesius 2015a.
[77] Article 1(3) of the GDPR. There are some exceptions: see Article 2(2) of the GDPR.
[78] Article 6(1) of the GDPR.
[79] See section 1.3 above.

ePrivacy Regulation (confidentiality of communications, Article 5) would apply. In such a situation, the ePrivacy Regulation does not function as a *lex specialis* to the GDPR. Hence, in such a situation the ePrivacy Regulation *complements* (but does not particularise) the GDPR. After all, the GDPR is irrelevant in this hypothetical situation. In sum, the ePrivacy Regulation often, but not always, functions as a *lex specialis* of the GDPR.[80]

**While we think that the basic idea of Article 1(3) and Recital 5 should be retained, we recommend that the EU lawmaker considers several amendments.**

**Recital 5 states: 'This Regulation (...) does not lower the level of protection enjoyed by natural persons under [the GDPR].' We recommend that the EU lawmaker adds a similar sentence to Article 1(3) of the ePrivacy Regulation.** The EDPS suggests using the following phrase (new words emphasised): 'This Regulation does not lower the level of protection enjoyed by natural persons under [the GDPR] – *to the contrary, where appropriate, it aims to provide additional, and complementary, safeguards considering the need for additional protection for the confidentiality of communications'.*[81]

**We also strongly recommend that the EU lawmaker considers the advice of the Article 29 Working Party and the EDPS with respect to the following legal particularities in the relationship between the general data protection regime and the new ePrivacy rules.** For instance, the Working Party calls upon the EU lawmaker to clarify that 'the prohibitions under the ePrivacy Regulation take precedence over permissions under the GDPR'.[82] Furthermore: 'Even if the processing is allowed under any exception (including consent) to the prohibitions under the ePrivacy Regulation, this processing, where it concerns personal data, still needs to comply with all relevant provisions in the GDPR'[83]

The EDPS adds that 'Personal data collected based on end-user consent or another legal ground under the ePrivacy Regulation must not be subsequently further processed outside the scope of such consent or exception on a legal ground which might otherwise be available under the GDPR, but not under the ePrivacy Regulation'[84] Therefore, the EDPS suggests that the EU lawmaker adds two provisions to the ePrivacy Regulation:

(i) 'Neither providers of electronic communications services, nor any third parties, shall process personal data collected on the basis of consent or any other legal ground under the ePrivacy Regulation, on any other legal basis not specifically provided for in the ePrivacy Regulation.'[85]

(ii) 'When the processing is allowed under any exception to the prohibitions under the ePrivacy Regulation, any other processing on the basis of Article 6 of the GDPR shall be considered as prohibited, including processing for another purpose on the basis of Article 6(4) of the GDPR. This would not prevent controllers from asking for additional consent for new processing operations'.'[86]

**We recommend that the EU lawmaker seriously considers adding the provisions suggested by the EDPS (or similar provisions) to the ePrivacy proposal.** Such provisions do not necessarily have to be added to Article 1.

**We also recommend that the lawmaker carefully considers whether additional specific exceptions are necessary.** The Article 29 Working Party and the EDPS suggest that it might be necessary to add narrow exceptions to the ePrivacy Regulation, for example, for processing for scientific purposes or to protect 'vital interests' of individuals.[87]

---

[80] The explanatory memorandum of the ePrivacy proposal says 'This proposal is lex specialis to the GDPR' (p. 2). That sentence seems to be phrased in an unfortunate manner, as, in our view, the ePrivacy proposal would *sometimes* function as a lex specialis to the GDPR.
[81] EDPS 2017/6, p. 15.
[82] Article 29 Working Party 2017 (WP247), p. 15-16.
[83] Article 29 Working Party 2017 (WP247), p. 15-16 (capitalisation adapted).
[84] EDPS 2017/6, p. 10. See also Article 29 Working Party 2017 (WP247), p. 16.
[85] EDPS 2017/6, p. 16 (capitalisation adapted).
[86] EDPS 2017/6, p. 16 (capitalisation adapted).

The EU lawmaker should also consider making more explicit, if possible, which provisions 'particularise' and which provisions 'complement' the GDPR.

**Lastly, the lawmaker should consider stating that certain principles of general data protection law (set out in Article 5 of the GDPR) should guide the interpretation of the ePrivacy rules to the extent that it covers the processing of personal data.** Relevant principles may include fairness and transparency,[88] data minimisation,[89] security,[90] and storage limitation.[91] However, other data protection principles may not be appropriate as guiding principles for the interpretation of the ePrivacy Regulation. For instance, the ePrivacy rules provide a stricter version of the purpose limitation principle than the statement of this principle in the GDPR.

**In sum, regarding Article 1(3), our main recommendations are: (i) state in Article 1 that the ePrivacy Regulation does not lower the level of protection enjoyed by natural persons under the GDPR; and (ii) consider the advice of the Article 29 Working Party and the EDPS with respect to the further clarification of the relationship between the GDPR and the ePrivacy regime.**

## 2.2. Article 2, material scope

### 2.2.1. Article 2(1), general scope

Article 2(1) of the ePrivacy proposal states:

'This Regulation applies to the processing of electronic communications data carried out in connection with the provision and the use of electronic communications services and to information related to the terminal equipment of end-users.'

See also Recitals 8-13.

**Comment**
Article 2 establishes the material scope of the ePrivacy Regulation. Recital 8 adds:

'This Regulation should apply to providers of electronic communications services, to providers of publicly available directories, and to software providers permitting electronic communications, including the retrieval and presentation of information on the internet. This Regulation should also apply to natural and legal persons who use electronic communications services to send direct marketing commercial communications or collect information related to or stored in end-users' terminal equipment.'

The current wording of Article 2(1) may be too narrow to cover the provisions on publicly available directories and unsolicited communications (Articles 15 and 16). It is unclear whether these provisions are covered by the phrase 'the processing of electronic communications data carried out in connection with the provision and the use of electronic communications services' (Article 2(1)). Hence, it seems the scope of Article 2(1) is narrower than the scope suggested in Recital 8. **We therefore recommend that the EU lawmaker clarifies the scope of Article 2(1); the scope of that provision seems to be too narrow.**

---

[87] Article 29 Working Party 2017 (WP247), p. 15-16. See also EDPS 2017/6, p. 16.
[88] Article 5(1)(a) of the GDPR.
[89] Article 5(1)(c) of the GDPR.
[90] Article 5(1)(f) of the GDPR.
[91] Article 5(1)(e) of the GDPR.

The wording of the scope in Article 2(1) of the ePrivacy proposal differs from that in Article 3(1) of the ePrivacy proposal, which stipulates the territorial scope. To describe activities to which the ePrivacy Regulation applies, Article 2(1) uses the phrase 'the processing of electronic communications data carried out in connection with the provision and the use of electronic communications services (…)'. Article 3(1) uses 'the provisions of electronic communications services to end-users' to describe activities to which the ePrivacy Regulation applies. **The EU lawmaker should consider whether the same wording could be used in both articles to increase certainty as to the ePrivacy Regulation's scope. It may also be advisable to split Article 2(1) into three paragraphs in a similar way to Article 3(1).**

**The EU lawmaker should also consider adding the wording 'irrespective of whether a payment of the end-user is required' to Article 2(1), in a similar fashion to Article 3(1).** The addition of these words would clarify that the ePrivacy Regulation applies to services that are offered to users, without requiring direct monetary payment from those users. For instance, many webmail providers do not charge a fee to their users.

### 2.2.2. Article 2(2), limitation of scope

Article 2(2) limits the scope of the ePrivacy Regulation, and reads as follows:

'This Regulation does not apply to:
(a) activities which fall outside the scope of Union law;
(b) activities of the Member States which fall within the scope of Chapter 2 of Title V [on 'specific provisions on the common foreign and security policy'] of the Treaty on European Union;
(c) electronic communications services which are not publicly available;
(d) activities of competent authorities for the purposes of the prevention, investigation, detection or prosecution of criminal offences or the execution of criminal penalties, including the safeguarding against and the prevention of threats to public security'.[92]

See also Recitals 8-13.

**Comment**
Exceptions (a), (b), and (d) are in line with the exceptions in the GDPR. The GDPR has an exception for processing by a natural person in the course of a purely personal or household activity.[93] The ePrivacy proposal does not contain such a general household exception. Since most of the proposal's provisions are not directed at individuals but at service providers, the ePrivacy Regulation does not need a general household exception. However, as this study notes, an exception could be useful for certain ePrivacy provisions that could have an unduly restrictive impact on how individuals interact with communications services in their private sphere.[94]

Exception (c) implies that the ePrivacy Regulation only applies to 'publicly available' electronic communications services. Recital 13 of the ePrivacy proposal discusses exception (c) and 'publicly available' services:

'The development of fast and efficient wireless technologies has fostered the increasing availability for the public of internet access via wireless networks accessible by anyone in public and semi-private spaces such as 'hotspots' situated at different places within a city, department stores, shopping malls and hospitals. To the extent that those communications networks are provided to an undefined group of end-users, the confidentiality of the communications transmitted through such networks should be protected. The fact that wireless electronic communications services may be ancillary to other services should not stand in the way of ensuring the protection of confidentiality of communications data and application of this Regulation. Therefore, this Regulation should apply to electronic communications data using electronic communications services and public communications networks. In contrast, this Regulation should not apply to closed groups of end-users such as corporate networks, access to which is limited to members of the corporation.'

---

[92] Amendments between square brackets by the authors.
[93] Article 2(2)(c) of the GDPR.
[94] See our comment on Article 6.

Recital 13 of the ePrivacy proposal partly solves the ambiguities regarding services 'which are not publicly available'. The exclusion of services 'which are not publicly available' has led to discussions in the past. The Article 29 Working Party noted in 2008 that the distinction between private and public networks and services is difficult to make.[95]

The Article 29 Working Party and the EDPS call for further clarification. The regulators say that the EU lawmaker should consider clarifying that secured wireless networks that are provided to the public fall within the scope of the ePrivacy Regulation. Recital 13 adds 'communications networks (…) provided to an undefined group of end-users' as the criterion for determining whether a specific network falls within the scope of the ePrivacy Regulation. **It should be clarified when networks that are secured via a password or similar authentication method fall within the scope of the ePrivacy Regulation.** Access to such a secured network can be provided to a group of end-users whose size and individual identities cannot be established in advance, for example in a café which gives out the network password to its customers.[96]

**The EU lawmaker could further reduce ambiguity regarding 'services which are not publicly available'.** For instance, many social network sites (that offer chat functions) require people to open an account to chat with other users. The lawmaker could clarify that such a service should be regarded as 'publicly available' if people have to open an account.

According to the EDPS, Recital 13 should 'also clarify what should be considered as "publicly accessible". For example, it should be made clear that a service remains considered publicly accessible even if the provider limits the service to registered users such as in the case of an organisation offering Wi-Fi access to its customers and visitors.'[97]

The EDPS also calls for further clarification and examples in Recital 13. Such examples 'should include Wi-Fi services in hotels, restaurants, coffee shops, shops, trains, airports and networks offered by universities to their students, as well as corporate Wi-Fi access offered to visitors and guests, and hotspots created by public administrations.'[98]

Regarding exception (d): it is important to clarify the relationship between Article 2(2)(d) and Article 11 of the ePrivacy proposal in line with the interpretation of the CJEU. One the one hand, under Article 2(2)(d) of the ePrivacy proposal, law enforcement activities and the execution of criminal penalties are exempted from the scope of the ePrivacy proposal. On the other hand, Article 11 of the ePrivacy proposal defines the conditions under which member states can deviate from the ePrivacy proposal and pass legislation, such as data retention laws. The CJEU ruled in its *Tele2 and Watson* judgment on the ePrivacy Directive:

---

[95] 'Services are increasingly becoming a mixture of private and public elements and it is often difficult for regulators and for stakeholders alike to determine whether the e-Privacy Directive applies in a given situation. For example, is the provision of internet access to 30.000 students a public electronic communication system or a private one? What if the same access is provided by a multinational company, to tens of thousands of employees? What if it is provided by a cybercafé?' Article 29 Working Party 2008 (WP150), p. 4. See also: European Commission, Time.Lex, and Spark legal network and consultancy ltd 2015, p. 24-32.
[96] Article 29 Working Party 2017 (WP247), p. 27; EDPS 2017/6, p. 8; p. 25. See also Article 29 Working Party 2016 (WP240), p 8.
[97] EDPS 2017/6, p. 25.
[98] EDPS 2017/6, p. 25.

'Article 15(1) [of the ePrivacy Directive] necessarily presupposes that the national measures referred to therein, such as those relating to the retention of data for the purpose of combating crime, fall within the scope of that directive, since it expressly authorises the Member States to adopt them only if the conditions laid down in the directive are met.'[99]

**The EU lawmaker should consider transposing the *Tele2 and Watson* judgment into the ePrivacy proposal. The EU lawmaker should consider adding the following phrase to Article 2(2)(d): 'without prejudice to Article 11' (or a similar phrase).** See also our comment on Article 11 (restrictions).

**In sum, we recommend that the EU lawmaker clarifies Article 2(2).**

### 2.2.3. Article 2(3), processing by EU bodies etc.

Article 2(3) states:

'The processing of electronic communications data by the Union institutions, bodies, offices and agencies is governed by [a new Regulation replacing Regulation 45/2001].'[100]

#### Comment
Regulation 45/2001 concerns the protection of individuals with regard to the processing of personal data by the EU institutions and bodies and the free movement of such data.[101] The proposal for the replacement to Regulation 45/2001 was published on 10 January 2017.[102] The EDPS will monitor the application of the ePrivacy Regulation by Union institutions, bodies, offices, and agencies. The European Commission aims to replace Regulation 45/2001 by 25 May 2018 (the date of application of the GDPR). We recommend that Article 2(3) be retained.

### 2.2.4. Article 2(4), eCommerce Directive

Article 2(4) of the ePrivacy proposal states:

'This Regulation shall be without prejudice to the application of [the eCommerce] Directive 2000/31/EC,[103] in particular of the liability rules of intermediary service providers in Articles 12 to 15 of that Directive.'[104]

#### Comment
A provision similar to Article 2(4) of the ePrivacy proposal is included in the GDPR.[105] The eCommerce Directive provides for safe harbours that limit the liability of certain types of internet intermediary.[106] Under certain circumstances, intermediaries may benefit from the liability exemption when they act as a 'mere conduit', use 'caching', or offer 'hosting' services. The eCommerce Directive determines the circumstances in which the liability of intermediaries should be limited. In many cases, companies (such as internet access providers) that are regulated under the ePrivacy proposal could rely on the 'mere conduit' exception in the eCommerce Directive for their activities that are regulated in the ePrivacy proposal. We recommend that Article 2(4) of the ePrivacy proposal be retained.

---

[99] CJEU, Tele2 and Watson, para. 73.
[100] Amendments between square brackets by the authors.
[101] Regulation (EC) No 45/2001 of the European Parliament and of the Council of 18 December 2000 on the protection of individuals with regard to the processing of personal data by the Community institutions and bodies and on the free movement of such data (OJ L 8, 12.1.2001, p. 1–22).
[102] Proposal for a Regulation of the European Parliament and of the Council on the protection of individuals with regard to the processing of personal data by the Union institutions, bodies, offices and agencies and on the free movement of such data, and repealing Regulation (EC) No 45/2001 and Decision No 1247/2002/EC, COM(2017) 8 final, 10 January 2017.
[103] Original footnote: Directive 2000/31/EC of the European Parliament and of the Council of 8 June 2000 on certain legal aspects of information society services, in particular electronic commerce, in the Internal Market ('Directive on electronic commerce') (OJ L 178, 17.7.2000, p. 1–16).
[104] Amendment between square brackets by the authors.
[105] Article 2(4) of the GDPR. See about the relation between data protection law and the eCommerce Directive: Van der Sloot 2015.
[106] See article 12-15 of the eCommerce Directive.

### 2.2.5. Article 2(5), Radio equipment Directive

Article 2(5) of the ePrivacy proposal states:

'This Regulation shall be without prejudice to the provisions of [the Radio Equipment] Directive 2014/53/EU.'[107]

See also Recital 10.

### Comment

Article 2(5) establishes the applicability of the ePrivacy Regulation alongside the Radio Equipment Directive.[108] Recital 10 of the ePrivacy proposal adds:

'Radio equipment and its software which is placed on the internal market in the Union, must comply with [the Radio Equipment Directive]. This [ePrivacy] Regulation should not affect the applicability of any of the requirements of [the Radio Equipment Directive] nor the power of the Commission to adopt delegated acts pursuant to [the Radio Equipment Directive] requiring that specific categories or classes of radio equipment incorporate safeguards to ensure that personal data and privacy of end-users are protected.'[109]

The Radio Equipment Directive regulates the making available on the market and putting into service in the Union of radio equipment.[110] In particular, that Directive 'requires that, before being placed on the market, radio equipment must incorporate safeguards to ensure that the personal data and privacy of the user are protected.'[111] The Commission is empowered to adopt measures under the Radio Equipment Directive and the European Standardisation Regulation.[112]

See on security also our comment on Article 17 and in section 4.7.

## 2.3. Article 3, territorial scope and representative

### 2.3.1. Article 3(1)(a)-(c), territorial scope

Article 3(1) of the ePrivacy proposal reads as follows:

'This Regulation applies to:
(a) the provision of electronic communications services to end-users in the Union, irrespective of whether a payment of the end-user is required;
(b) the use of such services;

---

[107] Amendment between square brackets by the authors.
[108] Directive 2014/53/EU of the European Parliament and of the Council of 16 April 2014 on the harmonisation of the laws of the Member States relating to the making available on the market of radio equipment and repealing Directive 1999/5/EC (OJ L 153, 22.5.2014, p. 62–106).
[109] Internal footnote omitted.
[110] Directive 2014/53/EU of the European Parliament and of the Council of 16 April 2014 on the harmonisation of the laws of the Member States relating to the making available on the market of radio equipment and repealing Directive 1999/5/EC, (OJ L 153, 22.5.2014).
[111] Explanatory memorandum to the ePrivacy proposal, p. 3; Article 3(3)(e) of the Radio Equipment Directive.
[112] Regulation (EU) No 1025/2012 of the European Parliament and of the Council of 25 October 2012 on European standardisation, amending Council Directives 89/686/EEC and 93/15/EEC and Directives 94/9/EC, 94/25/EC, 95/16/EC, 97/23/EC, 98/34/EC, 2004/22/EC, 2007/23/EC, 2009/23/EC and 2009/105/EC of the European Parliament and of the Council and repealing Council Decision 87/95/EEC and Decision No 1673/2006/EC of the European Parliament and of the Council (OJ L 316, 14.11.2012, p. 12–33).

(c) the protection of information related to the terminal equipment of end-users located in the Union'.

See also Recitals 8 and 9.

**Comment**

Article 3(1) establishes the territorial scope of the ePrivacy Regulation. Recital 9 adds:

'This Regulation should apply to electronic communications data processed in connection with the provision and use of electronic communications services in the Union, regardless of whether or not the processing takes place in the Union. Moreover, in order not to deprive end-users in the Union of effective protection, this Regulation should also apply to electronic communications data processed in connection with the provision of electronic communications services from outside the Union to end-users in the Union.'

For some provisions of the ePrivacy proposal, the territorial scope is unclear. For instance, it is unclear whether the territorial scope of article 3(1)(b) would cover a case in which a party located outside of the Union violates Articles 15 (on phone books and directories) or 16 (on unsolicited communications and spam).

To illustrate: suppose a spammer based outside the EU sends spam (unsolicited communications) to people in the EU. It is unclear whether Article 3(1) would cover that spammer. Another illustration: Recital 8 states that the ePrivacy Regulation should be applicable to 'software providers permitting electronic communications, including the retrieval and presentation of information on the internet'. It is unclear whether a browser vendor that is located outside the Union and offers browsers to people within the Union would fall under the territorial scope of the Regulation.

The EDPS recommends that the EU lawmaker states that the ePrivacy Regulation has the same territorial scope as the GDPR.[113] However, an exact copy of Article 3 GDPR would not suffice, as the terms 'controller' and 'processor' from the GDPR do not cover the parties concerned by the ePrivacy Regulation in all situations.[114]

There is a clear interest for the EU lawmaker to extend the territorial scope of certain rules to companies acting from outside of EU territory, to prevent loopholes. This is especially the case now that the ePrivacy rules would be extended to 'over the top' service providers, which by nature do not have the same physical connection to the territory as, for instance, access providers and telecommunications service operators.[115] At the same time, the EU lawmaker should observe accepted principles of international law and jurisdiction. And the EU lawmaker should recognise the importance of ensuring that the rules, when adopted, are also enforceable in practice.[116] **In sum, we recommend that the EU lawmaker clarifies Article 3; the scope of Article 3 appears to be too narrow.**

### 2.3.2.    Article 3(2)-(5), representative

Articles 3(2)-(5) of the ePrivacy proposal read as follows.

'(2) Where the provider of an electronic communications service is not established in the Union it shall designate in writing a representative in the Union.
(3) The representative shall be established in one of the Member States where the end-users of such electronic communications services are located.

---

[113] EDPS 2017/6, p. 26.
[114] EDPS 2017/6, p. 26.
[115] See generally on 'over the top' service providers: Godlovitch, W. et al 2015.
[116] The application of European data protection and privacy rules to companies that are based outside the EU is a topic that has been much discussed in literature. See for instance: European Commission, Time.Lex, and Spark legal network and consultancy ltd 2015; Kuner 2010; Kuner 2010a; Moerel 2011, chapter 1-4.

(4) The representative shall have the power to answer questions and provide information in addition to or instead of the provider it represents, in particular, to supervisory authorities, and end-users, on all issues related to processing electronic communications data for the purposes of ensuring compliance with this Regulation.

(5) The designation of a representative pursuant to paragraph 2 shall be without prejudice to legal actions, which could be initiated against a natural or legal person who processes electronic communications data in connection with the provision of electronic communications services from outside the Union to end-users in the Union.'

See also Recital 26.

## Comment

The rules on representatives for electronic communication service providers located outside the Union are established in Article 3. For example, an 'over the top' (OTT) service provider exclusively established in the United States offering an interpersonal communications service falls under the territorial scope of the proposal and needs to designate a representative according to Articles 3(2)-(5). An 'over the top' service provider that is established outside of the EU offering a service that enables interpersonal communication merely as a minor ancillary feature but intrinsically linked to the main service would also have to appoint a representative in the EU for the purposes of this regulation.

It is unclear whether 'representative' in the ePrivacy Regulation has the same meaning as in the GDPR. Problems might arise because the GDPR refers to the 'controller' and 'processor' definitions[117] in the definition of representative.[118] Applying the same definition of 'representative' to the ePrivacy Regulation might not cover all situations that could arise under the ePrivacy Regulation.

An example might be a company outside the Union providing secure electronic communications services designed for large businesses and government agencies. If the company does not process personal data, it is not clear whether it would be required to designate a representative in the Union.[119] We recommend that the EU lawmaker clarifies the rules on 'representatives'. The EU lawmaker could consider improving legal certainty by adding a phrase along the following lines: 'Article 27 of [the GDPR] applies *mutatis mutandis.'*

It is unclear why Article 3(1)(a) uses different wording than Article 3(4). Article 3(1)(a) uses 'the provision of electronic communications services to end-users in the Union', while Article 3(4) uses the wording 'processing electronic communications data'.

Furthermore, it is unclear whether Article 3(4) would cover the provisions of Chapter III, more specifically Articles 15 and 16. In cases where a party outside the Union provides publicly available directories to end-users in the Union or sends unsolicited communications to end-users in the Union, it is not clear whether Article 3(4) applies. After all, it is unclear whether that party 'process[es] electronic communications data'.

---

[117] Article 4(7) and Article 4(8) of the GDPR.
[118] Article 4(17) of the GDPR.
[119] We admit that the example is a tad far-fetched. In most cases, such a company would probably process personal data.

Article 3(5) uses different wording from Articles 3(1) and 3(4). Article 3(5) uses the wording 'processes electronic communications data in connection with the provision of electronic communications services'.

We recommend that the EU lawmaker clarifies whether these differences in wording are useful. If the differences are not useful, we recommend that the EU lawmaker amends or clarifies the provision. More generally, **we recommend addressing the ambiguities in Articles 3(2)-(5).**

## 2.4. Article 4, definitions

Article 4 of the ePrivacy proposal provides definitions. Article 4 refers to both the GDPR and the draft European Electronic Communications Code, and it provides some definitions itself.

### 2.4.1. Article 4(1)(a), GDPR

Article 4(1)(a) of the ePrivacy proposal reads as follows:

'For the purposes of this Regulation, following definitions shall apply:
(a) the definitions in [the GDPR]'.

See also Recital 5.

**Comment**
Article 4(1)(a) states that the definitions of the GDPR apply for the purposes of the ePrivacy Regulation. The preamble adds that the ePrivacy Regulation 'does not lower the level of protection enjoyed by natural persons under Regulation'.[120]

We recommend that Article 4(1)(a) be retained. Article 4(1)(a) helps to ensure consistency with the GDPR, and to ensure that the proposal does not lower the level of protection set by the GDPR. Article 4(1) can help to interpret the e-Privacy Regulation.

As noted in relation to Article 1(3), **we recommend that the EU lawmaker states in an article (rather than in a recital) that the ePrivacy Regulation does not lower the level of protection enjoyed by natural persons under the GDPR.[121]**

See also our comment on 'consent' and Article 9.

### 2.4.2. Article 4(1)(b), European Electronic Communications Code

For several definitions, the ePrivacy proposal refers to the 'Directive establishing the "European Electronic Communications Code"'.[122] These are the definitions for: 'electronic communications network', 'electronic communications service', 'interpersonal communications service', 'number-based interpersonal communications service', 'number-independent interpersonal communications service', 'end-user' and 'call'.[123]

---

[120] Recital 5 of the ePrivacy proposal.
[121] See Recital 5 of the ePrivacy proposal.
[122] Proposal for a Directive of the European Parliament and of the Council establishing the European Electronic Communications Code (Recast) COM/2016/0590 final/2 - 2016/0288 (COD), Corrigendum, <http://eur-lex.europa.eu/legal-content/EN/ALL/?uri=comnat:COM_2016_0590_FIN>. See for a concise introduction: European Parliamentary Research Service 2017.
[123] Article 4(1)(b) of the ePrivacy proposal.

## Comment

In autumn 2016, the European Commission published a proposal for a 'Directive establishing the European Electronic Communications Code' (hereafter: 'draft European Electronic Communications Code').[124] The European Electronic Communications Code is being drafted at the same time as the ePrivacy Regulation. As previously noted, for this study we rely on the draft text of the 'Directive establishing the European Electronic Communications Code', dated 12 October 2016.[125]

The European Electronic Communications Code aims to revise the regulatory framework for telecommunications, and to replace four telecommunications Directives.[126] The European Electronic Communications Code also aims to ensure the security of electronic communications services.[127]

**We recommend that, while drafting the ePrivacy Regulation, the EU lawmaker pay close attention to the development of the European Electronic Communications Code.** A slight change to a definition in the European Electronic Communications Code may have far-reaching effects on the scope of the ePrivacy Regulation.

According to the EDPS, there is no inherent reason why certain concepts should be defined in the European Electronic Communications Code, rather than in the ePrivacy Regulation. 'For independent ePrivacy provisions, it is no longer necessary to ensure that their scope is equivalent to an instrument enabling market regulation.'[128] **The EU lawmaker could consider defining the main concepts from Article 4(1)(b) in the ePrivacy Regulation itself, rather than in the European Electronic Communications Code (EECC)**: 'the EDPS recommends removing the unnecessary dependencies on the EECC Proposal and defining central terms in the ePrivacy Regulation itself, consistent with the EECC Proposal though not necessarily identical.'[129] The EDPS adds: 'even where definitions might be identical in the text of the two proposals, it would be preferable to include fully standalone definitions in the ePrivacy Regulation, where necessary particularised because of the specific context of the protection of fundamental rights.'[130]

There are two advantages of defining the concepts in the ePrivacy Regulation. First, drafting the ePrivacy Regulation would be easier, as the EU lawmaker would not have to focus on two proposals at the same time – two proposals that are both rather complicated and technical. Second, a delay in drafting the European Electronic Communications Code would not result in a delay for the ePrivacy Regulation.

---

[124] Proposal for a Directive of the European Parliament and of the Council establishing the European Electronic Communications Code (Recast) COM/2016/0590 final/2 - 2016/0288 (COD), Corrigendum, <http://eur-lex.europa.eu/legal-content/EN/ALL/?uri=comnat:COM_2016_0590_FIN>.

[125] Proposal for a Directive of the European Parliament and of the Council establishing the European Electronic Communications Code (Recast) COM/2016/0590 final/2 - 2016/0288 (COD), Corrigendum, http://eur-lex.europa.eu/legal-content/EN/ALL/?uri=comnat:COM_2016_0590_FIN.

[126] The European Electronic Communications Code aims to revise the following four directives:
(i) Directive 2002/19/EC of the European Parliament and of the Council of 7 March 2002 on access to, and interconnection of, electronic communications networks and associated facilities (Access Directive) (OJ L 108, 24.4.2002, p. 7).
(ii) Directive 2002/20/EC of the European Parliament and of the Council of 7 March 2002 on the authorisation of electronic communications networks and services (Authorisation Directive) (OJ L 108, 24.4.2002, p. 21).
(iii) Directive 2002/21/EC of the European Parliament and of the Council of 7 March 2002 on a common regulatory framework for electronic communications networks and services (Framework Directive) (OJ L 108, 24.4.2002, p. 33)
(iv) Directive 2002/22/EC of the European Parliament and of the Council of 7 March 2002 on universal service and users' rights relating to electronic communications networks and services (Universal Service Directive) (OJ L 108 24.4.2002, p. 51).

[127] See our comment on Article 17.

[128] EDPS 2016/5, p. 12

[129] EDPS 2017/6, p. 11.

[130] EDPS 2017/6, p. 11.

However, we recommend that the EU lawmaker aims for consistency between the European Electronic Communications Code and the ePrivacy Regulation. To ensure consistency, it may be best to define the concepts in the European Electronic Communications Code.

**If the EU lawmaker chooses to define relevant concepts in the European Electronic Communications Code, we recommend that, while working on the ePrivacy proposal, the EU lawmaker pays close attention to the development of that Code.[131]** The Article 29 Working Party calls upon 'all parties involved in the legislative process [to] ensure that both the Proposed Regulation and the EECC are discussed and voted on simultaneously, in order to allow stakeholders to correctly assess the scope and implications of the new instruments.'[132]

### 2.4.3. Article 4(1)(b), 'electronic communications network'

An 'electronic communications network' is defined as follows in Article (2)(1) of the draft European Electronic Communications Code:

'Electronic communications network' means transmission systems, whether or not based on a permanent infrastructure or centralised administration capacity, and, where applicable, switching or routing equipment and other resources, including network elements which are not active, which permit the conveyance of signals by wire, radio, optical or other electromagnetic means, including satellite networks, fixed (circuit- and packet-switched, including Internet) and mobile terrestrial networks, electricity cable systems, to the extent that they are used for the purpose of transmitting signals, networks used for radio and television broadcasting, and cable television networks, irrespective of the type of information conveyed'.[133]

See also Recital 7 of the draft European Electronic Communications Code.

**Comment**
An 'electronic communications network' is, in short, a transmission system. The phrase 'whether or not based on a permanent infrastructure or centralised administration capacity' is new,[134] compared to the current definition of 'electronic communications network'.[135]

Recital 7 of the draft European Electronic Communications Code emphasises that the Code does not, in short, regulate content, such as broadcasting content:

'It is necessary to separate the regulation of electronic communications networks and services from the regulation of content. This Code does not therefore cover the content of services delivered over electronic communications networks using electronic communications services, such as broadcasting content, financial services and certain information society services, and is therefore without prejudice to measures taken at Union or national level in respect of such services, in compliance with Union law, in order to promote cultural and linguistic diversity and to ensure the defence of media pluralism. The content of television programmes is covered by [the Audiovisual Media Services] Directive 2010/13/EU of the European Parliament and of the Council.'[136]

---

[131] See our comment on Article 4(1)(b).
[132] Article 29 Working Party 2017 (WP247), p. 25.
[133] See for the current definition: Article (2)(a) of Council Directive 2002/21 of 7 March 2002 on a common regulatory framework for electronic communications networks and services as amended by Directive 2009/140/EC and Regulation 544/2009 (Framework Directive) (OJ L 108, 24.4.2002, p. 33).
[134] See about that phrase: recital 12 of the draft European Electronic Communications Code.
[135] Currently, 'electronic communications network' is defined in Article 2(a) of Council Directive 2002/21 of 7 March 2002 on a common regulatory framework for electronic communications networks and services as amended by Directive 2009/140/EC and Regulation 544/2009 (Framework Directive).
[136] See on 'electronic communications network' also Recital 13 of the draft European Electronic Communications Code.

A similar limitation (regarding broadcasting content etc.) is set in the current regulatory framework for telecommunications.[137]

### 2.4.4. Article 4(1)(b), 'electronic communications service'

The draft European Electronic Communications Code defines an 'electronic communications service' as follows in Article 2(4):

'"electronic communications service" means a service normally provided for remuneration via electronic communications networks, which encompasses 'internet access service' as defined in Article 2(2) of [the Net Neutrality and Roaming] Regulation (EU) 2015/2[138] and/or 'interpersonal communications service'; and/or services consisting wholly or mainly in the conveyance of signals such as transmission services used for the provision of machine-to-machine services and for broadcasting, but excludes services providing, or exercising editorial control over, content transmitted using electronic communications networks and services'.

See also Recitals 11 and 12 of the ePrivacy proposal, and Recitals 15 and 16 of the draft European Electronic Communications Code.

#### Comment, scope of definition

Compared with the current definition of 'electronic communications service',[139] the definition in the draft European Electronic Communications Code has been amended significantly. The proposed definition has a much broader scope than the current definition. Because many provisions in the ePrivacy proposal refer to 'electronic communications service', the scope of many rules in the ePrivacy proposal is also much wider than the rules in the current ePrivacy Directive.[140]

In the current ePrivacy Directive, an electronic communications service is, in short, a service that consists wholly or mainly of the conveyance of signals on electronic communications networks. In practice, it is mainly internet access providers and phone companies that fall within the scope of the Directive's concept of electronic communications services. Webmail providers such as Gmail and messaging services such as WhatsApp fall outside the scope of the concept. From a user's privacy perspective, this difference in legal treatment does not always make sense.[141]

The definition of 'electronic communications services' in the draft European Electronic Communications Code aims to take a functional approach. As the preamble of the draft European Electronic Communications Code states, 'From an end-user's perspective it is not

---

[137] Recital 5 of Directive 2002/21/EC of the European Parliament and of the Council of 7 March 2002 on a common regulatory framework for electronic communications networks and services (Framework Directive) (OJ L 108, 24.4.2002, p. 33)
[138] Regulation (EU) 2015/2120 of the European Parliament and of the Council of 25 November 2015 laying down measures concerning open internet access and amending Directive 2002/22/EC on universal service and users' rights relating to electronic communications networks and services and Regulation (EU) No 531/2012 on roaming on public mobile communications networks within the Union (OJ L 310, 26.11.2015, p. 1–18).
[139] Article 2(c) of the Council Directive 2002/21 of 7 March 2002 on a common regulatory framework for electronic communications networks and services as amended by Directive 2009/140/EC and Regulation 544/2009 (Framework Directive) (OJ L 108, 24.4.2002, p. 33–50):
'"electronic communications service" means a service normally provided for remuneration which consists wholly or mainly in the conveyance of signals on electronic communications networks, including telecommunications services and transmission services in networks used for broadcasting, but exclude services providing, or exercising editorial control over, content transmitted using electronic communications networks and services; it does not include information society services, as defined in Article 1 of Directive 98/34/EC, which do not consist wholly or mainly in the conveyance of signals on electronic communications networks'
[140] See for instance Article 6 of the ePrivacy proposal.
[141] See Van Hoboken and Zuiderveen Borgesius 2015.

relevant whether a provider conveys signals itself or whether the communication is delivered via an internet access service.'[142] Recital 11 of the ePrivacy proposal adds:

'The services used for communications purposes, and the technical means of their delivery, have evolved considerably. End-users increasingly replace traditional voice telephony, text messages (SMS) and electronic mail conveyance services in favour of functionally equivalent online services such as Voice over IP, messaging services and web-based e-mail services.'[143]

In the draft European Electronic Communications Code, an 'electronic communications service' includes three type of service: (i) internet access services, (ii) services consisting wholly or mainly of the conveyance of signals, and (iii) interpersonal communications services. There are two types of 'interpersonal communications': (iii(a)) number-based, and (iii(b)) number-independent. These types of service may partly overlap.[144]

The illustration below may clarify the structure of the definition of 'electronic communications service':

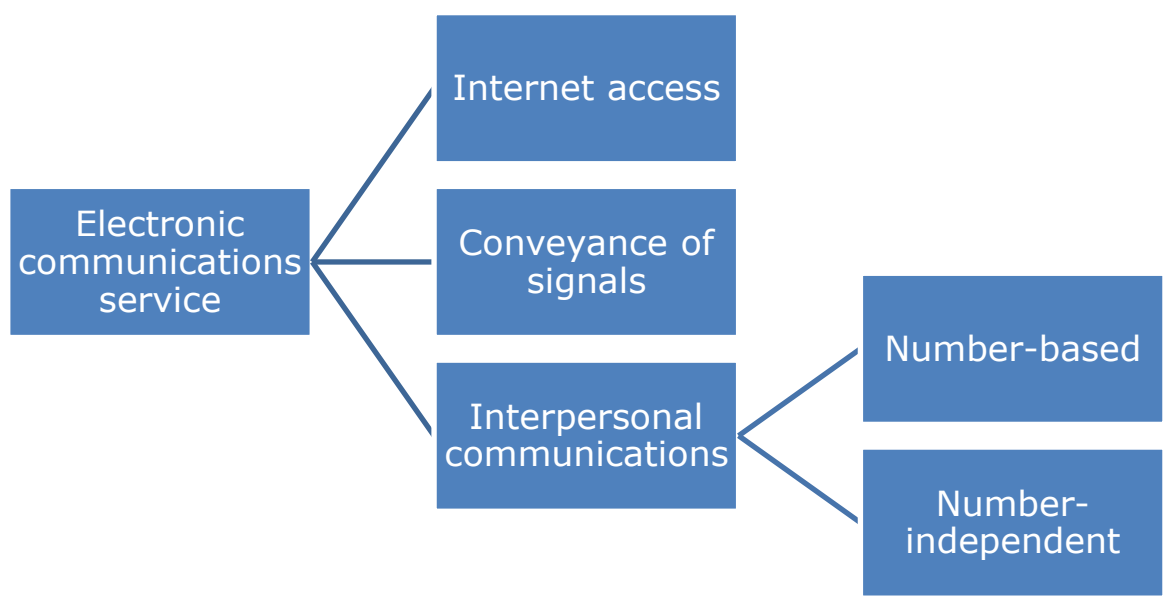

(i) The definition of 'electronic communications services' encompasses internet access services. An 'internet access service' is defined as 'a publicly available electronic communications service that provides access to the internet, and thereby connectivity to virtually all end points of the internet, irrespective of the network technology and terminal equipment used.'[145]

(ii) The definition of 'electronic communications services' encompasses 'services consisting wholly or partly in the conveyance of signals'.

---

(iii) The definition of 'electronic communications services' encompasses interpersonal communications services, which (a) may or (b) may not be number-based.

To illustrate: messaging services (such as WhatsApp), web-based email services (such as Gmail), and Voice over IP services (such as Skype) now fall within the scope of the definition of 'electronic communications services'.[146] (Those services are outside the scope of the current ePrivacy Directive.) Traditional voice telephony also falls within the scope of 'electronic communications services'.[147] A group chat, and a messaging or chatting function in a dating app or in an online game also fall within the scope of 'electronic communications services'.

The Impact Assessment to the ePrivacy proposal suggests there is widespread support for an extended scope:

'The public consultation showed that an overwhelming majority of citizens, civil society and public bodies finds that OTTs ['over the top' service providers] should provide the same level of protection when they provide communication services as ECS [electronic communication service] providers, while approximately a third of the industry respondents (including ECSs and OTTs) agree with this statement. National data protection authorities, BEREC and the EDPS also advocated for an extension of the scope of the ePD to OTTs. The International Working Group on Data Protection in Telecommunications reached similar views. This is also the predominant view of citizens according to a recent Eurobarometer survey (92%).'[148]

The Article 29 Working Party and the EDPS have both welcomed the extended scope.[149]

### So-called 'free' services

Many services that people can use without paying money also fall within the scope of 'electronic communications services'. A service must normally be provided in exchange for remuneration to fall within the scope of the definition of an electronic communications service.[150]

However, Recital 16 of the draft European Electronic Communications Code explains that the concept of remuneration should also encompass so-called 'free' services that are funded by advertising and services that monetise user data (usually for behavioural targeting):[151]

'In order to fall within the scope of the definition of electronic communications service, a service needs to be provided normally in exchange for remuneration. In the digital economy, market participants increasingly consider information about users as having a monetary value. Electronic communications services are often supplied against counter-performance other than money, for instance by giving access to personal data or other data. The concept of remuneration should therefore encompass situations where the provider of a service requests and the end-user actively provides personal data, such as name or email address, or other data directly or indirectly to the provider. It should also encompass situations where the provider collects information without the end-user actively supplying it, such as personal data, including the IP address, or other automatically generated information, such as information collected and transmitted by a cookie. In line with the jurisprudence of the Court of Justice of the European Union on Article 57 TFEU,[152] remuneration exists within the meaning of the Treaty also if the service provider is paid by a third party and not by the service recipient. The concept of remuneration should therefore also encompass situations where the end-user is exposed to advertisements as a condition for gaining access to the service, or situations where the service provider monetises personal data it has collected.'

---

[146] Recital 11 of the ePrivacy proposal.
[147] The preamble of the draft European Electronic Communications Code gives, as examples of electronic communications services, 'voice telephony, messaging services and electronic mail services' (Recital 10 of the draft European Electronic Communications Code).
[148] ePrivacy Impact Assessment 2017 Pt. 1, p. 7 (internal footnotes omitted).
[149] Article 29 Working Party 2017 (WP247), p. 8; EDPS 2017/6, p. 8. As the EDPS notes, 'Users' expectations are often similar with regard to the privacy and confidentiality of these messages and any breach of confidentiality may be equally intrusive' (p. 8).
[150] Recital 16 of the draft European Electronic Communications Code.
[151] Behavioural targeting involves monitoring people's online behaviour, and using the collected information to show people targeted advertisements. See our comment on Article 8(1).

Recital 18 of the ePrivacy proposal adds: 'In the digital economy, services are often supplied against counter-performance other than money, for instance by end-users being exposed to advertisements.'

The EDPS calls for an amendment to Recital 18 of the ePrivacy proposal. The EDPS wants the EU lawmaker to delete the sentence 'Electronic communications services are often supplied against counter-performance other than money, for instance by giving access to personal data or other data'. The EDPS proposes a different sentence: 'In the digital economy, services are often supplied with remuneration paid by a third party rather than by the recipient of the service'.[153]

**Indeed, the EU lawmaker should be careful not to legitimise situations in which personal data and communications data are treated as tradable goods.**

### Internet of Things

The ePrivacy proposal applies to many Internet of Things scenarios. Recital 12 of the ePrivacy proposal states:

'Connected devices and machines increasingly communicate with each other by using electronic communications networks (Internet of Things). *The transmission of machine-to-machine communications involves the conveyance of signals over a network and, hence, usually constitutes an electronic communications service.* In order to ensure full protection of the rights to privacy and confidentiality of communications, and to promote a trusted and secure Internet of Things in the digital single market, it is necessary to clarify that this Regulation should apply to the transmission of machine-to-machine communications. Therefore, the principle of confidentiality enshrined in this Regulation should also apply to the transmission of machine-to-machine communications.'

The Article 29 Working Party and the EDPS welcome the application to the Internet of Things. The EDPS notes that the Internet of Things 'includes sports trackers, health sensors, personal communications devices, smart TVs, intelligent cars and many other devices.'[154] Many such devices exchange communications with other devices, and such communications can be very sensitive. According to the EDPS, 'the Proposal [should] unambiguously cover machine-to-machine communications in the context of the Internet of Things, irrespective of the type of network or communication service used, on all networks and services which are otherwise within the scope.'[155] We agree that the confidentiality of machine-to-machine communications should generally be protected.

### Editorial control

**The definition of electronic communications service 'excludes services providing, or exercising editorial control over, content transmitted using electronic communications networks and services'. That exclusion deserves the lawmaker's attention, as there is significant discussion about the precise meaning of the term editorial control.**[156] Digital information and communications, including messenger services, allow for new types of algorithmic control, which sometimes resembles editorial control on content. Considering the inclusion of 'over the top' service providers, this exclusion could

---

[152] Original footnote: Case C-352/85 Bond van Adverteerders and Others vs The Netherlands State, EU:C:1988:196.
[153] EDPS 2017/6, p. 25.
[154] EDPS 2017/6, p. 9. See also Article 29 Working Party 2017 (WP247), p. 8.
[155] EDPS 2017/6, p. 9.
[156] See e.g. Moeller et al 2016; Van Hoboken 2012, chapter 8-10.

lead to significant debate as regards scope. Interpersonal 'over the top' service providers may include content and other filters and manage communication threads through ranking algorithms and other means.

**The EU lawmaker should consider whether additional rules are needed to protect the privacy of media users.** The Audiovisual Media Services Directive concerns the regulation of audiovisual media services but does not protect media users' privacy.[157] Monitoring individuals' media consumption can give a pertinent impression of someone's interests and habits.[158] See also our comment on Article 9, on 'tracking walls'.

### Conclusion
In summary, the scope of 'electronic communications service' is widened significantly, and also encompasses many 'over the top' services that enable communication. Because the scope of 'electronic communications service' is widened, many provisions in the ePrivacy proposal have a much wider scope than the current ePrivacy Directive.

**In principle we agree that it is a good idea to broaden the scope of the ePrivacy rules, especially the rules that protect communications confidentiality (see Article 5). However, we also point out that rules with a wider scope must often be more general, compared to rules with a narrow scope. Therefore, for rules with a wider scope, it is more difficult to write clear and specific rules. We recommend that the EU lawmaker keeps that possible effect of broadening the scope in mind. See also our comment on Article 6.**

### 2.4.5.   Article 4(1)(b) and 4(2), 'interpersonal communications service'

An 'interpersonal communications service' (a new regulatory concept) is defined as follows in Article 2(5) of the draft European Electronic Communications Code:

'"Interpersonal communications service" means a service normally provided for remuneration that enables direct interpersonal and interactive exchange of information via electronic communications networks between a finite number of persons, whereby the persons initiating or participating in the communication determine its recipient(s); *it does not include services which enable interpersonal and interactive communication merely as a minor ancillary feature that is intrinsically linked to another service'*.[159]

Article 4(2) of the ePrivacy proposal states:

'the definition of "interpersonal communications service" *shall* include services that enable interpersonal and interactive communication merely as a minor ancillary feature that is intrinsically linked to another service.'[160]

See also Recital 11 of the ePrivacy proposal and Recital 17 of the draft European Electronic Communications Code.

### Comment
An 'interpersonal communications service' is, in short, a service 'that enables direct interpersonal and interactive exchange of information via electronic communications networks between a finite number of persons, whereby the persons initiating or participating in the communication determine its recipient(s)'.[161] The scope of 'interpersonal communications services' is wide, and covers, for instance traditional voice calls between

---

[157] Directive 2010/13/EU of the European Parliament and of the Council.
[158] Cohen 1995; Richards 2014.
[159] Emphasis added.
[160] Emphasis added.
[161] Article 2(5) of the draft European Electronic Communications Code.

people, all types of email and messaging services, and group chats.[162] Recital 17 of the draft European Electronic Communications Code explains:

'Interpersonal communications services are services that enable interpersonal and interactive exchange of information, covering services like traditional voice calls between two individuals but also all types of emails, messaging services, or group chats. Interpersonal communications services only cover communications between a finite, that is to say not potentially unlimited, number of natural persons which is determined by the sender of the communication. Communications involving legal persons should be within the scope of the definition where natural persons act on behalf of those legal persons or are involved at least on one side of the communication. Interactive communication entails that the service allows the recipient of the information to respond. Services which do not meet those requirements, such as linear broadcasting, video on demand, websites, social networks, blogs, or exchange of information between machines, should not be considered as interpersonal communications services.'

The ePrivacy proposal's concept of 'interpersonal communications service' also applies to services that enable interpersonal and interactive communication merely as a minor ancillary feature.[163] Recital 11 of the ePrivacy proposal states: 'The protection of confidentiality of communications is crucial also as regards interpersonal communications services that are ancillary to another service; therefore, such type of services also having a communication functionality should be covered by this Regulation.' An example of a minor ancillary feature might be a chat function or a messaging function included in a computer game, or a chat function that is included in a social network site.[164]

For the purposes of the ePrivacy Regulation, the ePrivacy Regulation widens the scope of the 'interpersonal communications service' concept. In the draft European Electronic Communications Code, an 'interpersonal communications service' does *not* include such ancillary features.[165] Because the ePrivacy proposal *does* include such ancillary features in its definition of 'interpersonal communications service', providers of such services must, for instance, respect the confidentiality of communications.

**It makes sense that communications confidentiality should be respected when people use, for instance, a chat function in a game, a social network site, or a dating app. Indeed, the Article 29 Working Party and the EDPS both welcome the inclusion of ancillary services.[166]**

**However, the EU lawmaker could consider two amendments.** First, Recital 17 of the draft European Electronic Communications Code could be amended, to avoid misunderstandings. A chat function or a messaging function within a social network service (such as a site or an app) also falls within the scope of an 'interpersonal communications service'. However, Recital 17 of the draft European Electronic Communications Code gives 'social network' as an example of a service that is *not* an 'interpersonal communications service'. To avoid misunderstandings, it might be best to delete the 'social network' example from Recital 17 of the draft European Electronic Communications Code.

Second, the EU lawmaker should consider clarifying whether and under what conditions a person's 'timeline' on a social network page is also covered by the definition of 'interpersonal communications service'. In the ePrivacy proposal and the draft European Electronic Communications Code it is not clear whether messages exchanged over a timeline of a social network service are within the scope of 'interpersonal communications service'. Do such messages fall within the scope of the definition when the social network's settings are such that no one else than the finite number of contacts can see a posting? And how could that situation be legally distinguished from a more publicly oriented discussion, which would not be covered by the ePrivacy proposal?

---

[162] Recital 17 of the draft European Electronic Communications Code.
[163] Article 4(2) of the ePrivacy proposal.
[164] See Recital 17 of the draft European Electronic Communications Code.
[165] Article 2(5) of the draft European Electronic Communications Code.
[166] EDPS 2017/6, p. 9 ; p. 12; Article 29 Working Party 2017 (WP247), p. 8.

Snapchat, the popular communications platform, provides an example in which there are two types of communications, one of which is less private than the other. Would both the private snap function and the broadcasting function be covered the proposals, or merely the private communications function? And, what would be the reason for not providing some of the ePrivacy-specific protections for users (outside of the protection of communications confidentiality) when users publish information and ideas for a broader online audience?[167] **We recommend that the EU lawmaker considers these questions when deciding on the scope of the definitions as proposed by the Commission.**

### 2.4.6. Article 4(1)(b), 'number-based interpersonal communications'

A 'number-based interpersonal communications service' (a new regulatory concept) is defined as follows in Article (2)(6) of the draft European Electronic Communications Code:

'"Number-based interpersonal communications service" means an interpersonal communications service which connects with the public switched telephone network, either by means of assigned numbering resources, i.e. a number or numbers in national or international telephone numbering plans, or by enabling communication with a number or numbers in national or international telephone numbering plans'.[168]

See also Recitals 17 and 18 of the draft European Electronic Communications Code.

**Comment**
As noted, there are two types of 'interpersonal communications services': number-based and number-independent. First we discuss number-based interpersonal communications services.

Traditional phone calls fall within the scope of 'number-based interpersonal communications services'. Another example of a 'number-based interpersonal communications service' is Skype Number, which enables Skype users to connect to traditional phone numbers.[169] Recital 18 of the draft European Electronic Communications Code states:

'Interpersonal communications services using numbers from a national and international telephone numbering plan connect with the public (packet or circuit) switched telephone network. Those number-based interpersonal communications services comprise both services to which end-users numbers are assigned for the purpose of ensuring end-to-end connectivity and services enabling end-users to reach persons to whom such numbers have been assigned.'

According to the preamble of the draft European Electronic Communications Code, 'It is justified to treat number-based interpersonal communications services differently, as they participate in and hence also benefit from a publicly assured interoperable ecosystem.'[170]

Some new services, like WhatsApp, Signal, and Telegram, rely on phone numbers as identifiers. Recital 18 of the draft European Electronic Communications Code shows that a messaging service that uses phone numbers merely as identifiers does not automatically become a 'number-based interpersonal communications service'. Recital 18 states: 'The mere use of a number as an identifier should not be considered equivalent to the use of a number to connect with the public switched telephone network, and should therefore, in itself, not be considered sufficient to qualify a service as a number-based interpersonal communications service.' Many interpersonal communications services could be partially turned into number-based services, by connecting them to the public switched telephone network.

---

[167] See Van Hoboken & Zuiderveen Borgesius 2015.
[168] Relevant recitals in the draft European Electronic Communications Code are: Recital 18, 137, 253, 255, 256, and 264.
[169] 'A Skype Number is a phone number that you pay monthly for. People can call you from their mobile or landline and you pick the call up in Skype' (<https://support.skype.com/en/faq/FA331/what-are-the-different-types-of-skype-subscriptions-and-pay-as-you-go-options> accessed 5 May 2017). See also: European Parliamentary Research Service 2017, p. 6.
[170] Recital 18 of the draft European Electronic Communications Code.

### 2.4.7.  Article 4(1)(b), 'number-independent interpersonal communications'

A 'number-independent interpersonal communications service' (a new regulatory concept) is defined as follows in Article (2)(7) of the draft European Electronic Communications Code:

'"Number-independent interpersonal communications service" means an interpersonal communications service which does *not* connect with the public switched telephone network, either by means of assigned numbering resources, i.e. a number or numbers in national or international telephone numbering plans, or by enabling communication with a number or numbers in national or international telephone numbering plans'.[171]

**Comment**
Examples of such number-independent services are numerous. Various app services providing communication possibilities, such as WhatsApp, Signal and Telegram, have largely replaced traditional number-dependent communications such as SMS.

Recital 18 of the draft European Electronic Communications Code states: 'Number-independent interpersonal communications services should be subject only to obligations, where public interests require applying specific regulatory obligations to all types of interpersonal communications services, regardless of whether they use numbers for the provision of their service.' In the draft European Electronic Communications Code, the obligations for number-independent services are mostly limited to security requirements.[172]

### 2.4.8.  Article 4(1)(b), 'end-user'

An 'end user' is defined as follows in the draft European Electronic Communications Code (Article (2)(14)):

"End-user" means a user not providing public communications networks or publicly available electronic communications services'.

A 'user' is defined as follows in the draft European Electronic Communications Code (Article (2)(13)):

'"user" means a legal entity or natural person using or requesting a publicly available electronic communications service'.

**Comment**
In the ePrivacy proposal, a 'user' (and thus an end-user) can be a legal entity – a company for instance. A machine does not fall within the definitions of 'user' and 'end-user'. The definitions of 'user' and 'end-user' would remain the same as they are currently under the

---

[171] Emphasis added. Relevant recitals in the draft European Electronic Communications Code are: Recital 24, 91, 137, and 233.
[172] See Recital 91 of the draft European Electronic Communications Code. See also European Parliamentary Research Service 2017, p. 6.

Framework Directive.[173] However, the current ePrivacy Directive has a separate definition of 'user', which is limited to natural persons.[174]

Hence, the 'user' concept in the ePrivacy proposal is wider than the 'user' concept in the current ePrivacy Directive. The current ePrivacy Directive also uses the word 'subscriber'[175] whereas the ePrivacy proposal does not use that word. The deletion of the concept of a 'subscriber' improves clarity.[176]

The EDPS notes that the proposed definition of 'end-user' may cause problems in the context of ePrivacy rules that require companies to obtain consent. In some situations it is not clear who should give consent.[177] Therefore, the EDPS proposes an alternative definition of end-user specifically aimed at providing consent for processing of communications data. The definition should include four elements:
> (i) natural person,
> (ii) using a publicly available electronic communications service,
> (iii) for private or business purposes,
> (iv) without necessarily having subscribed to this service. [178]

**We recommend that the EU lawmaker considers clarifying and amending the definition of 'end-user', while taking into account the advice of the EDPS and the Article 29 Working Party.[179]**

Suppose that a company (a legal person) is an end-user, and that the company gives consent to a provider of an electronic communications service that asks whether it can analyse the end-users' communications content for marketing purposes.[180] However, the employees of the company might not want their communications to be analysed. **The EU lawmaker should consider ensuring that employees are protected in such situations.**

**We also recommend that generally, where the ePrivacy Regulation requires companies to obtain the end-user's consent, the relevant provision should require the consent of 'all end-users'.[181] (See also our comments on Article 6 and 8).**

## 2.4.9.    Article 4(1)(b), 'call'

A 'call' is defined as follows in Article (2)(21) of the draft European Electronic Communications Code:

'"Call" means a connection established by means of a publicly available interpersonal communications service allowing two-way voice communication'.

### Comment
Compared with the current definition of 'call' (in the Framework Directive),[182] the draft European Electronic Communications Code has changed 'electronic' to 'interpersonal'. The definition of 'call' in the draft European Electronic Communications Code excludes machine-to-machine communications. Hence, a machine that communicates with a machine does not make a 'call' in the sense of the draft European Electronic Communications Code, because a 'call', by definition, concerns 'interpersonal' communications.[183]

---

[173] Now, 'end user' (Article 2(n)) and 'user' (Article 2(h)) are both defined in Council Directive 2002/21 of 7 March 2002 on a common regulatory framework for electronic communications networks and services as amended by Directive 2009/140/EC and Regulation 544/2009 (Framework Directive) (OJ L 108, 24.4.2002, p. 33–50).
[174] Article 2(a) of the 2009 ePrivacy Directive: '"user" means <u>any natural person</u> using a publicly available electronic communications service, for private or business purposes, without necessarily having subscribed to this service'.
[175] Article 1(2) of the 2009 ePrivacy Directive.
[176] See Article 29 Working Party 2017 (WP247), p. 8.
[177] EDPS 2017/6, p. 14.
[178] EDPS 2017/6, p. 14.
[179] Article 29 Working Party 2017 (WP247), p. 26.
[180] See Article 6 of the ePrivacy proposal.
[181] EDPS 2017/6, p. 15.
[182] Currently, 'call' is defined in Article 2(s) of the Council Directive 2002/21 of 7 March 2002 on a common regulatory framework for electronic communications networks and services as amended by Directive 2009/140/EC and Regulation 544/2009 (Framework Directive) (OJ L 108, 24.4.2002, p. 33–50).

The scope of the definition of 'call' could be clearer. For instance, does a robocall (an automated call to a human) fall within the definition of 'call'? **We recommend that the EU lawmaker clarifies the scope of the definition of 'call'.**

### 2.4.10. Article 4(1)(c), 'terminal equipment'

For the definition of 'terminal equipment', the ePrivacy proposal refers to the Directive on competition in the markets in telecommunications terminal equipment (2008/63/EC).[184]

That Directive defines 'terminal equipment' as:

'a) equipment directly or indirectly connected to the interface of a public telecommunications network to send, process or receive information; in either case (direct or indirect), the connection may be made by wire, optical fibre or electromagnetically; a connection is indirect if equipment is placed between the terminal and the interface of the network;
(b) satellite earth station equipment'.

**Comment**
Examples of terminal equipment are cell phones, smart phones, routers, and internet-connected computers. But the terminal equipment definition also includes other devices connecting to public telecommunications networks, such as internet-connected personal computers, game consoles, and 'Internet of Things'-devices.

### 2.4.11. Article 4(2), 'interpersonal communications service'

See above; section 2.4.5, in relation to 'interpersonal communications service'.

### 2.4.12. Article 4(3)(a), electronic communications data

Article 4(3) of the ePrivacy proposal adds a number of terms that are defined in the ePrivacy proposal itself. We discuss each of these definitions below.

Article 4(3)(a) defines 'electronic communications data' as follows:

'electronic communications content and electronic communications metadata'.

See also Recitals 4, 12 and 14 of the ePrivacy proposal.

**Comment**
'Electronic communications data' include both content and metadata.[185] The definition of 'electronic communications data' can be illustrated as follows:

---

[183] Article (2)(21) of the draft European Electronic Communications Code.
[184] Article 1(1) of Commission Directive 2008/63/EC of 20 June 2008 on competition in the markets in telecommunications terminal equipment (OJ L 162, 21.6.2008, p. 20–26).
[185] The 2009 ePrivacy Directive uses different terminology.

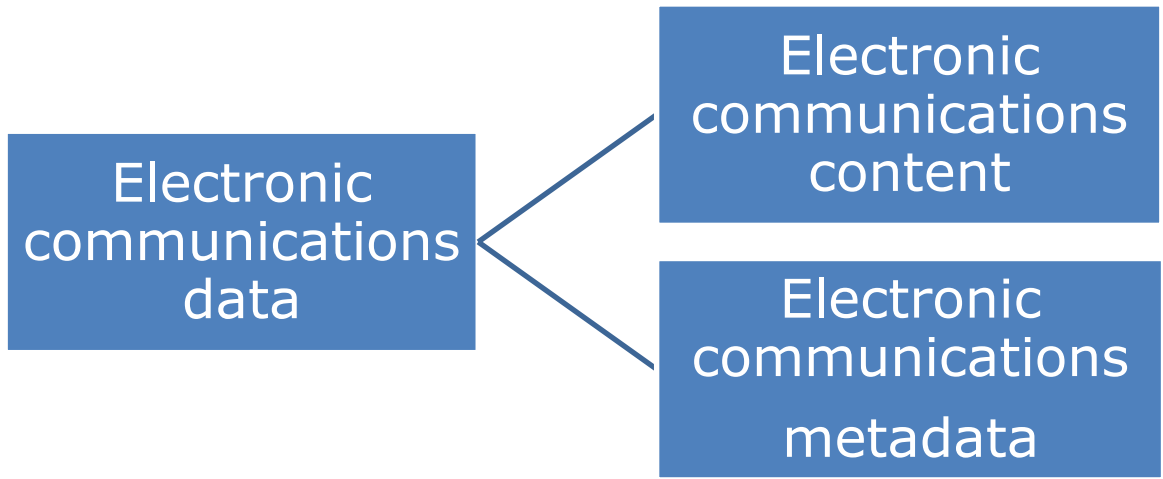

**We recommend that the EU lawmaker keeps in mind that the ePrivacy proposal does not only apply to typical communication services such as email, phone, Skype, and WhatsApp, but also applies to machine-to-machine communications.[186]** For instance, a computer might automatically send a security update to another computer. And metering equipment or sensors might regularly communicate with a central data storage facility.

**We also recommend that the EU lawmaker clarifies whether the data exchanged during machine-to-machine communications fall within the scope of the 'electronic communications data' definition.** Recital 12 seems to suggest that the 'electronic communications data' definition does apply to machine-to-machine communications, but the recital could be more explicit on that point.

An important question that can be asked is how the concept of electronic communications data relates to the concept of personal data (as defined in the GDPR). The Article 29 Working Party recommends amending Recital 4 to better reflect the relationship between these two concepts. Recital 4 currently states that 'Electronic communications data may include personal data as defined in [the GDPR]'. The Working Party recommends that the recital states that 'electronic communications data' are *generally* personal data.[187] **Indeed, in most cases 'electronic communications data' are themselves personal data and this should be clarified in the text of the proposal, to avoid confusion about these key concepts.**

---

[186] See Recital 12 of the ePrivacy proposal.
[187] Article 29 Working Party 2017 (WP247), p. 27.

### 2.4.13. Article 4(3)(b), 'electronic communications content'

The ePrivacy proposal defines 'electronic communications content' as:

'Content exchanged by means of electronic communications services, such as text, voice, videos, images, and sound'.

See also Recital 19.

#### Comment

Examples of 'electronic communications content' include the content of phone calls (what you say) and emails (what you write in or attach to an email). Recital 19 adds: 'The content of electronic communications pertains to the essence of the fundamental right to respect for private and family life, home and communications protected under Article 7 of the Charter.' The CJEU has arrived at a similar conclusion.[188]

**We recommend that the EU lawmaker clarifies the scope of 'electronic communications content'.** In the current ePrivacy Directive, web browsing and using IP-TV fall within the legal definition of 'communication'.[189] In the words of the ePrivacy Directive: 'in cases where the individual subscriber or user receiving such information can be identified, for example with video-on-demand services, the information conveyed is covered within the meaning of a communication.'[190] In the ePrivacy proposal and the draft European Communications Code, there is no separate definition of 'communication'. It is somewhat unclear whether information about people's web browsing and about internet-based TV usage fall within the ePrivacy proposal's 'electronic communications content' definition.

Recital 15 appears to suggest that 'electronic communications content' also includes web browsing behaviour: 'Other examples of interception include capturing payload data or *content data* from unencrypted wireless networks and routers, including browsing habits without the end-users' consent.'[191] Recital 2 of the ePrivacy proposal says that URLs of visited websites constitute metadata: 'metadata derived from electronic communications (...) includes (...) the websites visited'.[192] Hence, it appears that that 'electronic communications content' also covers web browsing behaviour. But the ePrivacy Regulation should be clearer and more explicit on this point. **Hence, we recommend that the EU lawmaker clarifies that 'electronic communications content' covers content that is exchanged between a website and a website visitor's browser.**

**It would also be useful to clarify that 'electronic communications content' encompasses information exchanged when people watch internet-based TV.** As the Article 29 Working Party argues, 'Data collected in the course of offering digital broadcasting services should be covered by the Proposed Regulation.'[193] The Article 29 Working Party states that viewing behaviour is sensitive, since it reveals personal interests and characteristics of viewers. Providers that offer both electronic communications services and content services should therefore not be excluded from the scope of the ePrivacy Regulation.[194]

**In sum, we recommend clarifying the scope of the definition of 'electronic communications content'.**

---

[188] See CJEU (Grand Chamber), Judgement of 6 October 2015, case C-362/14 (Schrems), ECLI:EU:C:2015:650, par 94.
[189] Article 2(d) of the 2009 ePrivacy Directive. See Steenbruggen 2009, p. 181; p. 354. Traung 2010, p. 227.
[190] Recital 16 of the 2009 ePrivacy Directive.
[191] Emphasis added.
[192] See also EDPS 2017/6, p. 28.
[193] Article 29 Working Party 2017 (WP247), p. 27.
[194] Article 29 Working Party 2017 (WP247), p. 27.

## 2.4.14. Article 4(3)(c), 'electronic communications metadata'

Article 4(3)(c) defines 'electronic communications metadata' as:

'Data processed in an electronic communications network for the purposes of transmitting, distributing or exchanging electronic communications content; including data used to trace and identify the source and destination of a communication, data on the location of the device generated in the context of providing electronic communications services, and the date, time, duration and the type of communication'.

See also Recitals 2, 14, 17 and 18.

### Comment

As Recital 2 notes, in a similar way to communications content, 'metadata derived from electronic communications may also reveal very sensitive and personal information.'[195] The CJEU confirmed the sensitivity of metadata in its recent *Tele 2 and Watson* judgment: 'that data provides the means (…) of establishing a profile of the individuals concerned, information that is no less sensitive, having regard to the right to privacy, than the actual content of communications.'[196]

Recital 14 says about metadata:

'Electronic communications data should be defined in a sufficiently broad and technology neutral way so as to encompass any information concerning the content transmitted or exchanged (electronic communications content) and the information concerning an end-user of electronic communications services processed for the purposes of transmitting, distributing or enabling the exchange of electronic communications content; including data to trace and identify the source and destination of a communication and the date, time, duration and the type of communication. *Whether such signals and the related data are conveyed by wire, radio, optical or electromagnetic means, including satellite networks, cable networks, fixed (circuit- and packet-switched, including internet) and mobile terrestrial networks, electricity cable systems, the data related to such signals should be considered as electronic communications metadata and therefore be subject to the provisions of this Regulation. Electronic communications metadata may include information that is part of the subscription to the service when such information is processed for the purposes of transmitting, distributing or exchanging electronic communications content*.'

For a phone call, such metadata include who you call and when. For calls from cell phones, metadata also include the location of the caller. For an email message, metadata include who you email and when. The ePrivacy proposal's preamble says: 'metadata derived from electronic communications (…) includes (…) the websites visited'.[197] Hence, tracking people's web browsing behaviour also concerns the processing of metadata.

Roughly speaking, the new concept of 'metadata' replaces the concepts of 'traffic data' and 'location data' from the current ePrivacy Directive.[198] It makes sense to merge 'traffic data' and 'location data' into one category of 'electronic communications metadata'. The regime in the current ePrivacy Directive, distinguishing traffic and location data, is unnecessarily complicated.[199]

---

[195] Recital 2 of the ePrivacy proposal.
[196] CJEU (Grand Chamber), Judgment of 21 December 2016, cases C-203/15 (Tele2 Sverige AB) and C-698/15 (Watson), ECLI:EU:C:2016:970, par. 99 (see also par. 98). See also CJEU, Judgement of 8 April 2014m C-293/12 (Digital Rights Ireland) and C-594/12 (Seitlinger a.o.), ECLI:EU:C:2014:238, par. 27.
[197] Recital 2 of the ePrivacy proposal.
[198] See Article 2(b) and 2(c) of the 2009 ePrivacy Directive.
[199] See on the complicated regime for location data in the current framework: Cuijpers and Pekárek 2011. See also Cuijpers, Roosendaal and Koops 2007.

**Metadata definition should be amended**

We recommend that the lawmaker adapts the definition of metadata. The Article 29 Working Party,[200] the EDPS,[201] and others[202] agree that the metadata definition is too narrow. In particular, it is insufficiently clear whether metadata generated by 'over the top' service providers are within the scope of the definition.

As the Article 29 Working Party notes, 'the definition of "electronic communications metadata" should be amended to include all data processed for the purposes of transmitting, distributing or exchanging electronic communications content.'[203] The EDPS further notes that the ePrivacy proposal only refers to 'data processed in an electronic communications network' in the definition of electronic communications metadata. The EDPS summarises: 'the definition of metadata in Article 4(3)(c) should encompass not only any data that is processed 'in an electronic communications network', but also any data that is processed by any other equipment for the provision of the service and which is not considered content.'[204] We recommend that the EU lawmaker amends the definition.

Recital 17 states: 'Location data that is generated other than in the context of providing electronic communications services should not be considered as metadata.' The intended meaning of this sentence is uncertain. The sentence might exclude too many types of location data. The Article 29 Working Party gives some examples of location data that *should* be regarded as metadata, but that might be excluded because of the sentence: location data from GPS functionality in smart devices, location data based on nearby Wi-Fi-routers and location data collected via on-board navigation.[205] We recommend amending that sentence in Recital 17. In relation to Recital 17 and location data, see also our comment on Article 6.

Distinguishing metadata and content is often difficult with communications over the internet. For example, the subject line of an email message could be seen as content or as metadata. Moreover, metadata can be as sensitive as content, and can reveal as much about somebody as content. Hence, both content and metadata deserve a high level of legal protection.

**In sum, we recommend amending the definition of metadata in line with the paragraphs above. We discuss the distinction between content and metadata in more detail in our comment on Article 6 ('Should metadata and content receive the same protection?').**

### 2.4.15. Article 4(3)(d), 'publicly available directory'

Article 4(3)(d) states:

'"Publicly available directory" means a directory of end-users of electronic communications services, whether in printed or electronic form, which is published or made available to the public or to a section of the public, including by means of a directory enquiry service.'

See also Recitals 8, 30, and 31 of the ePrivacy proposal.

**Comment**

The definition of 'publicly available directory' is used in Article 15. Recital 30 adds: 'Publicly available directories means any directory or service containing end-users information such as phone numbers (including mobile phone numbers), email address contact details and includes inquiry services.' Hence, 'publicly available directories' encompass phone books and electronic phone books.

---

[200] Article 29 Working Party 2017 (WP247), p. 16.
[201] EDPS 2017/6, p. 12.
[202] See for instance Rannenberg 2017.
[203] Article 29 Working Party 2017 (WP247), p. 16.
[204] EDPS 2017/6, p. 12.
[205] Article 29 Working Party 2017 (WP247), p. 29.

### 2.4.16.  Article 4(3)(e), 'electronic mail'

Article 4(3)(e) states:

'"Electronic mail" means any electronic message containing information such as text, voice, video, sound or image sent over an electronic communications network which can be stored in the network or in related computing facilities, or in the terminal equipment of its recipient.'

**Comment**

The phrase 'electronic mail' has a broad scope; it does not only cover what we call 'email' in daily life. The phrase 'electronic mail' also covers other types of messaging such as messaging in apps and games. Furthermore, the provision is technology neutral as far as the transmission process used is concerned. **We recommend that the EU lawmaker clarifies to what extent, for instance, Bluetooth messages are covered by the definition of 'electronic mail'.[206]** The EDPS calls for more examples in Recital 33 of ways in which electronic messages might be sent.[207]

The EDPS rightly recommends replacing the term 'electronic mail' with the more technology-neutral term 'electronic message'.[208] Recital 33 emphasises the need for technology-neutral protection against unsolicited communications. As Article 16(2) uses the definition of 'electronic mail', a sufficiently broad term is necessary to protect individuals against unsolicited communications in the future. **Indeed, the phrase 'electronic mail' may be confusing, and 'electronic message' would probably be better.**

### 2.4.17.  Article 4(3)(f), 'direct marketing communications'

Article 4(3)(f) states:

'"Direct marketing communications" means any form of advertising, whether written or oral, sent to one or more identified or identifiable end-users of electronic communications services, including the use of automated calling and communication systems with or without human interaction, electronic mail, SMS, etc.'

See also Recitals 32 and 33.

**Comment**

The phrase 'direct marketing communications' is a broad concept, defined in a reasonably technology-neutral way. Recital 33 adds: 'The degree of privacy intrusion and nuisance is considered relatively similar independently of the wide range of technologies and channels used to conduct these electronic communications, whether using automated calling and communication systems, instant messaging applications, emails, SMS, MMS, Bluetooth, etc.' Recital 32 adds, in relation to 'direct marketing communications':

'In addition to the offering of products and services for commercial purposes, [direct marketing] should also include messages sent by political parties that contact natural persons via electronic communications services in order to promote their parties. The same should apply to messages sent by other non-profit organisations to support the purposes of the organisation.'

The definition of 'direct marketing communications' is used in Article 16 of the ePrivacy proposal, on unsolicited communications. If an advertisement falls within the scope of the 'direct marketing communications' definition, Article 16 of the ePrivacy proposal ensures, in short, that such communications may only be sent following the consent of the individual concerned.

---

[206] See on Bluetooth messages and spam: Kosta, Valcke and Stevens 2009; Levallois-Barth 2012.
[207] EDPS 2017/6, p. 27.
[208] EDPS 2017/6, p. 27.

**We recommend that the EU lawmaker clarifies whether behavioural targeting falls within the scope of the definition of 'direct marketing communications'.**[209] If behavioural targeting were covered, the rules of Article 16 would apply. Article 16 generally requires the end-user's consent for direct marketing. Applying the GDPR's general rules on behavioural targeting would also lead to the conclusion that the individual's prior consent is generally required for behavioural targeting.[210] But it could improve legal clarify if the ePrivacy Regulation made that consent requirement more explicit.[211]

The Article 29 Working Party recommends amending Article 4(3)(f) and the accompanying Recitals to include all advertising *sent, directed or presented* to one or more identified or identifiable end-users.[212] We recommend, however, that careful consideration be given to that suggestion.

It may make sense if marketing communications sent to end-users via platform messaging capabilities, such as the chat functionality of social networks, fall within the scope of direct marketing communications. And it may make sense if behaviourally targeted marketing messages fall within the 'direct marketing' definition. However, *all* internet content could be said to be 'directed' to an identifiable end-user.[213] Should all online advertising fall within the scope of the 'direct marketing communications' definition? Probably not. For example, contextual advertising (such as ads for cars on a website about cars) should probably not be regarded as 'direct marketing'.

We also recommend that the lawmaker clarifies whether voice-to-voice direct marketing calls are within the scope of the 'direct marketing communications' definition.

The EDPS notes several categories of unsolicited communications that might not be included in the scope of the ePrivacy proposal's 'direct marketing communications' definition. Communications related to criminal activities such as (spear) phishing attacks and fraudulent financial proposals might not always be covered by the definition of Article 4(3)(f). The same goes for communications of a non-commercial nature such as those sent by political parties, religious or charitable organisations.[214] The EDPS suggests that Article 16 should also apply to such categories of communications.

**In sum, we recommend that the EU lawmaker clarifies the 'direct marketing communications' definition. Certain types of targeted marketing should probably fall within the definition. But the lawmaker should be careful not to extend the scope of the definition too much.**

---

[209] See on behavioural targeting also our comment on Article 8(1).
[210] See Zuiderveen Borgesius FJ 2015a.
[211] Article 8(1) of the ePrivacy proposal also applies to many behavioural targeting practices, and requires the end-user's consent.
[212] Article 29 Working Party 2017 (WP247), p. 21.
[213] As noted, under the current ePrivacy Directive, web surfing falls under the definition of communication. See our comment on Article 4(3)(b).
[214] EDPS 2017/6, p. 31.

### 2.4.18. Article 4(3)(g), 'direct marketing voice-to-voice calls'

Article 4(3)(g) defines 'direct marketing voice-to-voice calls' as:

'live calls, which do not entail the use of automated calling systems and communication systems'.

See also Recital 36.

#### Comment

The definition of 'direct marketing voice-to-voice calls' is used in Article 16(4) of the ePrivacy proposal, on unsolicited communications.

### 2.4.19. Article 4(3)(h), 'automated calling and communication systems'

Article 4(3)(h) defines 'automated calling and communication systems' as:

'systems capable of automatically initiating calls to one or more recipients in accordance with instructions set for that system, and transmitting sounds which are not live speech, including calls made using automated calling and communication systems which connect the called person to an individual'.

See also Recital 41.

#### Comment

Broadly speaking, the definition of 'automated calling and communication systems' covers systems for robocalls, and robocalls that connect the called person to a real person. The definition of 'automated calling and communication systems' is not used in the provisions of the ePrivacy proposal. But the definition is used in Recital 41 of the ePrivacy proposal. In relation to that recital, see Chapter 6 of this study and chapter VI of the ePrivacy proposal.

**We recommend that the EU lawmaker clarifies the scope of the phrase 'automated calling and communication systems'.** As suggested by the Article 29 Working Party, legal clarity can be improved by deleting the last sentence from the definition and including 'calls made with the help of semi-automated communication systems, such as for example automatic dialers, which connect the called person to an individual'[215] in the definition of Article 4(3)(h).[216]

---

[215] Article 29 Working Party 2017 (WP247), p. 34.
[216] Article 29 Working Party 2017 (WP247), p. 34.

# 3. PROTECTION OF ELECTRONIC COMMUNICATIONS OF NATURAL AND LEGAL PERSONS AND OF INFORMATION STORED IN THEIR TERMINAL EQUIPMENT (PROPOSAL CHAPTER II)

## 3.1. Article 5, confidentiality of electronic communications data

Article 5 of the ePrivacy proposal states:

'Electronic communications data shall be confidential. Any interference with electronic communications data, such as by listening, tapping, storing, monitoring, scanning or other kinds of interception, surveillance or processing of electronic communications data, by persons other than the end-users, shall be prohibited, except when permitted by this Regulation.'

See also Recitals 1, 12, 15, and 19.

### Comment

Article 5 requires that communications data (content and metadata) remain confidential. The proposed regime for electronic communications data (laid down in Articles 5 and 6 of the ePrivacy proposal) can be summarised as follows: unless a specified exception applies, consent of the end-user is required for the processing of electronic communications data by regulated services.

This regime for electronic communications data resembles the regime for special categories of data (sometimes called sensitive data) in the GDPR. In short, the GDPR prohibits the processing of special categories of data, unless a specified exception applies, or the data subject gave his or her explicit consent.[217] Electronic communications data are indeed very sensitive, so a similar regime – a prohibition with exceptions – is appropriate.

The right to communications confidentiality can be seen as the core of the ePrivacy rules.[218] **Generous protection for confidentiality of communications is in line with requirements from fundamental rights case law.** As noted previously, the right is also protected in the EU Charter of Fundamental Rights, the European Convention on Human Rights, and in many constitutions and human rights treaties.[219]

### Scope of Article 5

Recital 1 of the ePrivacy proposal states: 'Confidentiality of electronic communications ensures that information exchanged between parties and the external elements of such communication, including when the information has been sent, from where, to whom, is not to be revealed to anyone other than to the parties involved in a communication.'[220]

Recital 1 shows that Article 5 aims for broad protection, largely independent of the type of electronic communications service: 'The principle of confidentiality should apply to current and future means of communication, including calls, internet access, instant messaging applications, e-mail, internet phone calls and personal messaging provided through social media.'

---

[217] Article 9 of the GDPR.
[218] See Steenbruggen 2009. Article 5(1) of the ePrivacy proposal is the successor of Article 5(1) of the 2009 ePrivacy Directive.
[219] See our introduction on privacy, confidentiality of communications, and related rights in this study (chapter 1).
[220] Recital 1 of the ePrivacy proposal.

The EDPS rightly notes that the last sentence of Recital 5 could be misunderstood. That sentence reads: 'Processing of electronic communications data by providers of electronic communications services should only be permitted in accordance with this Regulation.' The EDPS is concerned that the sentence might be misread as implying that 'processing electronic communications data by parties other than providers of electronic communications services does not come under the scope of the ePrivacy Regulation.'[221]

But, says the EDPS, 'Processing of electronic communications data (…) should unambiguously come under the scope of the ePrivacy Regulation, irrespective of which entity processes such data.'[222] Therefore, the EDPS suggests that the sentence in Recital 5 be replaced with the following sentence: 'processing of electronic communications data should only be permitted in accordance with, and on a legal ground specifically provided under, this Regulation'.[223] **We recommend that the EU lawmaker considers amending Recital 5.**

### Machine-to-machine communications

Recital 12 shows that the right to communications confidentiality also applies to machine-to-machine communications and to many Internet of Things scenarios:

'Connected devices and machines increasingly communicate with each other by using electronic communications networks (Internet of Things). The transmission of machine-to-machine communications involves the conveyance of signals over a network and, hence, usually constitutes an electronic communications service. *In order to ensure full protection of the rights to privacy and confidentiality of communications, and to promote a trusted and secure Internet of Things in the digital single market, it is necessary to clarify that this Regulation should apply to the transmission of machine-to-machine communications. Therefore, the principle of confidentiality enshrined in this Regulation should also apply to the transmission of machine-to-machine communications.* (…)'[224]

The Article 29 Working Party and the EDPS agree that machine-to-machine communications should be protected against interference. The Working Party gives the example of Intelligent Transport Systems. In such systems, 'vehicles will continuously transmit data containing a unique identifier, via radio. Without the additional protection in the ePrivacy Regulation regarding communications data, this could lead to continuous tracking of the driving habits, itineraries and speed of the drivers.'[225] However, the Working Party suggests stating in an article (rather than a recital) that machine-to-machine communications deserve protection.[226]

The Working Party also suggests that for some exceptional cases, a specified exception should be added to the main rule (that requires machine-to-machine communications to remain confidential). According to the Working Party, 'a narrow category of pure machine-to-machine communication[s] should be exempted if they have no impact on either privacy or the confidentiality of communications, such as for example the cases where such communication is performed in execution of a transmission protocol between network elements (e.g. servers, switches,) to inform each other on their status of activity.'[227]

Furthermore, the Working Party discusses the example of self-driving cars and other devices that can 'warn each other about their vicinity or other risks.'[228] In such cases, consent from end-users is not workable. 'Providers should therefore be able to rely on a specific exception, allowing objects such as self-driving cars and devices to warn each other about their vicinity or other risks.'[229]

---

[221] EDPS 2017/6, p. 16.
[222] EDPS 2017/6, p. 16.
[223] EDPS 2017/6, p. 16.
[224] Emphasis added.
[225] Article 29 Working Party 2017 (WP247), p. 28.
[226] Article 29 Working Party 2017 (WP247), p. 28. See also EDPS 2017/6, p. 9 and p. 18.
[227] Article 29 Working Party 2017 (WP247), p. 28.
[228] Article 29 Working Party 2017 (WP247), p. 28.

**We recommend that the EU lawmaker seriously considers the suggestion of the Working Party regarding communications confidentiality and machine-to-machine communications. More generally, we recommend that the lawmaker considers whether other exceptions are necessary. Such exceptions should be narrowly defined, and should only apply when an exception is strictly necessary for a social good.**

### Stored communications

Article 5 seems to suggest that communications should be protected when they are in transit and when they are stored (in the cloud for instance): 'Any interference with electronic communications data, such as (…) *storing* (…) or other kinds of interception, surveillance or processing of electronic communications data, by persons other than the end-users, shall be prohibited (…).'

However, a sentence from Recital 15 could be interpreted as meaning that the right to communications confidentiality only applies during the transmission, or transit, of electronic communications.[230] Recital 15 states: 'The prohibition of interception of communications data should apply *during their conveyance*, i.e. until receipt of the content of the electronic communication by the intended addressee.'

That scope (limited to transit) would be too narrow; say both the Article 29 Working Party and the EDPS. They note that Recital 15 (focusing on communications in transit) is based on an outdated conceptual framework of communications. However, say the regulators, 'Most communications data remain stored with service providers, even after receipt.'[231] As the Working Party explains, 'communication between subscribers of the same cloud-based services (for instance webmail providers) will often entail only very little conveyance: sending a mail would mostly involve reflecting this in the database of the provider, rather then actually sending communications between two parties.'[232]

Moreover, it is not always easy for users to recognise whether communications are in transit or stored. Data can be in transit, or be stored on external servers (in the cloud), while the user assumes the data are on his or her device.[233]

The Article 29 Working Party and the EDPS say that the Regulation should give 'the same level of protection for communications stored on other equipment than user terminals, e.g. in mailboxes operated by a service provider or any cloud storage used as part of the communications service.'[234]

The GDPR can provide some protection to many stored communications data. However, the GDPR only protects communications that contain or are linked to personal data. Moreover, the general rules in the GDPR are not appropriate for electronic communications data, which are very sensitive.[235] Some stored communications are protected by Article 8 of the ePrivacy proposal. But Article 8 only applies to terminal equipment, so the protection of Article 8 may be insufficient. For instance, Article 8 does not apply to communications stored in the cloud.

---

[229] Article 29 Working Party 2017 (WP247), p. 28.
[230] See Article 29 Working Party 2017 (WP247), p. 26.
[231] Article 29 Working Party 2017 (WP247), p. 26. See also EDPS 2017/6, p. 26.
[232] Article 29 Working Party 2017 (WP247), p. 26. See also EDPS 2017/6, p. 26.
[233] See generally on personal records in the cloud: Irion 2015.
[234] EDPS 2017/6, p. 26. Protecting communications before and after conveyance fits 'the right to confidentiality of communications in the wide sense' (Steenbruggen 2009, p. 354).
[235] See our introduction on the need for ePrivacy rules in addition to the GDPR in this study (chapter 1).

Some might argue that the right to communications confidentiality should be limited to communications in transit.[236] Even if that argument were followed, stored communications often deserve protection on the basis of the general right to privacy, granted in Article 7 of the EU Charter and Article 8 of the European Convention on Human Rights.[237]

**We also recommend that the lawmaker further clarifies the scope of Article 5.** For instance, according to the Article 29 Working Party and the EDPS, the rules should also apply to 'all in-platform messages between users of a social network (such as Facebook or Twitter).'[238] Such a clarification could be included in the preamble. In addition, the EDPS suggests that Recital 1 should specify that 'the notion of communication does not only include electronic communication between two individuals (or machines) but also any communications within a defined group (e.g. a conference call, or messages sent to a defined group of recipients).'[239] **In sum, we recommend that the EU lawmaker considers ensuring that communications data are also protected when the data are stored in the cloud.**

### Interception and interference

Article 5 prohibits 'interception' of electronic communications data. Recital 15 elaborates on the meaning of the word 'interception':

'Interception of electronic communications data may occur, for example, when someone other than the communicating parties, listens to calls, reads, scans or stores the content of electronic communications, or the associated metadata for purposes other than the exchange of communications. Interception also occurs when third parties monitor websites visited, timing of the visits, interaction with others, etc., without the consent of the end-user concerned. As technology evolves, the technical ways to engage in interception have also increased. Such ways may range from the installation of equipment that gathers data from terminal equipment over targeted areas, such as the so-called IMSI (International Mobile Subscriber Identity) catchers, to programs and techniques that, for example, surreptitiously monitor browsing habits for the purpose of creating end-user profiles. Other examples of interception include capturing payload data or content data from unencrypted wireless networks and routers, including browsing habits without the end-users' consent.'

Hence, monitoring metadata, and tracking people's browsing behaviour fall within the scope of 'interception'.[240] Interception is in principle prohibited – but can be legitimised by an exception or end-user consent (see Articles 5 and 6).

**The phrase 'interference' should be clarified. We recommend that the lawmaker considers clarifying that injecting content in communications also constitutes an interference with electronic communications data.** For instance, a company could intercept internet traffic, and inject advertising or identifiers ('super cookies').[241] Injecting ads into internet traffic presumably falls within the scope of 'interception (…) or processing of electronic communications data' (Article 5 of the proposal). But **it would be useful if the lawmaker stated explicitly that injecting ads or other content violates communications confidentiality.**

---

[236] Such an approach could be called 'confidentiality of communications in the restricted sense' (Steenbruggen 2009, p. 354).
[237] That approach could be called 'confidentiality of communications in the wide sense (Steenbruggen 2009, p. 354). See also Koops and Smits 2014, p. 141.
[238] EDPS 2017/6, p. 26. See also Article 29 Working Party 2017 (WP247), p. 27-28.
[239] EDPS 2017/6, p. 26.
[240] Recital 15: 'Interception also occurs when third parties monitor websites visited, timing of the visits, interaction with others, etc., without the consent of the end-user concerned.' See also Recital 2, which mentions 'the websites visited' as an example of electronic communications metadata.
[241] See our comment on Article 6 ('stricter rules for telecom providers'), in particular on Phorm. See also Thomson 2015: certain telecom providers use a 'super cookie': a token unique to each subscriber that is injected into every HTTP request made through a telco's cellphone networks.'

## 3.2. Article 6, permitted processing of electronic communications data

### 3.2.1. Article 6(1), electronic communications data

Article 6(1), on electronic communications data reads as follows:

Article 6(1)
'Providers of electronic communications networks and services may process electronic communications data if:
(a) it is necessary to achieve the transmission of the communication, for the duration necessary for that purpose; or
(b) it is necessary to maintain or restore the security of electronic communications networks and services, or detect technical faults and/or errors in the transmission of electronic communications, for the duration necessary for that purpose.'

See also Recitals 16, 17, 18, and 19.

**Comment**
Article 5 says, in short, that electronic communications are confidential. Article 6(1) provides exceptions to the principle of communications confidentiality for electronic communications *data* (metadata and content). Article 6(2) provides exceptions regarding *metadata*; Article 6(3) regarding *content*.

Article 6(1) provides two exceptions to the confidentiality of communications data, for (a) transmission and (b) security. We discuss each exception below.

**Transmission**
Exception (a) states: 'Providers of electronic communications networks and services may process electronic communications data if: (a) it is necessary to achieve the transmission of the communication, for the duration necessary for that purpose.' Recital 16 adds: 'The prohibition of storage of communications is not intended to prohibit any automatic, intermediate and transient storage of this information insofar as this takes place for the sole purpose of carrying out the transmission in the electronic communications network.'

The Article 29 Working Party and the EDPS say that Article 6 should use the phrase 'strictly necessary' instead of 'necessary'. The addition of the word strictly would emphasise that 'the necessity-test in the context of this regulation should be interpreted narrowly'.[242] We agree with that advice.

**Security**
Regarding exception (b), security and detecting errors, Recital 16 adds: 'The prohibition of storage of communications (…) should not prohibit (…) the processing of electronic communications data to ensure the security and continuity of the electronic communications services, including checking security threats such as the presence of malware'.

According to the Article 29 Working Party and the EDPS, the security exception 'must be narrowly construed and limited to what is strictly necessary.[243]

Under Article 6(1)(b), 'providers of electronic communications networks and services' can rely on the security exception. Some have suggested that parties other than such providers should also be allowed to rely on the security exception.[244] However, it is not immediately apparent for which situations such an extended exception would be necessary. Presumably, no extension is needed for a situation in which a provider hires a security company as a sub-contractor.[245] The further processing of electronic communications data by such sub-contractors for their own purposes should generally be prevented. We recommend that the EU lawmaker carefully examines whether an extension of the security exception is appropriate.

---

[242] Article 29 Working Party 2017 (WP247), p. 20. See also EDPS 2017/6, p. 27.
[243] EDPS 2017/6, p. 27. See also Article 29 Working Party 2017 (WP247), p. 20.
[244] Chantzos 2017, slide 7.

**In sum, exceptions along the lines of Article 6(1) do indeed seem necessary. Hence, we recommend that the EU lawmaker retains the gist of Article 6(1), but takes into account the recommendations above. Because of the sensitivity of communications data, the lawmaker should use the phrase 'strictly necessary' to emphasise that the exceptions should be interpreted narrowly.**

### 3.2.2. Article 6(2)(a), metadata, quality of service

Article 6(2)(a) reads as follows:

'Providers of electronic communications services may process electronic communications metadata if:
(a) it is necessary to meet mandatory quality of service requirements pursuant to [the Directive establishing the European Electronic Communications Code] or [the Net Neutrality and Roaming] Regulation (EU) 2015/2120[246] for the duration necessary for that purpose'.

See also Recitals 16, 17, 18, and 19.

**Comment**
Article 6(2) provides three exceptions to the principle of communications confidentiality, regarding *metadata*. In brief, the exceptions concern (i) quality of service, (ii) billing, and (iii) consent. If one of the exceptions applies, a provider of electronic communications services may, as an exception to Article 5's prohibition, process or interfere with electronic communications metadata.

Article 6(2)(a) provides an exception for mandatory quality of service requirements. If a provider processes metadata for mandatory quality of service requirements, Article 6(2)(a) obliges the provider to restrict the processing to 'the duration necessary for that purpose'.

Quality of service is regulated in Article 97 of the draft European Electronic Communications Code, and in Article 4 of the Net Neutrality and Roaming Regulation.[247] Quality of service refers to the technical capabilities and performance of a connection. Quality of service management can help to prevent latency, jitter, packet loss, and network congestion.[248] End-users should be informed of any quality of service management applied to their connections, says the draft European Electronic Communications Code.[249] And quality of service management should be 'transparent, non-discriminatory and proportionate, and should not be based on commercial considerations.'[250]

---

[245] Such a security company would likely have the role of a 'processor' under the GDPR, as the provider determines the purposes and means of processing electronic communications data (personal data). The security provider would process electronic communications data (personal data) on behalf of the provider. See Article 4(7) and 4(8) of the GDPR.
[246] Original footnote: Regulation (EU) 2015/2120 of the European Parliament and of the Council of 25 November 2015 laying down measures concerning open internet access and amending Directive 2002/22/EC on universal service and users' rights relating to electronic communications networks and services and Regulation (EU) No 531/2012 on roaming on public mobile communications networks within the Union (OJ L 310, 26.11.2015, p. 1–18).
[247] Regulation (EU) 2015/2120 of the European Parliament and of the Council of 25 November 2015 laying down measures concerning open internet access and amending Directive 2002/22/EC on universal service and users' rights relating to electronic communications networks and services and Regulation (EU) No 531/2012 on roaming on public mobile communications networks within the Union (OJ L 310, 26.11.2015, p. 1–18). See also Recitals 9 and 17 of that Regulation.
[248] Recital 17 of Regulation 2015/2120. See also Recital 17 of the ePrivacy proposal.
[249] Recital 233 of the draft European Electronic Communications Code.

To optimise transmission quality, it might be necessary for providers of electronic communications services to differentiate between different categories of traffic. Hence, to meet mandatory quality of service requirements, it can indeed be useful, perhaps necessary, for such providers to process electronic communications metadata.

### 3.2.3.    Article 6(2)(b), metadata, billing etc.

Article 6(2)(b) reads as follows:

'Providers of electronic communications services may process electronic communications metadata if: (…)
(b) it is necessary for billing, calculating interconnection payments, detecting or stopping fraudulent, or abusive use of, or subscription to, electronic communications services'.

See also Recitals 16, 17, 18, and 19.

**Comment**
Metadata may be processed by providers of electronic communications services if that processing is 'necessary for billing, calculating interconnection payments, detecting or stopping fraudulent, or abusive use of, or subscription to, electronic communications services' (Article 6(2)(b)). An exception along the lines of Article 6(2)(b) is appropriate; a similar exception is included in the current ePrivacy Directive.[251]

It makes sense if the law allows, for instance, phone companies to process metadata for billing. A phone company must send a bill to its subscribers, and the subscribers want to know why they have to pay a certain fee. **Therefore, we recommend that the EU lawmaker retains the gist of Article 6(2)(b).**

**See also our comment below ('the same protection for content and metadata?'). And see Article 7(3), which discusses the retention period for metadata stored for billing.**

### 3.2.4.    Article 6(2)(c), metadata, consent

Article 6(2)(c) reads as follows:

'Providers of electronic communications services may process electronic communications metadata if: (…)
(c) the end-user concerned has given his or her consent to the processing of his or her communications metadata for one or more specified purposes, including for the provision of specific services to such end-users, provided that the purpose or purposes concerned could not be fulfilled by processing information that is made anonymous.'

See also Recitals 16, 17, 18, and 19.

**Comment, extended possibilities to process metadata**
Compared to the current ePrivacy Directive, Article 6(2) of the ePrivacy proposal extends the possibilities for telecom providers to interfere with the right to communications confidentiality. We use the phrase telecom providers as shorthand for companies that fall within the scope of 'electronic communications services' under the *current* ePrivacy

---

[250] Recital 9 of Regulation 2015/2120.
[251] Article 6(2) of the 2009 ePrivacy Directive.

Directive.[252] Such telecom providers are mostly internet access providers and phone companies.

Recital 17 states: 'The processing of electronic communications data can be useful for businesses, consumers and society as a whole. Vis-à-vis [the ePrivacy Directive], this Regulation *broadens* the possibilities for providers of electronic communications services to process electronic communications metadata, based on end-users consent.'[253]

**We recommend that the EU lawmaker carefully considers whether it accepts that the ePrivacy proposal lowers the protection of privacy and communications confidentiality in the context of telecom providers.**

### Location data
Recital 17 discusses the processing of electronic communications data and location data. Location data can be very sensitive. Location data could disclose, for instance, a visit to a mosque, church, hospital, rehab centre, or abortion clinic.[254] Moreover, in many circumstances people cannot prevent collection of location data concerning them. For instance, providers offering mobile phone services learn the end-user's location.[255]

Recital 17 contains several ambiguities and weaknesses. We recommend that the EU lawmaker addresses them. First, it is not clear whether the recital aims to discuss metadata and location data in the context of Article 6 or Article 8, or both.[256]

Second, the recital states that '[t]o display the traffic movements in certain directions during a certain period of time, an identifier is necessary to link the positions of individuals at certain time intervals.' Several commentators suggest that this sentence is incorrect, because anonymous data suffice for showing traffic movements. Hence, an identifier or pseudonym is not needed to show traffic movements.[257] The Article 29 Working Party recommends that the recital clarifies 'that most legitimate processing of location data and other metadata does *not* require a unique identifier.'[258]

Third, the recital states: 'Examples of commercial usages of electronic communications metadata by providers of electronic communications services may include the provision of heatmaps; a graphical representation of data using colors to indicate the presence of individuals.' It is unclear whether the proposal intends to allow such practices without end-users' consent.

Fourth: if the lawmaker chooses to allow the processing of location data without the individual's consent, the Article 29 Working Party recommends that Recital 17 'at the very least' specifies the following safeguards. The recital should clarify that providers that process metadata (for purposes other than the original purpose) should have to comply with the GDPR's data protection by design and by default requirements.[259] According to the Working Party, organisations that use location tracking should, at a minimum, take the following measures:

(i) the use of temporary pseudonyms;
(ii) deletion of any reverse look-up table between these pseudonyms and the original identifying data;

---

[252] Article 2(c) of the Council Directive 2002/21 of 7 March 2002 on a common regulatory framework for electronic communications networks and services as amended by Directive 2009/140/EC and Regulation 544/2009 (Framework Directive) (OJ L 108, 24.4.2002, p. 33–50).
[253] Emphasis added.
[254] See also Recital 2 of the ePrivacy proposal. See also EDPS 2017/6, p. 19.
[255] Article 29 Working Party 2017 (WP247), p. 29-30. See also Recital 14 of the ePrivacy proposal.
[256] See Article 29 Working Party 2017 (WP247), p. 29-30.
[257] Rannenberg 2017; Article 29 Working Party 2017 (WP247), p. 29-30.
[258] Article 29 Working Party 2017 (WP247), p. 30 (emphasis original).
[259] Article 25 of the GDPR.

(iii) aggregation to a level where individual users can no longer be identified through their particular itineraries, and;

(iv) the deletion of outliers with regard to which identification would still be possible (all of these measures need to be applied together).'[260]

**In sum, location data are exceedingly sensitive. We recommend that the EU lawmaker addresses Recital 17's weaknesses and ambiguities. We urge the lawmaker to carefully consider whether it wants to allow the use of anonymised metadata, without people's consent, for heatmaps or other purposes. From a privacy perspective, it would probably be better not to allow such practices. See also our comment below (section 3.2.6) on consent in Article 6.**

### 3.2.5. Article 6(3)(a) and 6(3)(b), content

Article 6(3) reads as follows:

Article 6(3)
'Providers of the electronic communications services may process electronic communications content only:
(a) for the sole purpose of the provision of a specific service to an end-user, if the end-user or end-users concerned have given their consent to the processing of his or her electronic communications content and the provision of that service cannot be fulfilled without the processing of such content; or
(b) if all end-users concerned have given their consent to the processing of their electronic communications content for one or more specified purposes that cannot be fulfilled by processing information that is made anonymous, and the provider has consulted the supervisory authority. Points (2) and (3) of Article 36 of [the GDPR] shall apply to the consultation of the supervisory authority.'

See also Recitals 16, 17, 18 and 19.

**Comment: clarify difference between (a) and (b)**
Article 6(3) concerns communications *content*. The provision provides two exceptions to the principle of confidentiality of communications regarding content. If one of the two exceptions applies, a provider of electronic communications services may, contrary to the main rule in Article 5, process the content.

It is not immediately apparent to which types of situations exceptions (a) and (b) apply. One way of interpreting the provision is roughly as follows. Exception (a) may apply if an *user* explicitly requests, from the provider, a service that can only be provided by analysing content. For instance, a user might explicitly ask a webmail provider to filter spam from his or her incoming emails. Filtering for spam often entails analysing the contents of emails.[261]

Exception (b) may apply when a *provider* wants to process communications content (and thus interfere with communications confidentiality). An example of situation (b) might be an email provider that analyses the content of emails to present the user with targeted advertising (a form of behavioural targeting).[262] Hence, under the proposed provision, an email provider could ask its users to consent to behavioural targeting practices based on analysing email content.

**If the EU lawmaker chooses to retain Article 6(3), we recommend clarifying the difference between situations (a) and (b). We discuss consent in the context of**

---

[260] Article 29 Working Party 2017 (WP247), p. 30.
[261] Some spam filtering based on metadata could perhaps be based on Article 6(2)(b). See also our comment on Article 6 below ('the same protection for content and metadata?').
[262] Some email providers analyse the content of emails for behavioural targeting. Gmail is the most well known example, but other providers do the same.

**Article 6 in more detail below.** See also our comments on the 'household exception' and 'the same protection for content and metadata' below.

### Extended possibilities to process content

While Recital 17 states that the ePrivacy proposal 'broadens the possibilities for providers of electronic communications services to process electronic communications *metadata',* the proposal also broadens the possibilities for such providers to process content. Meanwhile, as Recital 17 notes, 'end-users attach great importance to the confidentiality of their communications, including their online activities.' Moreover, people 'want to control the use of electronic communications data for purposes other than conveying the communication'.[263]

Hence, the ePrivacy proposal *reduces* the protection for privacy and confidentiality of communications, where telecom providers offer communication services. The current ePrivacy Directive contains important and strict rules to protect the confidentiality of communications.[264] Because of the limitation of the scope of the current ePrivacy Directive to the electronic communications sector, many companies (over the top service providers) do not have to comply with those important rules.[265] Companies outside the scope of the current ePrivacy Directive sometimes interfere with communications confidentiality in ways that would never be accepted (regardless of the law) from telecom providers. For instance, some webmail providers analyse the metadata and content of email messages for targeted marketing. Now the Commission aims to foster a level playing field for telecom providers and 'over the top' service providers, partly by lowering the requirements for telecom providers.

**The EU lawmaker should consider, very carefully, whether it accepts that the ePrivacy proposal reduces the protection of privacy and communications confidentiality in the context of telecom providers. See also our comments below, on 'stricter rules for telecom providers?' and 'the same protection for content and metadata?'.**

### 3.2.6.    Article 6, consent

### Consent of all end-users should be required

Article 6(2)(c) requires consent of the 'end-user concerned' (singular) for, in short, processing metadata. Article 6(3)(a) requires consent of the 'end-user or end-users concerned' for processing content, under certain circumstances.

However, communications usually concern two end-users, such as a sender and a recipient of an email.[266] The EDPS 'recommends that in each case where end-user consent is required, the same phrase, "all end-users" be consistently used throughout the Proposal.'[267] For that suggestion, the EDPS assumes that the definition of 'end-user' is amended as suggested by the EDPS.[268] (See our comment on Article 4(1)(b), 'end-user'.) The Working Party adds: 'the analysis of content and/or metadata for (…) analytics, profiling, behavioural advertising or other purposes for the (commercial) benefit of the provider, requires consent from all end-users whose data would be processed.'[269]

In addition, the Working Party rightly proposes the following clarification regarding consent:

'The Proposed Regulation should explain that the mere act of sending an e-mail or other kind of personal communication from another service to an end-user that has personally consented to the processing of his or her content and metadata (for example in the course of signing up to a mailservice), does not constitute valid consent from the sender.'[270]

---

[263] Recital 17 of the ePrivacy proposal.
[264] Article 5, 6, and 9 of the 2009 ePrivacy Directive.
[265] Or at least: it is unclear to what extent over the top service providers must comply with Article 5(1) of the 2009 ePrivacy Directive. See Zuiderveen Borgesius 2015, p. 174-175.
[266] Machine-to-machine communications are among the exceptions. See our comment on Article 5.
[267] EDPS 2017/6, p. 15.
[268] See our comment on Article 4(1)(b).
[269] Article 29 Working Party 2017 (WP247), p. 13. See also EDPS 2017/6, p. 15.

**We recommend that the EU lawmaker clarifies that the consent of all end-users is required for processing of electronic communications data (if the processing does not fall under the other exceptions). The lawmaker should also clarify that merely contacting another end-user does not signify consent.**

### Consent of non-users

Suppose that an email provider analyses the content of emails for targeted marketing. Suppose that Alice and Bob both use that email provider, and that both consented to the email provider analysing their emails. Alice emails Bob a message with detailed information about Carol. The email provider stores and analyses the message, and learns many details about Carol. But Carol has not consented to any analysis. Hence, Carol's privacy is invaded.

The Article 29 Working Party and the EDPS disapprove of such situations. As the Working Party notes, 'it should be clarified that the processing of data of persons other than the end-users (e.g. the picture or description of a third person in an exchange between two people) involved also needs to comply with all relevant provisions of the GDPR.'[271] The EDPS recommends that the EU lawmaker adds the following provision:

'Any processing based on end-user consent must not adversely affect the rights and freedoms of individuals whose personal data are related to the communication, in particular, their rights to privacy and the protection of their personal data'[272]

**We recommend that the EU lawmaker considers adding a provision along these lines, to ensure that the rights of non-end-users are respected.**

See also our comment on Article 1(3), on consent and the relationship between the ePrivacy Regulation and the GDPR, and on our comment on Article 9, on consent.

### 3.2.7.    Article 6, anonymous data

The references to anonymisation in Article 6 are somewhat ambiguous. For example, Article 6(2)(c) states that certain providers can process metadata if the end-user gave consent, 'provided that the purpose or purposes concerned could not be fulfilled by processing information that is made anonymous.'

First, the quoted phrase can be interpreted as follows: Even after the end-user has consented to processing for a certain purpose, the provider must assess whether it can fulfil that purpose by using only anonymous and aggregated data. That interpretation would be in line with the general data protection principles of data minimisation[273] and storage limitation,[274] and the general principle of proportionality.[275]

Another interpretation of the phrase could be as follows: as long as a provider processes anonymised data, the provider does not even have to ask consent of the end-user. Article 6(3)(b) states that content may be processed 'if all end-users concerned have given their consent to the processing of their electronic communications content for one or more specified purposes *that cannot be fulfilled by processing information that is made anonymous*, and the provider has consulted the supervisory authority.'

---

A provider might think that it can analyse people's communications content without their consent, as long as it aggregates and anonymises the data it collects. A provider might be tempted to train its machine learning system on the contents of people's emails, for instance to train the system's translation capabilities. A provider might even be tempted to analyse people's emails for targeted marketing without their consent, thinking that building group profiles entails processing 'information that is made anonymous'. (True, when the provider consults the supervisory authority, as required by the provision, that authority could warn against wrong interpretations.)

In addition, **it is difficult to see how communications content could be properly anonymised.** For instance, email messages often contain information about identifiable individuals.

If Article 6 is interpreted as not requiring consent for using anonymised content, Article 6 may fail to protect privacy, communications confidentiality, and related rights. Anonymisation does not guarantee that fundamental rights are respected. For instance, anonymous and aggregated group profiles could be used to make predictions about an individual, and could even be used for discriminatory purposes.[276] And as Gürses notes, anonymisation could prevent people 'from understanding, scrutinising, and questioning the ways in which these data sets are used to organise and affect their access to resources and connections to a networked world'.[277]

**We recommend that the lawmaker clarifies the meaning of references to anonymisation in Article 6. More generally, we recommend that the lawmaker keeps in mind that anonymising data does not take away all threats to end-users' fundamental rights.**

In this context, a suggestion by the Article 29 Working Party may be useful: 'even when anonymisation measures are to be applied, providers should always conduct a data protection impact assessment.'[278] The EU lawmaker should consider including such a requirement in the ePrivacy Regulation.

The Working Party also 'calls for an additional obligation to make public how the data are anonymised and aggregated.'[279] Re-identification researchers have made similar suggestions. For example, Narayanan, Huey and Felten say: 'wherever notice about data collection can be given, a short statement should be included that briefly describes what steps will be taken to protect privacy and notes whether records may be re-identified despite those steps.'[280] We recommend that the EU lawmaker seriously considers those recommendations.

**We seriously doubt whether it is acceptable to analyse the contents of people's communications without their consent, even if the communications are anonymised. Perhaps there could be, in some rare circumstances, an objective reason to introduce a specific legal obligation to do such content analysis. But the law should be clear on such obligations. And obligations to analyse content should always respect privacy, confidentiality of communications, freedom of expression, and other fundamental rights. In sum, the EU lawmaker should clarify Article 6 regarding anonymisation.**

---

[276] Barocas and Nissenbaum 2014; Gutwirth and Hildebrandt 2008; Oostveen 2016.
[277] Gürses 2014. See also Van Hoboken & Zuiderveen Borgesius 2015.
[278] Article 29 Working Party 2017 (WP247), p. 9. See about data protection impact assessments: Article 35 of the GDPR. See generally: Wright and De Hert 2011; Kloza 2014.
[279] Article 29 Working Party 2017 (WP247), p. 9.
[280] Narayanan, Huey and Felten 2016, p. 375.

### 3.2.8.    Article 6, a household exception?

We recommend that the EU lawmaker considers introducing a household exception for Article 6. Such an exception would make it possible, for instance, that somebody asks a provider of electronic communications services to analyse email messages for him or her, while no consent would have to be obtained from the sender of the email.

For instance, suppose Alice receives email messages from Bob. She can search her email inbox for emails by Bob, or for emails mentioning 'birthday'. Alice's webmail provider, a provider of electronic communications services, offers that functionality. Hence, the provider processes electronic communications content for Alice: the provider analyses the contents of emails. Article 6(3)(a) of the ePrivacy proposal might require Alice (or the provider) to ask Bob for consent for analysing his emails. But it does not seem necessary, from a privacy and confidentiality perspective, to bother Bob with such a consent request.

Therefore, a household exception could be useful. Under the household exception, Bob would not have to give consent before Alice (or her provider) can analyse his emails. If Alice would use an email service on her company phone, she should still be able to rely on the household exception, under certain circumstances.[281]

Another example. Say that a Voice over IP service (an internet phone service) offers real-time translation tools. Alice lets her provider of the electronic communications service (the Voice over IP service) automatically translate all incoming French phone calls into English. Under the household exception, Bob would not have to give consent before Alice (or her provider) translates his calls.

The Article 29 Working Party and the EDPS argue for the introduction of a type of household exception:  'It should be made possible to process electronic communications data for the purposes of providing services explicitly requested by an end-user, such as search or keyword indexing functionality, virtual assistants, text-to-speech engines and translation services.'[282]

According to the Working Party, the EU lawmaker should introduce 'an exemption for the analysis of such data for purely individual (household) usage, as well as for individual work related usage. This would thus be possible without the consent of all end-users, but may only take place with the consent of the end-user requesting the service. Such a specific consent would also preclude the provider from using these data for different purposes.'[283]

**In sum, a type of household exception (applicable to Article 6) could be useful. The EU lawmaker should consider introducing one. For such a household exception, inspiration could be drawn from the household exception in the GDPR.[284] Such an exception should only apply to processing specifically requested by the end-user, and the requested processing should not disproportionally affect the fundamental rights of other end-users. The lawmaker should ensure that such an exception does not create a loophole that enables further processing for other purposes.**

### 3.2.9.    Article 6, no legitimate interests provision should be added

We argue against adding a balancing provision, also called a legitimate interests provision, to the ePrivacy Regulation. Some have suggested that a new rule should be added to the ePrivacy Regulation (especially to Article 6 and 8).[285] Such a new rule should, so the argument goes, be comparable with the balancing provision in the GDPR (Article 6(1)(f)). In short, the GDPR's balancing provision allows personal data processing if a company's interests outweigh the data subject's interests or fundamental rights. If a company can rely on the balancing provision for personal data processing, it does not have to ask the individual concerned for prior consent.

---

[281] See our comment on Article 4(1)(b) on 'end-user', and the EDPS suggestions regarding that definition.
[282] Article 29 Working Party 2017 (WP247), p. 13. See also EDPS 2017/6, p. 15.
[283] Article 29 Working Party 2017 (WP247), p. 13.
[284] Article 2(2)(c) GDPR: 'This Regulation does not apply to the processing of personal data: (...) by a natural person in the course of a purely personal or household activity'.
[285] See for instance: Interactive Advertising Bureau Europe 2017; European Telecommunications Network Operators' Association (ETNO) 2017.

The Article 29 Working Party and the EDPS also argue against introducing a balancing provision: 'the introduction of open-ended exceptions along the lines of Article 6 GDPR, and in particular Art. 6(f) GDPR (legitimate interest ground), should be avoided.'[286]

There are several arguments against introducing a balancing provision. First, Article 6 concerns extremely sensitive and private situations. Article 6 covers, for instance, phone calls, Skype calls, emails, WhatsApp messages, and related metadata. For such situations, a regime along the following lines fits best: In principle it is prohibited to process communications data, subject to strictly interpreted limited exceptions (including the consent of the end-users). A similar regime is included in the GDPR for 'special categories' of personal data, sometimes called sensitive data.[287] With the Data Protection Directive's regime for special categories of data, there is more than twenty years of experience with a regime without a balancing provision. Such a prohibition-with-exceptions regime fits better when protecting communications data. Indeed, the ePrivacy proposal contains a similar regime for electronic communications data.

Second, to regulate such sensitive and private situations, an open norm such as the balancing provision is too vague. An open norm such as the balancing provision is useful for *general* data protection law, because general data protection law applies to many situations and sectors. In a more specific regime, such as the ePrivacy regime, there is less need for open norms. Indeed, such open norms would undermine the value of adopting a specific set of rules for communications.

Some might argue that technology develops too fast for a regime without a balancing provision. They might say that in a few years, a new situation might turn up with the following characteristics: (i) it is not feasible that providers of the electronic communications ask consent for processing electronic communications data; (ii) there is a serious public interest to allow such processing; (iii) the processing cannot be based on one of the exceptions. For such situations, so the argument goes, an open-ended balancing provision is needed. There is some merit to the argument that open norms are practical for quickly developing technology.

However, if such an unforeseen situation occurs, and another exception would be needed, the rules can be revised. The ePrivacy rules have been revised two times since 1997, excluding the current proposal. And Article 28 of the ePrivacy proposal requires that the ePrivacy Regulation is evaluated every three years. 'The evaluation shall, where appropriate, inform a proposal for the amendment or repeal of this Regulation in light of legal, technical or economic developments.' Hence, there is a system in place to amend the ePrivacy Regulation if needed.

**In conclusion, we recommend that the EU lawmaker does not add a 'legitimate interests provision' to the ePrivacy Regulation.**

### 3.2.10.    Article 6, stricter rules for telecom providers?

Providers of communications infrastructure (traditional telecom providers) are in a position to interfere with people's communications at a different level than services that merely use the infrastructure. For instance, internet networks *constitute* the electronic communications infrastructure, whereas over-the top services merely *use* such infrastructure to provide their services to end-users. Regulating the privacy conditions of infrastructure services (for instance under what conditions they can process content and metadata) also affects the privacy conditions of services that are available over such infrastructure, including 'over the top' services. Infrastructure providers could have a more significant impact on communications privacy than providers whose services do not qualify as infrastructure.

---

[286] Article 29 Working Party 2017 (WP247), p. 7.
[287] See Article 9 of the GDPR.

In the current ePrivacy Directive, many rules only apply to 'providers of publicly available electronic communications services', and 'providers of public communications networks', as they are *currently* defined.[288] In practice, mostly internet access providers and phone providers – telecom providers for short – fall in those legal categories. Hence, the current ePrivacy Directive has stricter rules for telecom providers than for 'over the top' service providers. That focus on communications infrastructure in the current ePrivacy Directive can be defended on the basis of the infrastructural nature of certain service providers.[289]

When drafting the new ePrivacy Regulation, the European Commission appears to have realised, to some extent, the special position of telecom providers vis à vis their customers. Recital 18 of the ePrivacy Regulation states:

'Basic broadband internet access and voice communications services are to be considered as essential services for individuals to be able to communicate and participate to the benefits of the digital economy. Consent for processing data from internet or voice communication usage will not be valid if the data subject has no genuine and free choice, or is unable to refuse or withdraw consent without detriment.'

**We agree with the gist of that rule Recital 18. However, we urge the EU lawmaker to include such a rule in Article 6, rather than in a recital.** Like that, the lawmaker could clarify that the rules as stipulated in Article 6 may have to be applied differently depending on whether they apply to basic broadband internet access and voice communications services, or to 'over the top' service providers.

What are the implications of Recital 18? For the definition of consent, the ePrivacy proposal refers to the GDPR.[290] Under the GDPR, consent requires a 'freely given, specific, informed and unambiguous indication of the data subject's wishes'.[291] Regarding the 'freely given' requirement, the GDPR's preamble explains: 'Consent should not be regarded as freely given if the data subject has no genuine or free choice or is unable to refuse or withdraw consent without detriment.'[292] (We discuss the requirements for valid consent in more detail in our comment on Article 9.)

Roughly speaking, the quoted sentences from Recital 18 could be interpreted as follows. If an internet provider or a phone company offers a take-it-or-leave-it choice regarding the processing of content or metadata, the individual's consent would generally not be 'freely given', and thus invalid. After all, Recital 18 states: 'Consent for processing data from internet or voice communication usage will not be valid if the data subject has no genuine and free choice, or is unable to refuse or withdraw consent without detriment.' Hence, Recital 18 suggests that a telecom provider should not offer a deal such as the following: 'If you subscribe to our internet access service, you allow us inspect all your internet traffic for marketing purposes.' Or, more precisely, the provider could offer such a deal, but the individual's consent will probably not be 'freely given', and thus invalid.

---

[288] 'Providers of publicly available electronic communications services' and 'providers of public communications networks' are *currently* defined in Article 2(a) and Article 2(c) of the Council Directive 2002/21 of 7 March 2002 on a common regulatory framework for electronic communications networks and services as amended by Directive 2009/140/EC and Regulation 544/2009 (Framework Directive) (OJ L 108, 24.4.2002, p. 33–50).
[289] See Van Hoboken & Zuiderveen Borgesius 2015.
[290] See Article 9 of the ePrivacy proposal.
[291] Article 4(11) of the GDPR.
[292] Recital 42 of the GDPR.

Recital 18 also explains the reason for that rule, namely that internet access and phone services are 'essential services for individuals to be able to communicate and participate to the benefits of the digital economy.' Hence, it seems that the Commission thinks that, to some extent, telecom providers should be subject to stricter rules.

**In sum, the stricter requirements for telecom providers and consent (in Recital 18) should be included in an article, rather than in a recital. In addition, we recommend that the EU lawmaker considers whether the rule from Recital 18 should refer to 'end-user', rather than to 'data subject'.**

A more general question arises: Regarding communications confidentiality, should telecom providers be subject to stricter rules than 'over the top' service providers? This is a difficult question. There are arguments in favour of and against stricter rules for telecom providers.

Arguments in favour of stricter rules for telecom providers include the following. First, internet providers can look deeper into internet traffic than many 'over the top' service providers. For instance, internet providers can use 'deep packet inspection'. Deep packet inspection entails opening the digital packets that are sent over the internet, to look at the contents.[293] To illustrate, the Phorm company contracted with internet providers in the United Kingdom to inspect their customers' internet traffic, to target subscribers with advertising (behavioural targeting). In 2006 a large access provider in the United Kingdom did tests with Phorm, without informing its subscribers. After media attention and parliamentary hearings, English access providers severed their business ties with Phorm.[294] Mobile operators can use deep packet inspection for behavioural targeting as well.[295] Deep packet inspection enables companies to access more data than web browsing behaviour. For instance, a company that uses deep packet inspection can read the contents of email messages. (The provider could learn less about encrypted traffic. We return to that point below.)

Second, people might have little choice regarding internet providers. To illustrate: people can use many webmail providers and messaging systems. But in some regions, people can only choose between a handful of internet providers.

Third, it seems plausible that people have different expectations from telecom providers, compared to their expectations from 'over the top' service providers that provide dedicated communications services. For instance, people may not expect their internet provider to analyse their traffic. The reactions to Phorm suggest that people do not want their internet provider to look at their internet traffic at all.[296] More generally, people rely on their basic internet connection for a wide variety of communicative needs, while their relationship to specific 'over the top' service providers tends to concern one type of service.

Fourth, by giving telecom providers more possibilities to analyse communications (with end-user consent), the ePrivacy Regulation might normalise surveillance of such communications.

But there are also arguments against stricter rules for telecom providers. First, some 'over the top' communication service providers could learn, by analysing communications, as much about their customers as telecom providers. Suppose that somebody uses the same webmail provider for ten years. That webmail provider might learn as much about that person as an internet provider that uses deep packet inspection.

Second, even when they use deep packet inspection, internet providers cannot see the contents of encrypted internet traffic. Many 'over the top' services now encrypt their traffic.

---

[293] See generally Asghari 2016, p. 123-138; Asghari, Mueller and Van Eeten 2012; Kuehn & Mueller 2012; Parsons 2013.
[294] See on Phorm McStay 2011, p. 15-42; Bernal 2011.
[295] Center for Democracy & Technology 2013, p. 6.
[296] See on Phorm McStay 2011, p. 15-42; Bernal 2011. See also Recital 17 of the ePrivacy proposal: 'end-users attach great importance to the confidentiality of their communications, including their online activities, and (…) they want to control the use of electronic communications data for purposes other than conveying the communication.'

If more 'over the top' providers encrypt their traffic, telecom providers can learn less from analysing traffic. However, an internet provider can still draw some conclusions about encrypted traffic.

Third, some might see it as unfair to regulate telecom providers stricter than other providers. For instance, in the public consultation about the revision of the ePrivacy Directive, trade associations of electronic communication service and electronic communication network providers said 'that special rules are not needed because (…) all actors are collecting and processing similar personal data.'[297]

**In sum, there are pro and contra arguments for stricter rules for telecom providers. We recommend that the EU lawmaker carefully considers whether stricter rules are necessary for such providers. However, we repeat that a rule such as the one included in Recital 18 (on telecom providers and consent) should be included in Article 6.**

**The EU lawmaker could also use more general and technology neutral rules to describe the characteristics that might make consent involuntary in the context of telecom providers.** For instance, rules along the following lines could be considered.

- If a provider can see and learn more by analysing people's communications, the requirements for 'freely given' consent should be interpreted stricter.

- If individuals have less choice between different providers, the requirements for 'freely given' consent should be interpreted stricter. Many scholars argue that dominant 'over the top' service providers could use their dominant position (and the high costs of leaving the service for users) to enforce unfair privacy terms and conditions on users.[298] Authorities have started to examine the dominant position of certain providers in this context.[299]

**More generally, take-it-or-leave-it choices regarding privacy and other fundamental rights are problematic. Rules that require an individual's consent for data processing do little to empower people if the law allows companies to offer take-it-or-leave-it choices regarding privacy. Many people are likely to consent if they encounter such take-it-or-leave-it choices. We recommend that the EU lawmaker considers prohibiting such take-it-or-leave-it choices, at the very minimum in certain circumstances. We discuss such take-it-or-leave-it choices in more detail in our comment on Article 9.**

## 3.2.11. Article 6, the same protection for content and metadata?

For at least fifteen years, there was lively discussion among legal scholars on the distinction between content and metadata.[300] Before the internet, it was reasonably simple to distinguish content from metadata. For instance, a written (snail mail) letter contains communications content; the address on an envelope is metadata. With traditional phone calls, the conversation between people is content; the date and time that one phone number called another number is metadata.[301] Courts usually assumed that content was more sensitive than metadata.[302]

---

[297] European Commission 2016, p. 5.
[298] See generally Ezrachi and Stucke 2016; Kerber 2016; Ritter 2016; Van Eijk et al 2015; Zuiderveen Borgesius 2015.
[299] See e.g. Autorité de la Concurrence and Bundeskartellamt 2016; EDPS 2016/8.
[300] See for a brief overview: Fischer 2010, p. 9 - p. 11. See also Breyer 2005; Felten 2013; Koops & Smits 2014; Kift & Nissenbaum 2016.
[301] Fischer 2014, p. 2. The two concepts, 'metadata' and 'traffic data', have similarities, but are not the same.
[302] See for instance: ECtHR, Malone v. United Kingdom, No. 8691/79, 2 August 1984, par. 83-84; CJEU (Grand Chamber), Judgement of 21 December 2016, Joined Cases C-203/15 (Tele2 Sverige AB) and C-698/15 (Watson), ECLI:EU:C:2016:970.

Regarding the distinction between content and metadata, Fisher provides a rule of thumb: First, communication providers should keep their hands off the content. Second, such providers are allowed to process metadata (traffic data) as far as necessary to offer the communication service. In the words of Fisher:

'(i). Communication content is confidential – the provider has nothing to do with it;
(ii). Traffic data from communication processes must be processed by the provider for the proper functioning of the services.'[303]

(Scholars, like the EU lawmaker, tended to speak of 'traffic data', rather than of metadata.[304])

However, the distinction between content and metadata is increasingly under pressure. First, it is becoming harder to distinguish content from metadata. For example, there is much discussion on whether the subject line of an email message should be regarded as content or metadata.[305] And regarding web browsing, URLs can be seen as content or as metadata. An URL can give much information about content. To illustrate, the URL <*https://ec.europa.eu/digital-single-market/en/proposal-eprivacy-regulation*> provides information about the content of the web page. And for many search engines, URLs can be very revealing. See for instance the URL when somebody searches in Google for 'eprivacy regulation': <*https://www.google.com/#q=**eprivacy+regulation***>. It is contentious when URLs should be seen as content or metadata.[306] In some situations, URLs should probably be regarded as content. In sum, there is a large grey area between content and metadata. The line between metadata and communications content becomes increasingly fuzzy.

Second, content and metadata can be equally sensitive and revealing. Sometimes, notes Felten, 'metadata is even more sensitive than the contents of a communication.'[307] For instance, metadata can show whether we communicate with a priest, an imam, or Alcoholics Anonymous. Metadata can reveal who our friends or business partners are, and whether people engage in adultery.[308] If somebody merely discusses 'an appointment for tomorrow' in a phone conversation with an abortion clinic, the metadata could be more revealing than the contents of the call. Moreover, metadata are easier to analyse than communications content.[309] And collecting metadata (rather than content) enables parties to capture data about millions of people, because storing metadata is usually cheaper than storing content. In sum, the distinction between communications content and metadata, as a distinction of how privacy-sensitive these data are, is largely passé.[310]

The Article 29 Working Party and the EDPS want a high level of protection for content and metadata.[311] We fully agree that both content and metadata can reveal intimate details about people's lives. We also agree that both content and metadata deserve a high level of legal protection.

### One regime for content and metadata?

The Article 29 Working Party and the EDPS do not only say that content and metadata are both sensitive; they go one step further. The Working Party and the EDPS 'recommen[d] that *the same rules* apply for consent for both content and metadata under Article 6.'[312]

---

[303] Fischer 2014, p. 214 (capitalisation adapted).
[304] See Article 2(b) of the 2009 ePrivacy Directive, which defines traffic data.
[305] Koops and Smits 2014, p. 38; p. 100-101; Fischer 2010, p. 117.
[306] Koops and Smits 2014, p. 100.
[307] Felten 2013, p. 9. See also Fischer, p. 5.
[308] Felten 2013, p. 9.
[309] Felten 2013, p. 4; Koops and Smits 2014, p. 141.
[310] This paragraph is based on, and includes sentences from: Arnbak & Zuiderveen Borgesius 2015.
[311] EDPS 2017/6, p. 28. See also Article 29 Working Party 2017 (WP247), p. 13.
[312] EDPS 2017/6, p. 28 (emphasis added). See also Article 29 Working Party 2017 (WP247), p. 13.

However, the conclusion that metadata are sensitive does not automatically imply that all metadata should be subject to the same rules as content. The law could provide a high level of protection to metadata, even if the law distinguishes between content and metadata.

There may be disadvantages to using one set of rules for content and metadata. For instance, some rules that are appropriate for metadata do not seem to be appropriate for content. The ePrivacy proposal's regime for metadata can be summarised as follows. It is prohibited to process metadata, unless an exception applies. The exceptions include necessity for mandatory quality of service requirements, necessity for billing, and the end-user's consent.[313] The billing exception, for instance, is appropriate for metadata, but does not seem appropriate for content.

On the other hand, there may be a need for more, specific, exceptions that enable providers to process content, without the consent of the end-users. We discuss two examples: (i) spam filtering, and (ii) billing and customer service.

### Spam filtering

Article 6(2)(b) allows providers of electronic communications services to process metadata, if that is 'necessary for (…) detecting or stopping fraudulent, or abusive use of, or subscription to, electronic communications services'. The Article 29 Working Party says 'that certain spam detection/filtering and botnet mitigation techniques may also be considered strictly necessary for the detection or stopping of abusive use of electronic communications services (Art. 6(2)(b)).'[314]

It does indeed make sense if providers of electronic communications services can process metadata to detect or filter spam, without the consent of the end-user. (In this case, the law should probably not require consent of *all* end-users, as a spammer would not consent to the analysis and blocking of the spam messages.) Hence, an exception may be needed to process metadata to combat spam, without the consent of the end-users. As far as providers process metadata to combat spam, they can probably rely on the exception of Article 6(2)(b).

But sometimes it may be useful, or even necessary, to analyse *content* for spam filtering. The Working Party appears to conclude that, therefore, providers of electronic communications services should sometimes be allowed to analyse content, based on the exception for detecting or stopping fraudulent or abusive use electronic communications services (Article 6(2)(b)). The Working Party adds that regarding 'spam filtering, end-users receiving spam should be offered, where technically possible, granular opt-out choices.'[315]

**In sum, to combat spam it may sometimes be necessary that providers analyse content. An exception is probably needed to allow such analysis. One way of providing that exception is making the exception in Article 6(2)(b) applicable to analysing content.**

**However, the EU lawmaker could also consider introducing a specific exception that allows the processing of content, if it is strictly necessary for combatting spam. Perhaps the same (or a similar) exception should apply to certain techniques to detect and combat botnets.**

### Billing

The ePrivacy proposal (and the current ePrivacy Directive) allows providers to process metadata for billing.[316] As noted, the law should allow, for instance, phone companies to process metadata for billing.[317] But if one set of rules applied to content and metadata, would the phone company also be allowed to store *content* (recorded conversations) for billing? Storing recorded phone conversations for billing is disproportionate.

---

[313] Article 6(2) of the ePrivacy proposal.
[314] Article 29 Working Party 2017 (WP247), p. 13.
[315] Article 29 Working Party 2017 (WP247), p. 13.
[316] See Article 6(2) of the 2009 ePrivacy Directive; Article 6(2)(b) of the ePrivacy proposal.
[317] See our comment on Article 6(2)(b).

Nevertheless, the Article 29 Working Party seems to suggest that the billing exception for metadata should also apply, in some circumstances, to content. The Working Party suggests that providers of electronic communications services should be able to rely on the billing exception for some types of customer service. The Working Party adds that, under certain circumstances, providers should be allowed to process communications *content and metadata* for customer service. In the words of the Working party: 'It should be clarified that the analysis of electronic communications data for customer service purposes may also fall under the "necessary for billing"- exception (cf. Art. 6(2) (b)).'[318] The Working Party adds:

'The relevant metadata may be kept until the end of the period during which a bill may lawfully be challenged or a payment may be pursued in accordance with national law. The relevant data (such as URL's) may only be retained at the request of the end-user, and then only for a period strictly necessary to resolve a dispute over a bill'.[319]

Presumably the Working Party thinks of cases such as the following. Alice subscribes to the cell phone services of a telecom provider. The telecom provider asked Alice for consent to store the URLs of her internet use for eight weeks. After Alice returns from a holiday abroad, she receives an large phone bill. When she calls customer service, the telecom provider can explain to her (on the basis of the stored URLs) which websites she used while abroad, and why she received the large bill. The telecom provider asked Alice for consent for storing the URLs she visited, so the provider stored them 'at the request of the end-user'.[320]

As noted, URLs should probably be seen as content in some situations. Hence, if URLs are seen as content, and if certain types of customer service should be allowed, it may be useful if providers of electronic communications services can rely on the billing exception for processing *content*.

On the other hand, perhaps it is not needed that the provider can rely on the 'billing' exception to process content for customer service. The provider could also ask the end-user (Alice) for consent to store the URLs. If the provider needs to store content for billing or customer service, maybe the appropriate legal basis is Article 6(3)(a). Or perhaps the EU lawmaker should add a specific exception that applies to processing content for billing and customer service.

It could also be argued that a well-drafted exception for billing does not bring too many privacy risks, even if the exception applies to content and metadata. For example, the ePrivacy proposal allows a provider (such as a phone company) to process metadata 'if (…) it is *necessary* for billing'.[321] If that exception would apply to both content and metadata, a phone company would still not be allowed to process the *contents* of phone calls for billing – storing the contents would not be necessary for billing. The ePrivacy Regulation would thus generally prohibit phone companies to store people's phone conversations for billing.

A necessity test (that applies to different types of providers, and that applies to content and metadata) enables providers and regulators to consider the difference between metadata processed by telecom providers for the proper functioning of his services, and metadata that are generated in other ways. Hence, such a necessity test enables a nuanced approach.

A counter argument is possible: such a necessity test is too vague, and does not provide sufficient legal clarity. The law should explicitly and specifically define under which circumstances providers could store different types of content and metadata. Hence, the legal clarity argument seems to imply that separate rules for content and metadata, with separate exceptions, are better.

---

[318] Article 29 Working Party 2017 (WP247), p. 13.
[319] Article 29 Working Party 2017 (WP247), p. 13.
[320] Article 29 Working Party 2017 (WP247), p. 13.
[321] Article 6(2)(b) of the ePrivacy proposal.

**Conclusion**

**In conclusion, electronic communications content and metadata both deserve a high level of protection. However, it does not follow that content and all metadata should be subject to the same rules.**

There are several arguments against using one regime for content and metadata. For instance, for metadata, certain exceptions to the processing prohibition are necessary and unavoidable. If one regime (with one set of exceptions) applied to content and metadata, in principle providers could rely on every exception for storing content. The only thing that would stop a provider from storing content might be phrases such as 'necessary' (or 'strictly necessary') in the exceptions. Hence: one set of rules for content and metadata could, accidentally, erode the protection of content to the level of protection of metadata.

Furthermore, the CJEU stated that the content of communications relates to the essence of the fundamental rights to privacy. The CJEU distinguished content and metadata, and saw metadata as less sensitive.[322] However, scholars have criticised the CJEU for making that distinction.[323] In any case, that CJEU judgment (based on rules that applied when the CJEU decided) does not prevent the EU lawmaker from providing the same protection to electronic communications content and metadata.

There may also be a danger that rules have to become too general, if they apply to telecom providers and 'over the top' providers, and to content and metadata. The wider the scope of a rule, the harder it is to write clear and specific rules. Hence, there might be a danger that widening the scope of rules comes at the cost of clarity and specificity. Broadly phrased open norms are probably not appropriate to protect privacy and communications confidentiality. It would be better to use a prohibition of processing content and metadata, combined with narrowly defined exceptions. Defining clear and specific rules and exceptions may be easier when different rules apply to metadata and content.

**In sum, we urge the EU lawmaker to consider, very carefully, the implications of adopting one set of rules for content and metadata. There are arguments in favour of and against using one set of rules for content and metadata. One thing is clear: the ePrivacy Regulation should offer a high level of protection to content and metadata.**

## 3.3. Article 7, storage and erasure of electronic communications data

### 3.3.1. Article 7(1), storage and erasure of content

Article 7(1) reads as follows:

'1. Without prejudice to point (b) of Article 6(1) and points (a) and (b) of Article 6(3), the provider of the electronic communications service shall erase electronic communications content or make that data anonymous after receipt of electronic communication content by the intended recipient or recipients. Such data may be recorded or stored by the end-users or by a third party entrusted by them to record, store or otherwise process such data, in accordance with [the GDPR].

---

[322] CJEU (Grand Chamber), Judgement of 6 October 2015, case C-362/14 (Schrems), ECLI:EU:C:2015:650, par 94. See also: CJEU (Grand Chamber), Judgment of 21 December 2016, cases C-203/15 (Tele2 Sverige AB) and C-698/15 (Watson), ECLI:EU:C:2016:970, par. 101.
[323] See e.g. Arnbak & Zuiderveen Borgesius 2015; Granger & Irion 2014.

See also Recital 16.

**Comment**
Article 7 concerns the storage and erasure of electronic communications content and metadata. Article 7 of the ePrivacy proposal sets different requirements for the erasure or anonymisation of content (Article 7(1)) and metadata (Article 7(2) and 7(3)).

Article 7(1) raises questions. First, it's not immediately apparent why providers of electronic communications services should be allowed to store electronic communications *content,* even in anonymised form. We recommend that the EU lawmaker clarify under which circumstances it is necessary to allow such providers to store anonymised content.

**The lawmaker could also consider describing specifically for which purposes such storage of anonymised communications content should be allowed. Or perhaps storage of anonymised content should not be allowed. Moreover, as mentioned previously, it is rarely possible to anonymise communications content such as email messages or phone conversations.[324]**

Second, Article 7(1) states that 'electronic communications content (…) may be recorded or stored by the end-users *or by a third party entrusted by them to record, store or otherwise process such data*, in accordance with [the GDPR].' It is not immediately apparent what types of situations the Commission had in mind with this phrase. When would end-users ask a third party to store their communications? Does the provision aim to enable people to store their received emails in a cloud service offered by a third party? **We recommend that the EU lawmaker clarify the goal and the meaning of the phrase regarding third parties.**

### 3.3.2.   Article 7(2) and 7(3), storage and erasure of metadata

Article 7(2) and 7(3) read as follows:

'2. Without prejudice to point (b) of Article 6(1) and points (a) and (c) of Article 6(2), the provider of the electronic communications service shall erase electronic communications metadata or make that data anonymous when it is no longer needed for the purpose of the transmission of a communication.
3. Where the processing of electronic communications metadata takes place for the purpose of billing in accordance with point (b) of Article 6(2), the relevant metadata may be kept until the end of the period during which a bill may lawfully be challenged or a payment may be pursued in accordance with national law.'

See also Recital 16.

**Comment**
Under Article 7(2), providers should erase or anonymise metadata when those data are no longer needed for transmission. Article 7(2) contains exceptions to that rule for, in short, metadata that providers store for security (Article 6(1)(b)), for quality of service requirements (Article 6(2)(a)), or if the end-user has consented to processing (Article 6(2)(c)).

Regarding Article 7(3): the Article 29 Working Party suggests that, under certain circumstances, providers should be allowed to process metadata for customer service under

---

[324] See our comment on Article 6 ('anonymous data').

the billing exception. See our comment on Article 6 ('the same protection for content and metadata?').[325]

**We recommend that the EU lawmaker considers clarifying Article 7. See also our comment on Article 6(2)(b).**

## 3.4. Article 8, protection of information stored in and related to end-users' terminal equipment

### 3.4.1. Article 8(1), terminal equipment

Article 8(1) reads as follows:

'The use of processing and storage capabilities of terminal equipment and the collection of information from end-users' terminal equipment, including about its software and hardware, other than by the end-user concerned shall be prohibited, except on the following grounds:
(a) it is necessary for the sole purpose of carrying out the transmission of an electronic communication over an electronic communications network; or
(b) the end-user has given his or her consent; or
(c) it is necessary for providing an information society service requested by the end-user; or
(d) if it is necessary for web audience measuring, provided that such measurement is carried out by the provider of the information society service requested by the end-user.'

See also Recital 6, 20, and 21.

**Comment**
Article 8 concerns 'protection of information stored in and related to end-users' terminal equipment'. Roughly speaking, Article 8(1) replaces Article 5(3) of the current ePrivacy Directive.[326] Article 5(3) of the current ePrivacy Directive is sometimes called the 'cookie provision', and is one of the most contentious provisions of the ePrivacy Directive. As the explanatory memorandum to the ePrivacy proposal notes:

'[T]he consent rule [in Article 5(3) of the ePrivacy Directive] to protect the confidentiality of terminal equipment failed to reach its objectives as end-users face requests to accept tracking cookies without understanding their meaning and, in some cases, are even exposed to cookies being set without their consent. The consent rule is over-inclusive, as it also covers non-privacy intrusive practices, and under-inclusive, as it does not clearly cover some tracking techniques (e.g. device fingerprinting) which may not entail access/storage in the device. Finally, its implementation can be costly for businesses.'[327]

**Rationales for Article 8(1)**
Several rationales for Article 8(1) of the ePrivacy proposal can be identified. These rationales can be summarised as: (i) people's devices and the contents of those devices are part of their private sphere; (ii) people should be protected against secretly installing software or other information on their devices; (iii) people should be protected against unwanted tracking and surveillance when they use the internet.

First, somebody's device and its contents are part of that person's private sphere, which deserves protection. In the words of Recital 20:

'Terminal equipment of end-users of electronic communications networks and any information relating to the usage of such terminal equipment, whether in particular is stored

---

in or emitted by such equipment, requested from or processed in order to enable it to connect to another device and or network equipment, are part of the private sphere of the end-users requiring protection under the Charter of Fundamental Rights of the European Union and the European Convention for the Protection of Human Rights and Fundamental Freedoms.'

Along similar lines, the German Bundesverfassungsgericht says people have a 'right to the guarantee of the confidentiality and integrity of information technology systems.'[328] Indeed, in a recent Europe-wide survey, '78% say it is very important that personal information on their computer, smartphone or tablet can only be accessed with their permission.'[329]

Recital 20 rightly emphasises that the contents of an end-user's device are extremely sensitive: 'such equipment contains or processes information that may reveal details of an individual's emotional, political, social complexities, including the content of communications, pictures, the location of individuals by accessing the device's GPS capabilities, contact lists, and other information already stored in the device'. Therefore, adds the recital, 'the information related to such equipment requires enhanced privacy protection.'

Article 8(1) could also help to protect communications stored by individuals. Article 5 (on communications confidentiality) protects the confidentiality of, for instance, emails. It would make sense if the ePrivacy Regulation also protected an email after somebody downloaded it to his or her phone. Article 8(1) provides such protection.[330] (Article 8 does not protect information stored in the cloud; see our comment on Article 5.) In sum, Article 8 aims to protect people's devices and the information on those devices.

A second rationale for Article 8(1) of the ePrivacy proposal is protecting people against secretly installing spyware, malware, or other information on their devices. Recital 20 states: 'so-called spyware, web bugs, hidden identifiers, tracking cookies and other similar unwanted tracking tools can enter end-user's terminal equipment without their knowledge in order to gain access to information, [or] to store hidden information and to trace the [end-user's] activities.'[331] (We recommend that the EU lawmaker change 'trace the activities' to 'trace the *end-user's* activities', or a similar phrase).

A third rationale for Article 8(1) is defending people against unwanted tracking, monitoring, and surveillance. Recital 20 says: 'Techniques that surreptitiously monitor the actions of end-users, for example by tracking their activities online or the location of their terminal equipment, (…) pose a serious threat to the privacy of end-users.'[332] The next section discusses online tracking and privacy in more detail.

### Online tracking and privacy

Article 8 (and its predecessor) lead to much discussion in the context of online tracking and behavioural targeting. Behavioural targeting involves monitoring people's online behaviour, and using the collected information to show people targeted advertisements.[333] This type of tracking-based advertising has grown into a large business over the past two decades.[334]

---

[328] Bundesverfassungsgericht, 27 February 2008, decisions, vol. 120, p. 274-350 (Online Durchsuchung). See also Arnbak 2016, p. 92-93; Koops et al 2017, section III, C, 3 (p. 30-32) See also ECtHR, Bernh Larsen Holding AS and others v. Norway, No. 24117/08, 14 March 2013, par. 106; par 163.

[329] Explanatory memorandum to the ePrivacy proposal, p. 6 (section 3.2).

[330] See also Steenbruggen 2009, p. 186.

[331] See also Recital 24 of the 2002 ePrivacy Directive.

[332] See also Recital 24 and 25 of the 2002 ePrivacy Directive, and Recital 65 and 66 of Directive 2009/136.

[333] Zuiderveen Borgesius 2015. The Interactive Advertising Bureau US provides this description: 'Behavioral targeting uses information collected on an individual's web browsing behavior such as the pages they have visited or the searches they have made to select which advertisements be displayed to that individual. Practitioners believe this helps them deliver their online advertisements to the users who are most likely to be influenced by them' (Interactive Advertising Bureau United States, Glossary).

[334] See Turow 2011; Angwin 2014.

However, **online tracking and behavioural targeting raise serious privacy concerns**.[335] Three such concerns are: chilling effects, a lack of control over personal information, and the risk of unfair discrimination and manipulation. First, behavioural targeting entails massive collection of information about people's online activities. Like other types of surveillance, this can cause chilling effects. People may adapt their behaviour if they suspect their activities may be monitored.[336] As Sir Tim Berners-Lee, Director of the World Wide Web Consortium, notes, 'widespread data collection (…) creates a chilling effect on free speech and stops the web from being used as a space to explore important topics, like sensitive health issues, sexuality or religion.'[337]

Second, people lack control over information concerning them. They rarely know or understand which information is collected, how it is used, and with whom it is shared. And large-scale personal data storage brings risks. For instance, a data breach could occur, or data could be used for unexpected purposes, such as identity fraud or price discrimination.[338] In addition, the feeling of lost control is a privacy problem.[339]

Third, behavioural targeting enables social sorting and discriminatory practices: companies can classify people as 'targets' and 'waste', and treat them accordingly.[340] For instance, an advertiser could use discounts to lure affluent people to become regular customers – but might exclude poor people from the campaign. Behavioural targeting could also be used for online price discrimination.[341] And some fear that behavioural targeting could be used to manipulate people. In theory personalised advertising could be used to target vulnerable people, thereby giving advertisers an unfair advantage over consumers.[342] To illustrate: a company reportedly 'showed advertisers how it has the capacity to identify when teenagers feel "insecure", "worthless" and "need a confidence boost"'.[343] In some contexts, undue influence would be more worrying than in others. For instance, behavioural targeting and personalised messages could bring serious risks if used in political campaigns.[344]

**We discuss consent to online tracking below, in our comment on Article 9.**

### Scope of Article 8(1)

Below, we sometimes speak about 'cookies', for ease of reading. But the scope of Article 8(1) is much wider. Article 5(3) of the current ePrivacy Directive already has a wide scope, and applies, in short to storing or accessing information on a user's device.[345] All activities that fall under Article 5(3) of the current ePrivacy Directive also fall under Article 8 of the ePrivacy proposal.

Article 8 of the ePrivacy proposal has a wider scope, and is phrased in a more technology neutral way than the current Article 5(3). The proposed Article 8 applies to '[t]he use of processing and storage capabilities of terminal equipment and the collection of information from end-users' terminal equipment, including about its software and hardware'.[346]

Article 8 of the ePrivacy proposal also applies to device fingerprinting, as Recital 21 explains: 'Information related to the end-user's device may also be collected remotely for the purpose of identification and tracking, using techniques such as the so-called "device fingerprinting", often without the knowledge of the end-user, and may seriously intrude upon the privacy of these end-users.'

---

[335] This section on privacy concerns is based on, and includes sentences from, Zuiderveen Borgesius 2015.
[336] See Richards 2008.
[337] Berners-Lee 2017. See also Berners-Lee 2009.
[338] Zuiderveen Borgesius 2015b. In cases such as identify fraud, Calo speaks of 'objective harms (Calo 2011).
[339] Calo calls such a feeling a 'subjective' harm (Calo 2011).
[340] Turow 2011. See also Turow 2017.
[341] Zuiderveen Borgesius 2015b.
[342] Calo 2014.
[343] Levin 2017.
[344] See generally on political behavioural targeting and online political microtargeting: Barocas 2012; Bennett 2015; Kreiss 2012; Zuiderveen Borgesius et al 2017b (forthcoming).
[345] Recital 65 and 66 of the Citizens' Rights Directive 2009/136. The Article 29 Working Party confirms that the provision applies, for instance, to apps that access information on a user's smartphone, such as location data or a user's contact list (Article 29 Working Party 2013 (WP202), p. 10).
[346] Article 8(1) of the ePrivacy proposal. See also Recital 6 of the ePrivacy proposal.

Device fingerprinting can rely, for instance, on looking at information that a device discloses, such as the type of browser, installed fonts, and other settings.[347] The device could send such information as a part of standard network traffic.[348] But people should indeed be protected against unwanted tracking and surveillance, regardless of the technology used. Therefore, **from a privacy and fundamental rights viewpoint, Article 8 should apply to device fingerprinting.**

The EDPS calls upon the EU lawmaker to ensure that Article 8 also applies to future tracking techniques, for instance in the Internet of Things.[349] Article 8 should also apply to 'all forms of "passive tracking", that is, the use of identifiers and other data broadcasted by devices.'[350] The EDPS adds that it might be best if Article 8 covered 'all information that can be obtained from the device.'[351] The EDPS suggests that future tracking techniques should generally be based on consent, subject to exceptions (along the lines of the exceptions in the proposed Article 8).[352]

Article 8(1) states: 'The use of processing and storage capabilities of terminal equipment and the collection of information from end-users' terminal equipment, including *about* its software and hardware, other than by the end-user concerned shall be prohibited (…)'[353] **The EU lawmaker should consider changing 'about' to 'from or about'. With that amendment, it would be clearer that Article 8(1) also applies if, for instance, an app turns on a microphone or a camera on a device to collect information.**

**In sum, we recommend that the EU lawmaker aims for a future-proof scoping of Article 8.**

### 3.4.2.    Article 8(1), exceptions

Article 8(1) of the ePrivacy proposal contains, in short, an in-principle prohibition of using processing and storage capabilities of a user's device. Article 8(1) provides four exceptions to this prohibition. Three exceptions resemble the exceptions in Article 5(3) of the current ePrivacy Directive: (a), transmission, (b), consent, and (c) requested service. Exception (d), web audience measurement, is new.

Under exception (a), placing a cookie on an end-user device is allowed if 'it is necessary for the sole purpose of carrying out the transmission of an electronic communication over an electronic communications network'. An example is a cookie or similar file that is used to route information over the network.[354]

Under exception (b), placing a cookie on an end-user device is allowed if the end-user has given his or her consent. **We return to the topic of consent below. (See our comment on Article 9.)**

---

[347] See: Acar et al 2013.
[348] To what extent Article 5(3) of the current ePrivacy Directive applies to various forms of device fingerprinting is a complicated question. See: Zuiderveen Borgesius 2015, p. 247; Article 29 Working Part 2014 (WP224).
[349] EDPS 2017/6, p. 28.
[350] EDPS 2017/6, p. 28.
[351] EDPS 2017/6, p. 28.
[352] EDPS 2017/6, p. 28. If the lawmaker uses the 'all information that can be obtained from the device' phrase, as suggested by the EDPS, there might be a need for an additional exception, 'for a very limited case of processing directly related to a service requested by the user and performed exclusively by the service provider' (p. 28).
[353] Emphasis added.
[354] For instance, 'load balancing' cookies. See Article 29 Working Party 2012 (WP194), p. 8.

Under exception (c), placing a cookie on an end-user device is allowed if 'it is necessary for providing an information society service requested by the end-user'. An example is a session cookie that enables an end-user to log in to a service, such as an email service.[355] Recital 21 explains: 'consent should not be requested for authorizing the technical storage or access which is strictly necessary and proportionate for the legitimate purpose of enabling the use of a specific service explicitly requested by the end-user. This may include the storing of cookies for the duration of a single established session on a website to keep track of the end-user's input when filling in online forms over several pages.'

We agree that exceptions to Article 8's consent requirement are needed, roughly along the lines of the exceptions in the proposal. But amendments are needed.

First, exception (a), (c) and (d) say that consent is not required if the processing is 'necessary' for (a) transmission, (c) a requested service, or (d) audience measurement. **As noted previously, we recommend that the EU lawmaker uses the phrase 'strictly necessary' instead of 'necessary'.[356]**

**Second, Article 8(1) should state explicitly that the exceptions (a), (c) and (d) only apply if there is no, or only very minor, interference with the privacy or other fundamental rights of the end-user.** As Recital 21 notes, 'Exceptions to the obligation to obtain consent (…)** *should be limited to situations that involve no, or only very limited, intrusion of privacy*.' **We recommend that the EU lawmaker includes that, or a similar, sentence in Article 8 (rather than in a recital). The sentence could be improved by adding the words 'or related fundamental rights' at the end.**

**Regarding exception (c) (requested service), we recommend that the EU lawmaker follow the advice of the EDPS: 'a recital should explicitly confirm that "processing of data for purposes of providing targeted advertisements cannot be considered as necessary for the performance of a service"'.[357] A correct interpretation of the rules of the GDPR (and the Data Protection Directive) would lead to a similar conclusion.[358] Nevertheless, such a recital could improve legal clarity and help to avoid misunderstandings.**

### Audience measurement exception

Exception (d) is new. The use of cookies is allowed if 'if it is necessary for web audience measuring, provided that such measurement is carried out by the provider of the information society service requested by the end-user.' A service of the information society roughly means: a service provided over the internet.[359]

In principle we agree that an exception is needed for, in short, privacy-friendly analytics. As the European Commission notes, the current ePrivacy Directive's cookie 'consent rule is over-inclusive, as it also covers non-privacy intrusive practices'.[360] The Article 29 Working Party has arrived at a similar conclusion, and has called upon the Commission to include an exception for privacy-friendly analytics cookies.[361]

---

[355] See Article 29 Working Party 2012 (WP194).
[356] See our comment on Article 6(1) of the ePrivacy Proposal. See also Article 29 Working Party 2017 (WP247) p. 19-20; EDPS 2017/6, p. 27.
[357] EDPS 2017/6, p. 17.
[358] See Zuiderveen Borgesius 2015a.
[359] An information society service is defined in EU law as:
'any service normally provided for remuneration, at a distance, by electronic means and at the individual request of a recipient of services.
For the purposes of this definition:
(i) "at a distance" means that the service is provided without the parties being simultaneously present;
(ii) "by electronic means" means that the service is sent initially and received at its destination by means of electronic equipment for the processing (including digital compression) and storage of data, and entirely transmitted, conveyed and received by wire, by radio, by optical means or by other electromagnetic means;
(iii) "at the individual request of a recipient of services" means that the service is provided through the transmission of data on individual request.'
(Article 1(1)(b) of Directive (EU) 2015/1535 of the European Parliament and of the Council of 9 September 2015 laying down a procedure for the provision of information in the field of technical regulations and of rules on Information Society services (OJ L 241, 17.9.2015, p. 1–15)).

However, amendments to the proposed text are necessary. The EDPS warns 'that the exception must not create a loophole for long-term storage or further processing of personal data for additional purposes.'[362] The EDPS suggests that the following phrase should be added to the exception (the emphasised phrase below):

'(d) if it is necessary for web audience measuring, provided that such measurement is carried out by the provider of the information society service requested by the end-user *and further provided that no personal data is made accessible to any third parties*.'[363]

The EU lawmaker could also draw inspiration from the Dutch exception for innocuous analytics cookies. Under Dutch law, the consent requirement for cookies and similar files does not apply:

'if it concerns storage or access which (…) is strictly necessary (…) – provided that this has no or minor impact on the privacy of the user or subscriber concerned – to obtain information about the quality or effectiveness of a delivered information society service.'[364]

If the EU lawmaker wants to retain (parts of) the text of the proposed Article 8(1)(d), the lawmaker should consider clarifying another aspect. Article 8(1)(d) suggests that cookies for web audience measuring are allowed 'provided that such measurement is carried out by the provider of the information society service requested by the end-user.' A website publisher could employ an analytics company to do the web audience measurement. The publisher could require the analytics company to sign a contract (a processor agreement) that stipulates that the company processes the data only on instructions from the publisher.[365] The lawmaker could clarify whether such a processor agreement could make web audience measurement by an analytics company legal, even if Article 8(1)(d) requires that the 'measurement is carried out *by the provider of the information society service* requested by the end-user.'[366]

Lastly, the term 'web audience measuring' may be too narrow. The Working Party says: 'The term "web audience measuring" should (…) be redefined in a technology neutral manner, in order to also include similar analytical usage information retrieved from apps, wearables and internet of things devices.'[367] **Indeed, the lawmaker should consider phrasing the analytics exception in a more technology neutral way.**

**In sum, an exception for, in short, privacy-friendly analytics seems appropriate. But the EU lawmaker should amend Article 8(1)(d), to ensure that privacy and related rights are respected.**

### Exception needed for security updates?

If a company installs a security update on somebody's device, the company uses 'storage capabilities of terminal equipment' (Article 8(1)). Such updates are important to keep the security of devices up-to-date.

On some devices, such as phones and laptops, the end-user can give informed consent to security updates. (Even though the user could react to a request to install such an update, it may be useful if a vendor updates, for instance, a browser without asking for prior consent.) On other devices, such as 'things' in the Internet of Things, the end-user cannot easily give informed consent to a security update. Some 'things' may not even have a screen. On such devices it could be desirable to let manufacturers install security updates without consent from the end-user.

---

[360] Explanatory memorandum to the ePrivacy proposal, p. 5, section 3.1.
[361] Article 29 Working Party 2012 (WP194), p. 10-11. See also Article 29 Working Party 2017 (WP247), p. 10.
[362] EDPS 2017/6, p. 29.
[363] EDPS 2017/6, p. 29.
[364] Translation by Kosta 2016.
[365] Article 28(3) of the GDPR.
[366] See also Article 29 Working Party 2017 (WP247), p. 19.
[367] Article 29 Working Party 2017 (WP247), p. 18-19.

We recommend that the EU lawmaker clarifies that installing security updates does not require the end-user's consent. The Article 29 Working Party adds that such security updates should only be allowed without user consent under the following conditions:

'(i) the security updates are discretely packaged and do not in any way change the functionality of the software on the equipment (including the interaction with other software or settings chosen by the user),
(ii) the end-user is informed in advance each time an update is being installed, and
(iii) the end-user has the possibility to turn off the automatic installation of these updates.'[368]

**In sum, an exception for necessary security updates is needed. The EU lawmaker has at least two options. First, the lawmaker could add a specific exception for security updates. That first option is probably the best option. Second, the lawmaker could clarify that security updates can fall under exception (c), requested service.[369]** More generally, we recommend that the EU lawmaker examines whether EU legislation on security of devices needs to be improved. See our comment on Article 17 and in section 4.7.

## Exception needed for employment relationships?

An exception for certain activities in employment relationships should be considered. Consent must be 'freely given' to be valid. If an employer asks an employee for consent, the consent might not be sufficiently free, because of the imbalance of power.[370] But there might be situations in which interference with terminal equipment is legitimate, while Article 8(1) does not provide a suitable exception, and where freely given consent is impossible to obtain for an employer. The Article 29 Working Party says:

'One example is where an employer wants to update a company-issued phone. A second example is where an employer offers employees lease cars, and for administrative purposes lets a third party collect location data via the onboard unit of a car. In both cases, the employer has an interest in interfering with these devices.'[371]

Therefore, says the Working Party, the EU lawmaker should consider adding an additional exception to the prohibition of Article 8(1). Such a new exception should only apply to situations 'where (i) the employer provides certain equipment in the context of an employment relationship, (ii) the employee is the user of this equipment, and (iii) the interference is strictly necessary for the functioning of the equipment by the employee'.[372]

**We recommend that the lawmaker considers adding such an exception for employment relationships. As noted previously, we do not think an open-ended exception (along the lines of a legitimate interests provision) should be added to Article 8.[373]**

---

[368] Article 29 Working Party 2017 (WP247), p. 20. Ideally, security updates are sent via secured communication channels. Such a secured channel could prevent man-in-the-middle attacks and tampering with security updates by hackers. A possible downside of such a requirement would be that older devices that cannot communicate via secured communication channels cannot be updated any longer.
[369] The second option (exception c) seems to be the option preferred by the Working Party: Article 29 Working Party 2017 (WP247), p. 20.
[370] See Article 29 Working Party (WP 187), p. 13-14. See also Recital 43 of the GDPR.
[371] Article 29 Working Party 2017 (WP247), p. 29.
[372] Article 29 Working Party 2017 (WP247), p. 29.
[373] See our comment on Article 6.

**Adblocking**

A growing number of people use an adblocker, such as a browser plug-in.[374] Such an adblocker could block online ads. But adblockers are also one of the few measures people can take against security risks and online tracking. Some research suggests that security is one of the main reasons that people use adblockers.[375] We think, by the way, that people should be allowed to use ad blockers.

A website publisher can often detect whether a website visitor uses an adblocker. Some website publishers have denied access to their websites to adblock users. Sometimes a publisher asks adblock users to make an exception for their website: to whitelist that website. Other publishers have asked adblock users to pay money for a website visit.

Article 8(1) is often applicable if a website publisher detects whether a visitor uses an adblocker. For instance, a website publisher may store a script on a user's device to detect whether that user has installed an adblocker. Hence, the website publisher uses 'processing and storage capabilities of terminal equipment', or 'collect[s] information from end-users' terminal equipment, including about its software' (Article 8(1)). Generally, in such cases only the end-user's consent could provide an exception to the prohibition in Article 8(1).

However, Recital 21 may provide another exception (to the consent requirement) that is relevant for detecting adblockers. Recital 21 states: 'the mere logging of the fact that the end-user's device is unable to receive content [such as ads] requested by the end-user should not constitute access to such a device or use of the device processing capabilities.' The sentence is ambiguous, but could be interpreted as an extra exception to the consent requirement. In that interpretation, a website publisher does not need the website visitor's consent for checking whether he or she uses an adblocker. However, another interpretation is that the website visitor did not 'request' the ads, and that thus the sentence from Recital 21 is not applicable.

The EDPS calls for an 'explicit prohibition on the practice of excluding users who have ad-blocking or other applications and add-ons installed to protect their information and terminal equipment.'[376] A possible counter argument is the following. Suppose a website only features contextual ads (ads for cars on a site about cars, for instance), and suppose that no data are collected through the site. It is not obvious that the ePrivacy Regulation should require that website, which is funded by advertising, to allow adblock users to visit its site.

**We recommend that the EU lawmaker is aware of the possible implications of the adblock sentence in Recital 21, and that the lawmaker clarifies the recital as appropriate.**

### 3.4.3. Article 8(2)-(4), information emitted by terminal equipment

Article 8(2)-(4) concern information emitted by terminal equipment. Article 8(2)-(4) read as follows:

'(2) The collection of information emitted by terminal equipment to enable it to connect to another device and, or to network equipment shall be prohibited, except if:
(a) it is done exclusively in order to, for the time necessary for, and for the purpose of establishing a connection; or
(b) a clear and prominent notice is displayed informing of, at least, the modalities of the collection, its purpose, the person responsible for it and the other information required under

---

[374] Some estimate that 11% of all internet users worldwide, and 20% of internet users in Western Europe use an adblocker (Pagefair 2017).
[375] Pagefair 2017, p. 4.
[376] EDPS 2017/6, p. 17.

Article 13 of [the GDPR] where personal data are collected, as well as any measure the end-user of the terminal equipment can take to stop or minimise the collection. The collection of such information shall be conditional on the application of appropriate technical and organisational measures to ensure a level of security appropriate to the risks, as set out in Article 32 of [the GDPR], have been applied.

(3) The information to be provided pursuant to point (b) of paragraph 2 may be provided in combination with standardized icons in order to give a meaningful overview of the collection in an easily visible, intelligible and clearly legible manner.

(4) The Commission shall be empowered to adopt delegated acts in accordance with Article 27 determining the information to be presented by the standardized icon and the procedures for providing standardized icons.'

See also Recital 17 and 25.

## Comment

Article 8(2) applies, for instance, to location tracking is that happens on the basis of Wi-Fi or Bluetooth signals emitted by people's smart phones. Such tracking techniques enable organisations to follow people's movements.[377] As discussed previously, location data are extremely sensitive and revealing.[378] Recital 25 rightly notes that some types of location tracking involve 'high privacy risks', for instance when they entail 'the tracking of individuals over time, including repeated visits to specified locations'.

But Article 8(2) falls short in protecting people against secretive or unwanted location tracking. Article 8(2)(b) could be interpreted as follows. Article 8(2)(b) allows location tracking without people's consent. A company that wants to follow people's movements (based on Wi-Fi- or Bluetooth tracking) only has to put up posters in town that say: '*In this city we track your location based on the Wi-Fi and Bluetooth signals of your devices. Turn off your phone or other device, or your Wi-Fi and Bluetooth, if you don't want to be tracked*'.

The effect of the proposed article 8(2)(b) could be that people never feel free from surveillance when they walk or drive around. People would always have to look around whether they see a sign or poster that informs them of location tracking. Moreover, it would be unacceptable if people could only escape location tracking by limiting the functionalities of their phones and other devices.[379]

Article 8(2)(2) does not even require companies to offer people an opt-out possibility. In contrast: if, in rare cases, the GDPR allowed location tracking without consent, the data subject would generally have the right to opt out.[380]

In sum, the proposed Article 8(2)(b) is incompatible with the goal of not lowering the protection of the GDPR. As the ePrivacy Regulation's preamble states, the Regulation should not 'lower the level of protection enjoyed by natural persons under [the GDPR].' The Article 29 Working Party and the EDPS confirm that the proposed Article 8(2) fails to protect privacy, and fails to provide a GDPR-compliant level of protection.[381]

**Article 8(2)(b) should be improved. In brief, (i) the law should allow the collection of information emitted by user devices after the user's informed consent. (ii) The EU lawmaker could consider introducing another exception, with sufficient safeguards, for anonymous people counting.**

---

[377] See generally on location tracking: Turow 2017.
[378] See our comments on Article 6 of the ePrivacy Proposal.
[379] EDPS 2017/6, p. 20.
[380] See Articles 6(1)(f) and 21(1) of the GDPR. See also: Autoriteit Persoonsgegevens (Dutch DPA) 2015; Datatilsynet (DPA Norway) 2016.
[381] Article 29 Working Party 2017 (WP247), p. 11; EDPS 2017/6, p. 20.

(i) If somebody gives prior and fully informed consent to location tracking (the collection of information emitted by terminal equipment), that tracking should be allowed. For instance, a company could ask people for consent for location tracking in a certain area, such as an airport or a shopping mall. The company could ask consent, for instance, when people install an app.[382] Sufficient information and transparency are required.

(ii) The lawmaker could consider adding an exception for anonymous people counting. Estimating the size of crowds or counting people could be useful, for instance, during large music festivals. If a certain area becomes too crowded, the organisation could intervene. For such purpose, it's not necessary to follow specific individuals.

For anonymous people counting, the data can generally be anonymised and aggregated immediately.[383] According to the Working Party and the EDPS, such people counting without consent is only acceptable when the following conditions are met:

(1) Organisations using location tracking should provide sufficient information and transparency.[384]
(2) 'The purpose of the data collection must be restricted to mere statistical counting'[385]
(3) 'The tracking is limited in time and space to the extent strictly necessary for this purpose'[386]
(4) 'The data is deleted or anonymised immediately afterwards'[387]
(5) 'The data 'should not be processed to support any measures or decisions that are taken with regard to the individual concerned'. [388]
(6) There should be 'an effective horizontal opportunity to opt-out of the processing (similar to "do not call" registers in the context of unsolicited communications or "do not track" in the context of online tracking).'[389]

Conditions along these lines could be listed in the ePrivacy Regulation. As the EDPS suggests, it could be useful if the ePrivacy Regulation included a possibility for the European Data Protection Board to provide guidance regarding safeguards in this context.[390]

The EDPS recommends more amendments to Article 8. First, the beginning of Article 8(2) should be amended. The proposed Article 8(2) states: 'The collection of information emitted by terminal equipment *to enable it to connect to another device and, or to network equipment* shall be prohibited, except if (…)'. The emphasised phrase would be deleted. Deletion of that phrase would 'ensure a technologically neutral coverage and full protection of all data emitted by terminal equipment irrespective of the purpose.'[391]

Second, in Article 8(2)(a), a phrase should be added after 'for the purpose of establishing a connection'. The following phrase (or a similar phrase) should be added: 'which the end-users concerned have authorised'. The EDPS recommends that addition to 'ensure that the connection established is the one the user actually is aware of and has given his or her prior consent to.'[392]

The EDPS also suggests that the EU lawmaker could consider whether additional, narrow, exceptions are appropriate, for instance for scientific research, and to protect 'vital interests' of individuals.[393]

More generally, **the EU lawmaker could consider encouraging the development of a Do Not Track-like system for location tracking, or for tracking in the Internet of Things in general.**[394] Ideally, people should be able to make privacy settings through one interface (on their phone for instance). For instance, somebody might want to signal to all parties and Internet of Things-devices in the area: 'Do not collect or store any information about me or this device'.[395] Further research and perhaps further legislation would be needed for such a system.

**In sum, the proposed Article 8(2) should be significantly amended to ensure reasonable protection of privacy and related rights. Especially Article 8(2)(b), which allows location tracking without consent and without an opt-out option, reduces the protection that people enjoy under the GDPR. The proposed Article 8(2)(b) would also violate reasonable privacy expectations.**

## 3.5. Article 9, consent

### 3.5.1. Article 9(1), reference to GDPR's consent definition

Article 9(1) states:

'The definition of and conditions for consent provided for under Articles 4(11) and 7 of [the GDPR] shall apply.'

See also Recital 3 and 18.[396]

**Comment**

For the definition and the requirements for valid consent, Article 9(1) refers to the GDPR. Recital 18 of the ePrivacy proposal adds: 'For the purposes of this Regulation, consent of an end-user, regardless of whether the latter is a natural or a legal person, should have the same meaning and be subject to the same conditions as the data subject's consent under [the GDPR].'[397]

It makes sense that the ePrivacy Regulation refers to the GDPR for the consent definition. The current ePrivacy Directive also refers to general data protection law for the consent definition.[398] The GDPR defines consent as follows: 'any freely given, specific, informed and unambiguous indication of the data subject's wishes by which he or she, by a statement or by a clear affirmative action, signifies agreement to the processing of personal data relating to him or her'.[399] Article 7 of the GDPR elaborates on the requirements for valid consent. The requirements for valid consent in the GDPR are well-written.

However, as the Article 29 Working notes, there are practical problems with the reference to the GDPR's consent provision.[400] For instance, under the GDPR only 'informed' consent is valid. But when is a legal person (such as a company) informed? The GDPR requirements for an indication of wishes also raise questions when applied to a legal person. For example, how could a company signify agreement, and especially agreement 'to the processing of personal data relating to him or her'?[401] **We recommend that the EU lawmaker clarifies these ambiguities.**

---

[394] See on Do Not Track: our comment on Article 9(2).
[395] Article 29 Working Party 2017 (WP247), p. 11-12.
[396] Consent is also mentioned in Recital 19, 22, 24, and 25.
[397] See also Recital 3 of the ePrivacy proposal.
[398] Article 2(f) of the 2009 e-Privacy Directive.
[399] Article 4(11) of the GDPR.
[400] Article 4(11) of the GDPR.
[401] See Article 29 Working Party 2017 (WP247), p. 28-29.

**In addition, we recommend that the EU lawmaker clarifies that the end-user's consent can never legitimise a disproportionate interference with privacy, communications confidentiality, or related rights.[402]**

### 3.5.2.    Article 9(2), software settings

Article 9(2) states:

'Without prejudice to paragraph 1, where technically possible and feasible, for the purposes of point (b) of Article 8(1), consent may be expressed by using the appropriate technical settings of a software application enabling access to the internet.'

See also Recital 22, 23, and 24.

**Comment**

In principle, it may be a good idea to enable people to give or withhold consent through their browsers and similar software. As the explanatory memorandum to the ePrivacy proposal notes regarding tracking cookies: 'The implementation of the ePrivacy Directive has not been effective to empower end-users. Therefore the implementation of the principle by centralising consent in software and prompting users with information about the privacy settings thereof, is necessary to achieve the aim.'[403] Recital 22 adds:

'Given the ubiquitous use of tracking cookies and other tracking techniques, end-users are increasingly requested to provide consent to store such tracking cookies in their terminal equipment. As a result, end-users are overloaded with requests to provide consent. (…) Therefore, this Regulation should provide for the possibility to express consent by using the appropriate settings of a browser or other application.'

Indeed, a common complaint about the ePrivacy Directive is that clicking 'I agree' to hundreds of separate cookie consent requests is not user-friendly. And people might click 'I agree' without realising the consequences.[404]

While making consent procedures more user-friendly is a laudable goal, the ePrivacy proposal should be significantly amended to attain that goal. The Working Party notes that 'the Proposed Regulation incorrectly suggests that valid consent can be given through non-specific browser settings.'[405] Indeed, most browsers do not provide granular controls to users. Browsers often give people limited choices such as: 'accept all cookies / accept only first party cookies / accept no cookies'. It would be better to require browsers and other parties to comply with Do Not Track or a similar standard. We discuss that approach in the next section.

### 3.5.3.    Do Not Track

According to the Working Party and the EDPS, **the EU lawmaker should make compliance with Do Not Track and similar standards obligatory.**[406] In brief, the Do Not Track standard should enable people to signal with their browser that they do not want to be tracked. A website publisher or another company that receives a 'Do not track me' signal should refrain from tracking that person.[407]

---

[402] A similar conclusion can be reached under the GDPR, regarding personal data processing. The overarching data protection principles, including the principles of 'lawfulness, fairness and transparency' and 'data minimization' (Article 5), always apply, also after the data subject's consent (Article 1(a) of the GDPR). The (i) Dutch and the (ii) Polish Supreme Court have confirmed that consent cannot legitimate disproportionate data processing. See: (i) Naczelny Sąd Administracyjny [Supreme Administrative Court], 1 December 2009, I OSK 249/09 (Inspector General for Personal Data Protection), unofficial English translation at <www.giodo.gov.pl/417/id_art/649/j/en/> accessed 28 April 2017; (ii) Hoge Raad [Dutch Supreme Court], 9 September 2011, ECLI:NL:HR:2011:BQ8097 (Santander), English summary by Valgaeren & Gijrath 2011.
[403] Explanatory memorandum to the ePrivacy proposal, p. 4 (section 2.3).
[404] Zuiderveen Borgesius 2015.
[405] Article 29 Working Party 2017 (WP247), p. 20.
[406] Article 29 Working Party 2017 (WP247), p. 20; EDPS 2017/6, p. 19.

It may not immediately apparent how Do Not Track could help companies to comply with European privacy and data protection rules. But an arrangement along the following lines could be envisioned. Companies should refrain from tracking people that have not set a Do Not Track preference. Only if somebody signals to a specific company 'Yes, you can track me' after receiving sufficient information, that company may track that user. Hence, in Europe not setting a preference would have the same legal effect as setting a preference for 'Do not track me.' Do Not Track could thus be a system to opt in to tracking.

**A Do Not Track system (or a similar standard) could have several advantages. For instance, people could signal through a simple setting that they do not want to be tracked. Do Not Track could apply to all types of web tracking (based on cookies or device fingerprints for example).**

**Furthermore, the Do Not Track standard does not rely on a distinction between 'first parties' and 'third parties'.** (Article 10(1) of the ePrivacy proposal uses the phrase 'third party'.) **The Do Not Track Standard distinguishes 'site-wide' and 'internet-wide' tracking.**[408] In addition: If the EU lawmaker phrases a provision on technical consent standards in sufficiently technology neutral way, the provision could also enable Do Not Track-like systems in other contexts, for instance for the Internet of Things.

A Do Not Track system could reduce the transaction costs of (not) consenting to tracking companies separately. In that way, the Do Not Track standard is somewhat comparable with a centralised Do Not Call registry where people can opt out of telemarketing.

However, Do Not Track does not stop website publishers from asking for an exception from a website visitor. Hence, a publisher could ask somebody who signals 'Do not track me' for an exception, roughly as follows: 'We see your Do Not Track signal. But do you make an exception for me and my ad network partners so we can track you?'.

If a website asks for such an exception, consent should *not* easily be assumed. For instance, a website should not be able to obtain consent through a banner that says: 'if you continue to use this website you make an exception to your "Do not track me" setting for us'.

The Working party says that the ePrivacy Regulation should protect people against an overwhelming amount of consent requests (or requests for exceptions). The Working Party suggests that 'the ePrivacy Regulation should ensure that a refusal to accept internet-wide tracking from a specific organisation (…) blocks that organisation from making future consent requests, for at least 6 months.'[409] The Working Party adds that the organisation could still ask for consent for tracking within its own website (for site-wide tracking).[410]

**In sum, we recommend that the EU lawmaker makes compliance with Do Not Track and similar standards obligatory.** Do Not Track backed with law and proper enforcement could lead to a major improvement, compared to the situation under the current ePrivacy Directive. Do Not Track could lead to better privacy protection and a user-friendlier browsing experience.

---

### 3.5.4. Article 9(3), withdrawing consent

Article 9(3) reads as follows:

'End-users who have consented to the processing of electronic communications data as set out in point (c) of Article 6(2) and points (a) and (b) of Article 6(3) shall be given the possibility to withdraw their consent at any time as set forth under Article 7(3) of [the GDPR] and be reminded of this possibility at periodic intervals of 6 months, as long as the processing continues.'

**Comment**
Article 9(3) concerns the withdrawal of consent, where the user's consent concerns the processing of electronic communications data (metadata and content). Article 9(3) of the ePrivacy proposal refers to Article 7(3) of the GDPR. Article 7(3) of the GDPR states that the data subject shall have the right to withdraw his or her consent at any time. Article 7(3) adds: 'It shall be as easy to withdraw as to give consent.'[411]

Article 9(3) of the ePrivacy proposal fails to mention Article 8 of that proposal. Article 8(1)(b) states, in short, that cookies and similar technologies can be used with the end-user's consent. The Article 29 Working Party and the EDPS note that a reference to Article 8(1)(b) must be added to Article 9(3).[412] Additionally, the ePrivacy Regulation should clarify that 'the reminder of the possibility to withdraw consent also applies to consent through browser settings.'[413] **We recommend that the EU lawmaker amends Article 9(3), while taking the above advice into account.**

### 3.5.5. Tracking walls and take-it-or-leave-it choices

On the internet, we encounter take-it-or-leave-it choices regarding our privacy on a daily basis.[414] Some websites use a tracking wall (or 'cookie wall'), a barrier that visitors can only pass if they agree to tracking by third parties. When confronted with such take-it-or-leave-it choices, many people click 'I agree'.[415] It is debatable whether people have meaningful control over personal information if they have to consent to tracking to be able to use services or websites. We recommend that the EU lawmaker adds rules on tracking walls and similar take-it-or-leave-it choices.

The ePrivacy proposal does not include specific rules on tracking walls.[416] What should the lawmaker do? We discuss four options: (i) no specific rules for tracking walls; (ii) ban tracking walls in certain circumstances; (iii) fully ban tracking walls; (iv) ban all third party tracking. Option (i) and option (iv) are probably not appropriate at this moment. **The lawmaker should seriously consider options (ii) and (iii): a partial or complete ban of tracking walls.**

**Option (1): no rules on tracking walls**
A first option for the lawmaker is: not including specific rules on tracking walls in the ePrivacy Regulation. We argue against this option. If the lawmaker does not add specific rules, the voluntariness of consent would have to be assessed separately for each tracking wall. The main question would be, in each case, whether people can give 'freely given' consent. As noted, the ePrivacy proposal refers to the GDPR for the definition of consent. The GDPR gives some guidance on when consent is 'freely given'. Article 7 of the GDPR states:

---

[411] Article 7(3) of the GDPR.
[412] Article 29 Working Party 2017 (WP247), p. 33. See also EDPS 2017/6, p. 30.
[413] Article 29 Working Party 2017 (WP247), p. 34.
[414] This section is based on, and includes sentences from, Zuiderveen Borgesius et al 2017b.
[415] See Helberger 2013; Leenes and Kosta 2015; Zuiderveen Borgesius 2015.
[416] The ePrivacy proposal does mention take-it-or-leave-it choices in the context of Article 6, in Recital 18. See our comment on Article 6.

'When assessing whether consent is freely given, utmost account shall be taken of whether, *inter alia*, the performance of a contract, including the provision of a service, is conditional on consent to the processing of personal data that is not necessary for the performance of that contract.'[417]

Hence, under article 7 of the GDPR, to assess whether consent is freely given (and therefore valid), it must be considered whether a service is made conditional on consent. This rule can be applied to tracking walls. The GDPR does not say that take-it-or-leave-it choices *always* lead to invalid consent. Rather, 'utmost account shall be taken' of whether a contract or service is made conditional on consent.[418]

But the GDPR's consent rules remain open for conflicting interpretations. More guidance on the legality of tracking walls would improve legal clarity. True, case law could clarify the conditions under which consent should be considered to be 'freely given'. But it may take a long time until there is enough case law to have clarity. Therefore, specific rules on tracking walls would be better.

## Option (ii): ban tracking walls in certain circumstances

A second option is banning tracking walls under certain circumstances. Indeed, **in some circumstances, tracking walls should not be allowed at all**.[419] (Or, to be more exact: in some circumstances, companies can never use a tracking wall to obtain valid consent.)

For instance, in 2016, the Article 29 Working Party mentioned five circumstances in which tracking walls should be banned:

'1. Tracking on websites, apps and or locations that reveal information about special categories of data (health, political, sexual, trade union etc.). Even if visits to services providing information about such special categories of data do not disclose in themselves special categories of data about these users, there is a high impact on the private life of those users if they are labelled as being interested in such information.

2. Tracking by unidentified third parties for unspecified purposes. This is for example the case when a website or app auctions its advertising space, and unknown third parties may actually start to track the users through the website or app;[420]

3. All government funded services;

4. All circumstances identified in the GDPR that lead to invalid consent, such as for example an unequal balance of power, if there is no equivalent alternative, or forced consent is part of a contract;

5. Bundled consent for processing for multiple purposes. Consent should be granular.'[421]

Regarding point (1): the EU lawmaker should phrase the prohibition carefully. It should be examined to what extent, for instance, news sites should fall in the category of sites that reveal information about special categories of data. Reading about certain topics could reveal ones political opinion. Moreover, information about people's media use and reading habits is generally sensitive.[422]

---

[417] Article 7(4) of the GDPR.
[418] Article 7(4) of the GDPR. See also Recitals 42 and 43 of the GDPR.
[419] See Helberger 2013; Zuiderveen Borgesius 2015, chapter 9; Zuiderveen Borgesius et al 2017b.
[420] Our footnote: presumably the Working Party refers to 'real time bidding'. See Olejnik, Minh-Dung and Castelluccia 2013.
[421] Article 29 Working Party 2016 (WP240), p. 17.
[422] See Helberger 2013; Zuiderveen Borgesius 2015, chapter 9; Zuiderveen Borgesius et al 2017b.

Regarding point (3): Dutch law already contains such bans on tracking walls on state-funded websites.[423] Indeed, state-funded websites, paid by tax money, should not interfere with people's privacy by forcing commercial third party tracking (internet-wide tracking) upon them.

The five circumstances mentioned by the Working Party are a good start. But tracking walls should probable be banned in more circumstances. For instance, a ban should also be considered for websites in sectors with confidentiality requirements. To illustrate: websites of hospitals, banks, and lawyers should generally refrain from enabling third party tracking, and from installing tracking walls.

**Such a black list of circumstances in which tracking walls are banned, should be non-exhaustive.** In other words: depending on the circumstances, a tracking wall can make consent involuntary (and thus invalid), even if the situation is not explicitly included on the black list.

The lawmaker should consider supplementing the black list with a grey list. If a situation is on the grey list, there is a legal presumption that a tracking wall makes consent involuntary, and therefore invalid. Hence, the legal presumption of the grey list shifts the burden of proof. For situations on the grey list, it's up to the company employing the tracking wall to prove that people can give 'freely given' consent, even though the company installed a tracking wall. Some consumer protection laws use a similar system, with a black list (illegal practices) and a grey list (practices presumed to be illegal).[424] A grey list could also be called a circumstance catalogue. The catalogue contains a non-exhaustive list of circumstances in which it is particularly questionable whether consent is valid if a company offers a take-it-or-leave-it choice, for instance with a tracking wall. Situations that could be included on the grey list include the following.

- If a company has a monopoly position, or a position resembling a monopoly, there is more chance of imbalance between the contract parties as individuals have little negotiation power vis-à-vis a monopolist. As Bygrave notes in the context of data protection law, 'fairness implies a certain protection from abuse by data controllers of their monopoly position.'[425]

- Sometimes, for other reasons it is not a realistic option for people to go to a competitor, for instance because of a lock-in situation. In a lock-in situation, it is difficult or costly to leave a service.[426] To illustrate, when all one's friends are on social network X, joining another social network site makes little sense.

- It is dubious whether consent is still 'freely given' if a company offers a take-it-or-leave-it choice and there are no competitors that offer a similar, more privacy-friendly service.[427] Again, it is questionable how much freedom people have. After all, somebody who does not want to be tracked would not have the possibility to use that type of service.[428]

- Other circumstances can also be considered when assessing whether consent is 'freely given' when a company offers a take-it-or-leave-it choice. For example, some people may be more vulnerable to pressure – such situations call for a more privacy-protective interpretation of the 'freely given' requirement. To illustrate: a take-it-or-leave-it choice is questionable when a service is aimed at, or often used by, children. Children are less likely to fully understand the implications of that choice, or might feel more readily pressured into accepting what they may perceive as a non-choice. Similarly, people with a medical condition might be more easily pressured into consenting to data collection if they believe that access to a particular website will give them important health information. Hence, whether consent can be considered to be freely given may also depend, in part, on personal circumstances, or on personal characteristics (such as the level of media literacy).

We mention those examples of situations for the grey list as starting points for a discussion. We do not intend to give a complete list.

**In sum, tracking walls could be banned in certain circumstances (a black list). Such a black list should be complemented with a grey list, with circumstances in which a tracking wall is presumed to be illegal.**

### Option (iii): ban tracking walls
**A third option is banning tracking walls completely.** Both the Article 29 Working Party and the EDPS call for a complete ban.[429] The EDPS proposes text for such a ban:

'No one shall be denied access to any information society services (whether these services are remunerated or not) on grounds that he or she has not given his or her consent under Article 8(1)(b) to the processing of personal data that is not necessary for the provision of those services'.[430]

It appears that a complete ban on tracking walls would be popular among the general public. A survey we did in the Netherlands shows that most people think tracking walls are unfair and unacceptable.[431] An EU-wide Eurobarometer survey found in 2017 that 64% of the respondents finds it unacceptable to 'hav[e] your online activities monitored (for example what you read, the websites you visit) in exchange for unrestricted access to a certain website.'[432] Civil society organisations also argue for a complete ban on tracking walls.[433]

If tracking walls were banned, a website publisher could still ask visitors whether they want to be tracked for targeted advertising. The main effect of the ban would be that the publisher couldn't offer people a take-it-or-leave-it choice regarding third party tracking (internet-wide tracking). We emphasise that a ban on tracking walls would not interfere with cookies that can be set without consent (for instance for a service requested by the end-user). Generally speaking, the ban on tracking walls should probably only apply to third party tracking (internet-wide tracking).

### Option (iv): ban third party tracking
A fourth option is: ban all third party tracking (internet-wide tracking). Some have argued for such a prohibition. For instance, according to security technologist Schneier it is 'vital' to adopt such a 'ban on third party ad tracking'. He adds: 'it's the companies that spy on us from website to website, or from device to device, that are doing the most damage to our privacy.'[434] Along similar lines, Ceglowski argues for these rules:

'Sites showing ads may only use two criteria in ad targeting: the content of the page, and whatever information they have about the visitor. Sites may continue to use third-party ad networks to serve ads, but those third parties must be forgetful; they may not store any user data across ad requests.'[435]

---

[429] Article 29 Working Party 2017 (WP247), p. 15; EDPS 2017/6, p. 17.
[430] EDPS 2017/6, p. 17.
[431] Zuiderveen Borgesius et al 2017b.
[432] Eurobarometer 2016, Question 5.1 (p. 93/T.24).
[433] European Digital Rights 2017; La Quadrature du Net 2017.
[434] Schneier 2016.
[435] Ceglowski 2015.

Such a ban on third party tracking would probably improve the protection of privacy and personal data. An additional advantage of a complete ban is its simplicity and legal clarity. However, we do not discuss the option of banning third party tracking in detail here. A disadvantage of a complete ban is that some people might prefer targeted (tracking-based) ads to non-targeted ads.

For the longer term, there is a need for a broader debate about the best way to regulate the collection and use of people's data. Are there practices that society should not accept, regardless of whether people consent to the practices? For instance, should online tracking be allowed on websites aimed at children? Is it acceptable if data are used for online price discrimination? Such questions warrant further research and debate.

### Economic effects

It is sometimes claimed that rules that require consent for online tracking, or that limit online tracking, will hurt website publishers because advertising income will go down.[436] This is not the place for a full discussion of the economics of online advertising and publishing.[437] But a couple of points need to be kept in mind.

First, advertising and online tracking are not the same. This distinction is important. Advertising is possible without tracking and personal data collection, and without interfering with privacy. For instance, even if third party tracking were completely banned, websites could still use contextual ads, such as ads for cars on a website about cars. Ad-funded media, such as commercial radio and 'free' newspapers, have existed for over a century – without tracking individuals.[438]

Second, in the long term, behavioural targeting may lead to less income for certain website publishers. With traditional advertising models, advertisers had to advertise in certain media to reach certain people. By way of illustration, a printed newspaper with many wine lovers among its readers could be a good place for a wine manufacturer to advertise. The newspaper assembles an audience, and provides the advertiser access to this audience. The price of an ad is based, among other things, on the number of readers.[439] Similarly, with contextual online advertising, advertisers aim to reach people by showing ads on certain webpages.

With behavioural targeting, an ad network can show a wine ad anywhere on the web to people whose profile suggests that they like wine. An ad network does not have to buy expensive ad space on a large professional website to advertise to an individual. The ad network can reach that individual when she visits a small unknown website, where advertising space is cheaper.[440] In sum, in the long term, behaviourally targeted advertising may reduce income for certain website publishers.

Third, some suggest that a rule that limits behavioural targeting would be bad for the economy and innovation. Even if this were true, the argument would not be sufficient to refrain from adopting that rule. Economic growth and innovation are very important. But so are fundamental rights. Moreover, if regulation pushes companies towards developing new and privacy preserving technologies, including technologies for privacy-preserving behavioural targeting, this is also innovation.[441]

Fourth: even if fundamental rights were ignored and only economic effects were considered, the most relevant question would be whether society *as a whole* gains or loses. From an economic perspective, it is unclear whether more or less legal privacy protection is better.

---

[436] See for instance: Interactive Advertising Bureau Europe 2017.
[437] See for more details: Zuiderveen Borgesius 2015, chapter 2 and 7; Mayer 2011; Strandburg 2013; Turow 2011.
[438] See generally: Wu 2016.
[439] Bermejo 2007.
[440] Idem. See also Turow 2011; Madrigal 2013.
[441] In principle, behavioural targeting would be possible without large-scale data collection, because behavioural targeting systems exist that don't involve sharing one's browsing behaviour with a firm. See e.g. Toubiana et al 2010. See also Zuiderveen Borgesius 2015, p. 271-272.

As Acquisti, the leading scholar on the economics of privacy, puts it, 'economic theory shows that, depending on conditions and assumptions, the protection of personal privacy can increase aggregate welfare as much as the interruption of data flows can decrease it.'[442]

Companies, such as 'ad tech' firms and ad networks, can also invoke a fundamental right. Companies could invoke their right to conduct a business, as protected by the EU Charter of Fundamental Rights: 'The freedom to conduct a business in accordance with Union law and national laws and practices is recognised.' But this right is not absolute and must be balanced against other fundamental rights, such as the right to privacy and the right to data protection.[443] Lawmaking often involves making difficult choices on balancing rights and interests – the ePricacy Regulation is no exception.

### Take-it-or-leave-it choices in other contexts

We focused our discussion on one specific type of take-it-or-leave-it choices: tracking walls. But take-it-or-leave-it choices regarding privacy also occur in other situations. For instance, chat and email services often require users to agree to a data use policy – if people do not agree, they cannot use the service. An app might require access to the camera or the contact list on an end-user's phone, while that access is unnecessary for providing the service. A 'smart' TV might listen to the sounds in people's living rooms, and might only work when people 'consent' to that.

We recommend that the EU lawmaker considers adopting rules regarding take-it-or-leave-it choices regarding privacy that do not concern tracking on websites. The arguments used above, for circumstances in which tracking walls should be prohibited for websites, can partly be applied to take-it-or-leave-it choices in other situations, such as email and chat services. However, for some situations, the rules for websites that were suggested above do not fit well. For example, a ban on take-it-or-leave-it choices for situations in which unknown third parties do the tracking may not fit well when an internet provider asks consent for analysing internet traffic.[444]

As noted, the EDPS suggests text for a ban on tracking walls. In brief, information society service providers should not offer take-it-or-leave-it choices regarding privacy.[445] That draft text by the EDPS may be too narrow for some situations. For example, providers of electronic communications networks would be outside the scope of the rule proposed by the EDPS.[446] The EDPS also suggests a specific provision that bans take-it-or-leave-it choices regarding privacy and personal data in the context of the Internet of Things:

'No one shall be denied any functionality of an IoT device (whether use of a device is remunerated or not) on grounds that he or she has not given his or her consent under Article 8(1)(b) for processing of any data that is not necessary for the functionality requested'.[447]

### Conclusion

**In conclusion, we recommend that the EU lawmaker seriously considers option (ii), a partial ban of tracking walls, and option (iii), a complete ban on tracking walls.**

**An advantage of option (ii) is that banning tracking walls only under certain circumstances is a more nuanced approach than completely banning them. The partial ban (the black list) should be complemented with a grey list (circumstances in which tracking walls are presumed to be illegal). A disadvantage of option (ii) is that the nuance comes at the cost of legal clarity.**

---

[442] Acquisti 2010, p. 19. See also Acquisti 2014, p. 90.
[443] The Google Spain case suggests that a firm's economic interests have less weight than the data subject's privacy rights (CJEU, C-131/12, Google Spain, 13 May 2014, par. 81, dictum, 4).
[444] See the second point made by Article 29 Working Party 2016 (WP240), p. 17.
[445] EDPS 2017/6, p. 17.
[446] Recital 18 of the ePrivacy proposal addresses take-it-or-leave-it choices offered by telecom providers. See our comment on Article 6.
[447] EDPS 2017/6, p. 18.

**A major advantage of option (iii), a complete ban of tracking walls, is that such a rule could be phrased in a relatively clear and straightforward way. Hence, a complete ban, the option preferred by the Article 29 Working Party and the EDPS, would provide more legal clarity than a partial ban.**

**We also recommend that the EU lawmaker considers adopting rules for take-it-or-leave-it choices regarding other situations than websites.** For instance, Article 6 requires providers to obtain consent for analysing people's metadata and content for targeted marketing. In that context, rules on take-it-or-leave-it choices are appropriate. **The lawmaker should also consider rules regarding take-it-or-leave-it choices in Internet of Things scenarios.**

## 3.6. Article 10, information and options for privacy settings to be provided

Article 10 states:

'1. Software placed on the market permitting electronic communications, including the retrieval and presentation of information on the internet, shall offer the option to prevent third parties from storing information on the terminal equipment of an end- user or processing information already stored on that equipment.
2. Upon installation, the software shall inform the end-user about the privacy settings options and, to continue with the installation, require the end-user to consent to a setting.
3. In the case of software which has already been installed on 25 May 2018, the requirements under paragraphs 1 and 2 shall be complied with at the time of the first update of the software, but no later than 25 August 2018.'

See also Recitals 20, 22, 23 and 24.

### Comment

Article 10 does not offer sufficient privacy protection. Article 10(1) of the ePrivacy proposal imposes an obligation on providers of software allowing electronic communications, including internet browsers, to offer people the 'option' to prevent third parties from storing information on their terminal equipment, or processing information already stored on their equipment. This type of software would include not only web browsers, but also other types of applications, such as messaging apps and apps providing route guidance.[448] According to the Article 29 Working Party, such applications could include mobile operating systems and software interfaces for connected devices in the Internet of Things.[449]

The proposed Article 10 reflects a 'privacy options', or 'prompted choice' approach,[450] where people are presented with a set of privacy settings, and must consent to one such setting. This approach contrasts radically with the previous version of Article 10 contained in the December 2016 draft. In the December 2016 draft, Article 10 was entitled 'Privacy by design', and provided that the settings of all components of terminal equipment 'shall be configured to, *by default*' to prevent third parties from storing information, processing information already stored on the equipment, and preventing third-party use of the equipment's processing capabilities.[451]

Thus, the Commission has rejected a 'privacy by design' approach, where privacy settings are set to privacy-protecting setting by default. The 'privacy options' approach is hard to reconcile with the GDPR. Article 25 of the GDPR requires data protection 'by design and by default', and imposes an obligation on data controllers to ensure that 'by default' only personal data that are necessary for each specific purpose of the processing are processed. Recital 23 of the ePrivacy proposal mentions this fact. But curiously Recital 23 fails to explain how an obligation to offer only privacy setting 'options' follows from this principle, rather than an obligation to provide a privacy-by-default setting.

---

[448] Recital 22 of the ePrivacy proposal.
[449] Article 29 Working Party 2017 (WP247), p. 14.
[450] See generally Sunstein 2014.
[451] Emphasis added.

Indeed, the Article 29 Working Party and the EDPS argue that the proposed Article 10(1) 'undermines' the GDPR principles, while a 'conscious policy choice' had been made to introduce the principles of data protection and privacy by design and by default in the GDPR.[452] The Working Party expresses 'grave concern' over the proposed Article 10.[453] The Working Party recommends that terminal equipment and software must by default offer privacy protective settings, and guide users through configuration menus to deviate from these default settings upon installation.[454]

As noted in Recital 20, information stored on a user's phone or other device 'may reveal details of an individual's emotional, political, social complexities, including the content of communications, pictures, the location of individuals by accessing the device's GPS capabilities, contact lists, and other information already stored in the device, the information related to such equipment requires enhanced privacy protection.'

A privacy-by-default principle better provides this 'enhanced privacy protection', rather than a mere privacy options approach. The Working Party points out that 'the "option" to prevent certain interferences already currently exists, and to date this has not resulted in sufficiently addressing the problem of unwarranted tracking'.[455]

In the Commission's public consultation, 81.2% of citizens and 63% of public authorities support imposing obligations on manufacturers of terminal equipment to market products with privacy-by-default settings activated, while 58.3% of industry favour the option to support self/co-regulation. Similarly, in a recent Europe-wide survey, 89% agree with the suggested option that the default settings of their browser should stop the sharing of their information.[456]

The Radio Equipment Directive requires that certain categories of radio equipment incorporate safeguards to ensure that personal data and privacy of users are protected,[457] and the Commission is empowered to specify which categories or classes of radio equipment are concerned. Recital 10 of the ePrivacy proposal mentions that the Regulation 'should not affect the applicability of any of the requirements' of the Radio Equipment Directive. However, as the Working Party points out, the Radio Equipment Directive only provides 'for a very limited security obligation', and it 'cannot replace specific privacy by default settings'.[458]

In our comment on Article 9, we discussed the possibility to make compliance with Do Not Track and similar standards obligatory. Even if such an obligation were included in the Regulation, **the EU lawmaker should require privacy-friendly defaults.**

---

[452] Article 29 Working Party 2017 (WP247), p. 14. See also EDPS 2017/6, p. 19.
[453] Article 29 Working Party 2017 (WP247), p. 3.
[454] Article 29 Working Party 2017 (WP247), p. 14. See also European Digital Rights 2017.
[455] Article 29 Working Party 2017 (WP247), p. 14.
[456] Explanatory memorandum to the ePrivacy proposal, p. 6 (section 3.2).
[457] Article 3(3)(e) of Directive 2014/53/EU of the European Parliament and of the Council of 16 April 2014 on the harmonisation of the laws of the Member States relating to the making available on the market of radio equipment and repealing Directive 1999/5/EC (OJ L 153, 22.5.2014, p. 62–106).
[458] Article 29 Working Party 2017 (WP247), p. 14.

## 3.7. Article 11, restrictions

### 3.7.1. Article 11(1), restricting rights and obligations, data retention

Article 11(1) of the ePrivacy proposal reads:

'Union or Member State law may restrict by way of a legislative measure the scope of the obligations and rights provided for in Articles 5 to 8 where such a restriction respects the essence of the fundamental rights and freedoms and is a necessary, appropriate and proportionate measure in a democratic society to safeguard one or more of the general public interests referred to in Article 23(1)(a) to (e) of [the GDPR] or a monitoring, inspection or regulatory function connected to the exercise of official authority for such interests.'[459]

See also Recital 26.

**Comment**

Article 11 concerns restrictions on some of the ePrivacy proposal's rights and obligations, including the right to communications confidentiality. Article 11 is the successor of article 15 of the current ePrivacy Directive.

Article 11 of the ePrivacy proposal provides a legal basis for the European Union and the Member States to adopt legislation that restricts the rights conferred by Articles 5 to 8 of the ePrivacy proposal. However, such legislation would be subject to conditions. By reference to Articles 23(1)(a) to (e) of the GDPR, the ePrivacy proposal effectively adds two new purposes for interfering with users' communications rights.[460] This means a substantive change from the ePrivacy Directive and a lower threshold for interventions.[461] **We recommend keeping the original wording in Article 15 of the ePrivacy Directive.**

The CJEU, in its *Tele 2 and Watson* judgment,[462] decided that national legislation, such as that relating to the retention of data for the purpose of combating crime, falls within the scope of European law, and must meet the conditions laid down in (the predecessor to) Article 11 of the ePrivacy proposal. Any such restriction laid down in Union or Member State law must respect the essence of the fundamental rights and freedoms and be a necessary, appropriate and proportionate measure in a democratic society to safeguard one or more of the general public interests referred to in Articles 23(1)(a) to (e) of the GDPR or a monitoring, inspection or regulatory function connected to the exercise of official authority for such interests.

In its *Digital Rights Ireland* ruling,[463] with which it invalided the Data Retention Directive (2006/24), the CJEU further specified how these conditions must be interpreted. According to the CJEU, data retention measures do not violate per se the essence of the fundamental rights to privacy and data protection. The distinction the CJEU maintains between communications content and traffic data[464] has been criticised because it reverts to an outdated perspective, according to which metadata are less sensitive that communications content.[465] Blanket data retention measures, according to the CJEU, are deemed a particularly serious interference with the fundamental rights to privacy and data protection as guaranteed by the EU Charter. The CJEU subjects such measures to a strict scrutiny test and applies a rigorous proportionality test under the EU Charter. In its later *Tele2* judgment[466] the CJEU confirms that legislation prescribing blanket data retention, i.e. the indiscriminate collection of traffic data, exceeds the limits of what is strictly necessary and cannot be considered to be justified within a democratic society.

---

[459] See also Recital 26 of the ePrivacy proposal.
[460] I.e. "the execution of criminal penalties, including the safeguarding against and the prevention of threats to public security" (Art. 23(1)(d) GDPR) and "other important objectives of general public interest of the Union or of a Member State, in particular an important economic or financial interest of the Union or of a Member State, including monetary, budgetary and taxation a matters, public health and social security" (Art. 23(1)(e) GDPR).
[461] Article 29 Working Party 2017 (WP247), p. 23-24.
[462] CJEU (Grand Chamber), Judgement of 21 December 2016, Joined Cases C-203/15 (Tele2 Sverige AB) and C-698/15 (Watson), ECLI:EU:C:2016:970, para. 73.
[463] CJEU (Grand Chamber), Judgement of 8 April 2014, Joined Cases C-293/12 (Digital Rights Ireland) and C-594/12 (Seitlinger a.o.), ECLI:EU:C:2014:238.
[464] Ibid., para. 40.
[465] See Granger and Irion 2014, p. 847. See also our comment on Article 6.

The wording of Article 11(1) of the ePrivacy proposal would not alter the interpretations rendered by the CJEU. Article 11(1) thus preserves the status quo according to which national data retention measures must comply with the EU Charter.[467] The proposed Article 11(1) would only clarify that the exhaustive list of general public interests which can justify a restriction of users' fundamental rights and freedoms includes the monitoring, inspection, or regulatory function connected to the exercise of official authority for such interests. This clarification inter alia strengthens the right to independent supervision guaranteed in Article 8(3) of the EU Charter.

The EDPS notes that the extension of the scope of the ePrivacy Regulation compared with the ePrivacy Directive should not be understood as a mandate for Member States to extend current data retention schemes to services which are not included in Article 15(1) ePrivacy Directive but are included in the scope of the ePrivacy Regulation.[468]

Furthermore, the EDPS recommends that prior judicial authorisation should be required for access to electronic communications content or metadata in cases in which Article 23(1)(e) of the GDPR applies.[469] **We recommend that the EU lawmaker considers that advice.**

### 3.7.2. Article 11(2), internal procedures etc.

Article 11(2) of the ePrivacy proposal states:

'Providers of electronic communications services shall establish internal procedures for responding to requests for access to end-users' electronic communications data based on a legislative measure adopted pursuant to paragraph 1. They shall provide the competent supervisory authority, on demand, with information about those procedures, the number of requests received, the legal justification invoked and their response.'

See also Recital 26.

**Comment**
Article 11(2) of the ePrivacy proposal carries obligations for providers of electronic communications services to implement measures adopted pursuant to paragraph 1 and to comply with access requests for end-users' electronic communications data. The notion of internal procedures is potentially very wide and in need of clarification that it does not, for example, include a duty to provide law enforcement access to encryption technology.

The information duties in the second sentence are limited to requests from public authorities. **However, considering the important role of transparency reporting, we recommend that the EU lawmaker adds a duty for providers of electronic communications services to publish, on an annual basis, statistics about the number of requests it has received, including the legal basis for such requests.** Transparency reporting has evolved into a good practice backed by large providers of public electronic communications services.[470] Such reporting of aggregate level data on law enforcement requests for user data is important to gauge the magnitude to which such powers are used. Some Member States' legacy legislation still prohibits providers from issuing transparency reporting. Since only statistics are periodically reported, transparency reporting would not interfere with particular law enforcement activities.

---

[466] CJEU (Grand Chamber), Judgement of 21 December 2016, Joined Cases C-203/15 (Tele2 Sverige AB) and C-698/15 (Watson), ECLI:EU:C:2016:970, para. 73.
[467] CJEU (Grand Chamber), Judgement of 21 December 2016, Joined Cases C-203/15 (Tele2 Sverige AB) and C-698/15 (Watson), ECLI:EU:C:2016:970, para. 89.
[468] EDPS 2017/6, p. 21.
[469] EDPS 2017/6, p. 21.

According to the EDPS, it would be advisable to consider whether such transparency reports should be sent to the supervisory authorities periodically, instead of being only made available on demand.[471] We recommend that the EU lawmaker takes this advice into account.

---

[470] Including numerous European companies, such as Deutsche Telekom, Telenor and Vodafone. See for an overview <https://www.accessnow.org/transparency-reporting-index/> accessed 5 May 2017. See also: <http://www.telecomindustrydialogue.org/about/guiding-principles/> accessed 5 May 2017. See also European Digital Rights 2017, p. 11.
[471] EDPS 2017/6, p. 22.

# 4. NATURAL AND LEGAL PERSONS' RIGHTS TO CONTROL ELECTRONIC COMMUNICATIONS (PROPOSAL CHAPTER III)

## 4.1. Article 12, presentation and restriction of calling and connected line identification

Article 12 reads as follows:

'Where presentation of the calling and connected line identification is offered in accordance with Article [107] of the [Directive establishing the European Electronic Communication Code], the providers of publicly available number-based interpersonal communications services shall provide the following:
(a) the calling end-user with the possibility of preventing the presentation of the calling line identification on a per call, per connection or permanent basis;
(b) the called end-user with the possibility of preventing the presentation of the calling line identification of incoming calls;
(c) the called end-user with the possibility of rejecting incoming calls where the presentation of the calling line identification has been prevented by the calling end-user;
(d) the called end-user with the possibility of preventing the presentation of the connected line identification to the calling end-user.
(2) The possibilities referred to in points (a), (b), (c) and (d) of paragraph 1 shall be provided to end-users by simple means and free of charge.
(3) Point (a) of paragraph 1 shall also apply with regard to calls to third countries originating in the Union. Points (b), (c) and (d) of paragraph 1 shall also apply to incoming calls originating in third countries.
(4) Where presentation of calling or connected line identification is offered, providers of publicly available number-based interpersonal communications services shall provide information to the public regarding the options set out in points (a), (b), (c) and (d) of paragraph 1.'

See also Recital 27.

**Comment**
Article 12 of the ePrivacy Proposal, the successor to Article 8 of the current ePrivacy Directive, concerns 'presentation and restriction of calling and connected line identification'.
Recital 27 of the ePrivacy Proposal adds:

'As regards calling line identification, it is necessary to protect the right of the calling party to withhold the presentation of the identification of the line from which the call is being made and the right of the called party to reject calls from unidentified lines. Certain end-users, in particular help lines, and similar organisations, have an interest in guaranteeing the anonymity of their callers. As regards connected line identification, it is necessary to protect the right and the legitimate interest of the called party to withhold the presentation of the identification of the line to which the calling party is actually connected.'

Article 12 of the ePrivacy proposal does not seem to cover abuse of call line identification such as 'spoofing'. Caller ID spoofing is the practice of causing the telephone network to indicate to the receiver of a call that the originator of the call is a station other than the true originating station. For example, a Caller ID display might display a phone number different from that of the telephone from which the call was placed. **We recommend that the EU lawmaker considers whether specific rules for ID spoofing are needed.** See also Article 16(6), on informing end-users of the identity of the party on behalf of whom direct marketing communication is transmitted.

## 4.2. Article 13, exceptions to presentation and restriction of calling and connected line identification

Article 13 reads as follows:

'(1) Regardless of whether the calling end-user has prevented the presentation of the calling line identification, where a call is made to emergency services, providers of publicly available number-based interpersonal communications services shall override the elimination of the presentation of the calling line identification and the denial or absence of consent of an end-user for the processing of metadata, on a per- line basis for organisations dealing with emergency communications, including public safety answering points, for the purpose of responding to such communications.

(2) Member States shall establish more specific provisions with regard to the establishment of procedures and the circumstances where providers of publicly available number-based interpersonal communication services shall override the elimination of the presentation of the calling line identification on a temporary basis, where end-users request the tracing of malicious or nuisance calls.'

See also Recital 28.

**Comment**

Article 13 concerns 'exceptions to presentation and restriction of calling and connected line identification'. Recital 28 adds:

'There is justification for overriding the elimination of calling line identification presentation in specific cases. End-users' rights to privacy with regard to calling line identification should be restricted where this is necessary to trace nuisance calls and with regard to calling line identification and location data where this is necessary to allow emergency services, such as eCall, to carry out their tasks as effectively as possible.'

**The provision does not indicate the criteria for overriding the call line identification. This might result in a threshold not compliant with privacy standards. The EU lawmaker could consider introducing several criteria.** For instance, the law could require that the interference should be subject to a proportionality test, or should be described by law or ordered by a court.

## 4.3. Article 14, incoming call blocking

Article 14 reads as follows:

'Providers of publicly available number-based interpersonal communications services shall deploy state of the art measures to limit the reception of unwanted calls by end-users and shall also provide the called end-user with the following possibilities, free of charge:

(a) to block incoming calls from specific numbers or from anonymous sources;

(b) to stop automatic call forwarding by a third party to the end-user's terminal equipment.'

See also Recital 29.

**Comment**

Article 14 concerns the blocking of incoming calls. Recital 29 adds:

'Technology exists that enables providers of electronic communications services to limit the reception of unwanted calls by end-users in different ways, including blocking silent calls and other fraudulent and nuisance calls. Providers of publicly available number-based interpersonal communications services should deploy this technology and protect end-users against nuisance calls and free of charge. Providers should ensure that end-users are aware of the existence of such functionalities, for instance, by publicising the fact on their webpage.'

Article 14 states that certain providers must 'deploy state of the art measures to limit the reception of unwanted calls'. It is not clear who interprets the 'state of the art' criterion. Is this a responsibility of the provider or regulators? Finally, 'state of the art measures' do not necessarily equal user-friendly solutions. **The EU lawmaker should consider stating that measures should be user-friendly.**

Incoming call blocking can also be used to block unsolicited communications, which should be prefixed by a code as required by Article 16(3)(b) of the ePrivacy Proposal. To further enhance the possibility of blocking unsolicited communications, Article 14(1)(a) should also include blocking unsolicited communications that are prefixed with the required code. The EDPS recommends adding the phrase 'or having a specific code/prefix identifying the fact that the call is a marketing call, as foreseen in Article 16(3)(b)' after 'to block incoming calls from specific numbers'.[472] **We recommend that the EU lawmaker considers amending the provision.**

## 4.4.  Article 15, publicly available directories

Article 15 reads as follows:

(1) The providers of publicly available directories shall obtain the consent of end-users who are natural persons to include their personal data in the directory and, consequently, shall obtain consent from these end-users for inclusion of data per category of personal data, to the extent that such data are relevant for the purpose of the directory as determined by the provider of the directory. Providers shall give end-users who are natural persons the means to verify, correct and delete such data.
(2) The providers of a publicly available directory shall inform end-users who are natural persons whose personal data are in the directory of the available search functions of the directory and obtain end-users' consent before enabling such search functions related to their own data.
(3) The providers of publicly available directories shall provide end-users that are legal persons with the possibility to object to data related to them being included in the directory. Providers shall give such end-users that are legal persons the means to verify, correct and delete such data.
(4) The possibility for end-users not to be included in a publicly available directory, or to verify, correct and delete any data related to them shall be provided free of charge.'

See also Recitals 30 and 31.

### Comment

Article 15 concerns publicly available directories, which are defined in Article 4(3)(d).[473] Contrary to the current ePrivacy Directive, the ePrivacy proposal does not distinguish search and reverse search of publicly available directories. Article 12(3) of the current ePrivacy Directive provides the option for Member States to require separate consent for reverse search. The Article 29 Working Party and the EDPS recommend introducing the option for granular consent for different modes of searching publicly available directories.[474] And according to the EDPS, the phrase 'as determined by the provider of the directory' should be

---

deleted from Article 15(1).[475] **We recommend that the EU lawmaker considers this advice.**

## 4.5. Article 16, unsolicited communications

### 4.5.1. Article 16(1), direct marketing communications

Article 16(1) reads as follows:

'Natural or legal persons may use electronic communications services for the purposes of sending direct marketing communications to end-users who are natural persons that have given their consent.'

See also Recitals 32, 33, 34 and 35.

**Comment**
Article 16 deals with unsolicited communications, or, as they are sometimes called, spam. The main rule of Article 16 is, in short, that direct marketing communications are only allowed following the consent of the individual concerned. Article 16 uses the phrase 'direct marketing communications' (see our comment on Article 4(3)(f)). Recital 33 adds:

'Legal certainty and the need to ensure that the rules protecting against unsolicited electronic communications remain future-proof justify the need to define a single set of rules that do not vary according to the technology used to convey these unsolicited communications, while at the same time guaranteeing an equivalent level of protection for all citizens throughout the Union. (…)'.

Article 16(1) should be amended to improve clarity. As the Working Party notes, Article 16(1) 'does not contain an explicit prohibition on sending (directing, or presenting) direct marketing without consent.'[476] The Working Party says that Article 16(1) should be phrased as follows:

'The use by natural or legal persons of electronic communications services, including voice-to-voice calls, automated calling and communication systems, including semi-automated systems that connect the called person to an individual, faxes, electronic mail or other use of electronic communication services for the purposes of presenting direct marketing communications to end-users *may be allowed only in respect of end-users who have given their prior consent*.'[477] We agree that such an amendment would improve clarity.

The Article 29 Working Party notes that Article 16 might unwantedly include communications concerning commercial concerns or interests sent to elected (political) representatives. The Article 29 Working Party recommends clarifying that the ePrivacy Regulation does not prevent such communications.[478] The EDPS notes that the word 'citizen' in Recital 33 should be changed to 'individual'.[479] **Indeed, the ePrivacy Regulation should protect not only EU citizens, but all people in EU territory.**[480]

The scope of Article 16(1) is limited to natural persons, and excludes legal persons. The EDPS suggests that it should be made clear that the provisions applicable to natural persons should apply when a direct marketing communication is sent to a natural person working for a legal person.[481]

---

[475] EDPS 2017/6, p. 31.
[476] Article 29 Working Party 2017 (WP247), p. 31.
[477] Article 29 Working Party 2017 (WP247), p. 31.
[478] Article 29 Working Party 2017 (WP247), p. 32.
[479] EDPS 2017/6, p. 26.
[480] See also Article 8 of the EU Charter of Fundamental Rights, which says that the personal data of 'everyone' (not of 'citizens') deserve protection.
[481] EDPS 2017/6, p. 32.

But the EU lawmaker can also consider another approach. Various member states have extended the protection to legal persons because they see no difference between natural and legal persons when it comes to being protected against unsolicited communications. Small companies are often in a similar situation as natural persons. **We recommend that the EU lawmaker considers extending the scope of Article 16(1), to protect legal persons against direct marketing communications without consent. More generally, we recommend that the lawmaker considers amending the provision, taking into account the above advice.**

### 4.5.2.   Article 16(2), electronic mail

Article 16(2) reads as follows:

'Where a natural or legal person obtains electronic contact details for electronic mail from its customer, in the context of the sale of a product or a service, in accordance with [the GDPR], that natural or legal person may use these electronic contact details for direct marketing of its own similar products or services only if customers are clearly and distinctly given the opportunity to object, free of charge and in an easy manner, to such use. The right to object shall be given at the time of collection and each time a message is sent.'

See also Recital 33.

**Comment**

Article 16(2) provides, in short, an opt-out system for direct marketing emails to existing customers. Recital 33 adds: that 'it is reasonable to allow the use of e-mail contact details within the context of an existing customer relationship for the offering of similar products or services. Such possibility should only apply to the same company that has obtained the electronic contact details in accordance with [the GDPR].' The current ePrivacy Directive has a similar exception to the opt-in rule for marketing email: within the context of an existing customer relationship, the directive allows, in short, a company to send marketing email offering similar products or services, if each email includes a clear opt-out possibility.[482]

Article 16(2) of the ePrivacy proposal needs further clarification. 'In the context of the sale of a product or a service' might be understood to exclude contact details that a company has obtained in a context other than the sale of a product or a service.

The Article 29 Working Party has further concerns regarding Article 16(2) as it is currently drafted. The direct marketing provisions apply to non-commercial activities such as those conducted by charities and political parties. Article 16(2) should therefore also include those organisations when promoting their aims or ideals to previous supporters.[483]

Furthermore, the Article 29 Working Party calls for a time limit to be set for the validity of existing customer contacts. The Article 29 Working Party suggests a one- or two-year timespan in which companies or organisations can contact their previous customers.[484] **We recommend that the EU lawmaker considers amending Article 16(1), taking into account the above suggestions.**

### 4.5.3.   Article 16(3), direct marketing calls

Article 16(3) reads as follows:

---

[482] Article 13(2) and Recital 41 of the 2009 ePrivacy Directive.
[483] Article 29 Working Party 2017 (WP247), p. 31.
[484] Article 29 Working Party 2017 (WP247), p. 31.

'(3) Without prejudice to paragraphs 1 and 2, natural or legal persons using electronic communications services for the purposes of placing direct marketing calls shall:
(a) present the identity of a line on which they can be contacted; or
(b) present a specific code/or prefix identifying the fact that the call is a marketing call.'

See also Recital 34.

### Comment
Article 16(3) concerns marketing calls. Unsolicited marketing calls are a serious problem in some Member States. A Europe-wide survey shows 'that a significant majority of responding citizens (61%) believe that they receive too many unsolicited calls offering them goods or services.[485] The percentages of citizens receiving too many communications are particularly high in three large MS, such as UK, Italy and France where it is on average around 75%.'[486]

The way in which Article 16(3) is currently drafted allows for the contact line identification requirement to be read as an alternative to the prefix requirement. These requirements have different aims and should therefore be cumulative instead of alternative.[487] The prefix requirement allows the recipient to identify calls as direct marketing calls upfront and creates the possibility of blocking them. The line identification requirement allows the recipient to identify the sender of direct marketing calls and to contact the sender. We recommend that Article 16(3)(a) should be reworded from 'can be contacted; <u>or</u>' to 'can be contacted; <u>and</u>'.[488]

Recital 34 states:

'To facilitate effective enforcement of Union rules on unsolicited messages for direct marketing, it is necessary to prohibit the masking of the identity and the use of false identities, false return addresses or numbers while sending unsolicited commercial communications for direct marketing purposes. *Unsolicited marketing communications should therefore be clearly recognizable as such and should indicate the identity of the legal or the natural person transmitting the communication or on behalf of whom the communication is transmitted and provide the necessary information for recipients to exercise their right to oppose to receiving further written and/or oral marketing messages*.'

**The EU lawmaker should consider adding the emphasised (italics) requirements to Article 16(3). In sum, we recommend that the lawmaker considers amending Article 16(3). See also our comment on Article 12.**

### 4.5.4.    Article 16(4), voice-to-voice calls

Article 16(4) reads as follows:

'Notwithstanding paragraph 1, Member States may provide by law that the placing of direct marketing voice-to-voice calls to end-users who are natural persons shall only be allowed in respect of end-users who are natural persons who have not expressed their objection to receiving those communications.'

See also Recital 36.

### Comment
Article 16(4) concerns voice-to-voice direct marketing calls. Recital 36 states that 'Member States should therefore be able to establish and or maintain national systems only allowing such calls to end-users who have not objected.'

---

[485] Original footnote: SMART 2016/079.
[486] ePrivacy Impact Assessment 2017 Pt. 1, p. 9.
[487] Article 29 Working Party 2017 (WP247), p. 22.
[488] See Article 29 Working Party 2017 (WP247), p. 22-23; EDPS 2017/6, p. 32.

In some Member States, centralised opt-out systems (Do Not Call lists) have been very successful in protecting people against unwanted calls. We recommend that Member States be *required* to set up a centralised Do Not Call system.[489] We recommend that the EU lawmaker requires that people should be able to join a Do Not Call register in at least two ways. First, people should be able to join the Do Not Call register by contacting the centralised Register in a straightforward manner. Second, each party making a direct marketing voice-to-voice call should offer the called person the option to join the Do Not Call register immediately.[490]

The EU lawmaker could consider whether a Union-wide Do Not Call register should be implemented.[491] However, such a Union-wide Do Not Call register might take a long time to implement, because it involves cooperation and administration. Therefore, at least in the short and medium term, national Do Not Call registers may be preferable. In any case, as the EDPS notes, 'It is crucial that situations could no longer exist where a user would have to opt-out with each individual communication provider, instead of simply registering via a Do Not Call register.'[492]

**The EU lawmaker could consider adopting an opt-in system for direct marketing voice-to-voice calls – that seems to be the preference of the EDPS.[493] On the other hand, as Recital 36 notes, voice-to-voice direct marketing calls 'are more costly for the sender and impose no financial costs on end-users'. Therefore, an easy and centralised opt-out system might suffice.**

A final minor point: there appears to be a typo in Recital 36; we recommend amending the first sentence.

### 4.5.5.    Article 16(5), legal persons

Article 16(5) reads as follows:

'Member States shall ensure, in the framework of Union law and applicable national law, that the legitimate interest of end-users that are legal persons with regard to unsolicited communications sent by means set forth under paragraph 1 are sufficiently protected.'

### Comment
The Article 29 Working Party states that a distinction should be made between natural persons working for legal persons and generic contact details that legal persons made public. In the first situation consent would be required for sending direct marketing; in the second situation consent would not be required for sending direct marketing.[494] The EDPS has made similar comments.[495]

The Article 29 Working Part also notes that the wording of article 16(5) is different from the comparable Article 13(5) of the ePrivacy Directive. The effect of the difference in wording is unclear. We agree that it should be made clear that this difference does not lower the level of protection.[496]

**We recommend that the EU lawmaker considers the above suggestions, and amends the provision as appropriate. As mentioned, we also recommend that the**

---

[489] EDPS 2017/6, p. 33.
[490] See: Article 29 Working Party 2017 (WP247), p. 22.
[491] See: Article 29 Working Party 2017 (WP247), p. 22; EDPS 2017/6, p. 33.
[492] EDPS 2017/6, p. 33.
[493] EDPS 2017/6, p. 33.
[494] Article 29 Working Party 2017 (WP247), p. 32.
[495] EDPS 2017/6, p. 32.
[496] Article 29 Working Party 2017 (WP247), p. 32.

**lawmaker considers providing the same level of protection, in the context of unsolicited communications, to legal persons and natural persons.**

### 4.5.6. Article 16(6), withdrawing consent

Article 16(6) reads as follows:

'Any natural or legal person using electronic communications services to transmit direct marketing communications shall inform end-users of the marketing nature of the communication and the identity of the legal or natural person on behalf of whom the communication is transmitted and shall provide the necessary information for recipients to exercise their right to withdraw their consent, in an easy manner, to receiving further marketing communications.'

See also Recitals 34 and 35.

**Comment**

Article 16(6) concerns withdrawing consent. Recital 35 adds:

'In order to allow easy withdrawal of consent, legal or natural persons conducting direct marketing communications by email should present a link, or a valid electronic mail address, which can be easily used by end-users to withdraw their consent. Legal or natural persons conducting direct marketing communications through voice-to-voice calls and through calls by automating calling and communication systems should display their identity line on which the company can be called or present a specific code identifying the fact that the call is a marketing call.'

**In a similar way to the GDPR, Article 16(6) should make explicit that withdrawing consent should be at least as easy as giving consent.[497] We also recommend that the EU lawmaker considers making it clear that withdrawing consent should lead to the immediate termination of the marketing.**

Furthermore, withdrawing consent should be free of charge. This requirement is included in Article 16(2) but it is not included in Article 16(6). Recital 70 of the GDPR states that people should be able to object to processing for direct marketing purposes free of charge.[498]

There is no explicit prohibition of the use of false identities in Article 16(6), whilst this is noted in Recital 34. A mere obligation to inform the end-user of 'the identity of the legal or natural person on behalf of whom the communication is transmitted' may not be enough to ban the use of false identities for direct marketing purposes. **We recommend that the lawmaker considers adding an explicit ban on the use of false identities for direct marketing purposes.[499]**

### 4.5.7. Article 16(7), implementing measures

Article 16(7) reads as follows:

'The Commission shall be empowered to adopt implementing measures in accordance with Article 26(2) specifying the code/or prefix to identify marketing calls, pursuant to point (b) of paragraph 3.'

---

[497] See Article 7(3) of the GDPR.
[498] Article 29 Working Party 2017 (WP247), p. 21; EDPS 2017/6, p. 32.
[499] Article 29 Working Party 2017 (WP247), p. 22; EDPS 2017/6, p. 33.

**Comment**

As unsolicited marketing has a high impact on end-users, **it might be advisable to replace 'shall be empowered to' with 'shall'**.

Article 16(7) of the ePrivacy proposal refers to Article 26(2) of the ePrivacy proposal. Article 16(6) of the December 2016 draft referred to Article 28(2) of the December 2016 draft, which stated that for the implementation measures, the Committee shall be the Communications Committee established under the Directive establishing the European Electronic Communications Code. But Article 16(7) of the ePrivacy proposal refers to Article 26(2) of the ePrivacy proposal, stating that Article 5 of Regulation 182/2011 is applicable.

It is unclear whether Article 16(7) of the ePrivacy proposal should refer to Article 26(1) or 26(2) of the ePrivacy proposal. **We recommend that the EU lawmaker clarifies Article 16(7).**

## 4.6. Article 17, Information about detected security risks

Article 17 reads as follows:

'In the case of a particular risk that may compromise the security of networks and electronic communications services, the provider of an electronic communications service shall inform end-users concerning such risk and, where the risk lies outside the scope of the measures to be taken by the service provider, inform end-users of any possible remedies, including an indication of the likely costs involved.'

See also Recital 37.

**Comment**

Article 17 introduces a duty of care for the providers of electronic communications services. Such providers must alert end-users in case of a particular risk that may compromise the security of networks and services. A similar requirement is included in the current ePrivacy Directive.[500]

The title of Article 17 is 'Information about detected security risks', whilst the text of the Article actually concerns 'risk[s] that may compromise the security of networks and electronic communication services'. The title of Article 17 seems to imply a narrower scope than the text of Article 17. **It should be considered whether the title of Article 17 should be changed to 'Information about possible security risks'.[501]**

In the next section, we discuss security and encryption in the context of the ePrivacy proposal.

## 4.7. Security and encryption

### 4.7.1. Security

Recital 37 of the ePrivacy proposal states that security is appraised in the light of Article 32 of the GDPR. Article 32 of the GDPR concerns security of processing (personal data). In the explanatory memorandum to the ePrivacy proposal, the Commission adds that '[t]he security obligations in the GDPR and in the EECC [European Electronic Communications Code] will apply to the providers of electronic communications services.'[502] The European Commission adds that the draft European Electronic Communications Code complements the ePrivacy Regulation 'by ensuring the security of electronic communications services.'[503]

---

[500] Article 4(2) of the 2009 ePrivacy Directive.
[501] See Article 29 Working Party 2017 (WP247), p. 24.
[502] Explanatory memorandum to the ePrivacy proposal, section 5.2.

The draft European Electronic Communications Code defines 'security' as follows:

'"Security" of networks and services means the ability of electronic communications networks and services to resist, at a given level of confidence, any action that compromises the availability, authenticity, integrity or confidentiality of stored or transmitted or processed data or the related services offered by, or accessible via, those networks or services.'[504]

Article 40 of the draft European Electronic Communications Code establishes the main duty for undertakings providing public communications networks or publicly available electronic communications services to secure networks and service:

'Member States shall ensure that undertakings providing public communications networks or publicly available electronic communications services take appropriate technical and organisational measures to appropriately manage the risks posed to security of networks and services. Having regard to the state of the art, these measures shall ensure a level of security appropriate to the risk presented. In particular, measures shall be taken to prevent and minimise the impact of security incidents on users and on other networks and services.'

Furthermore, Article 40 of the draft European Electronic Communications Code establishes a duty to notify the competent authority of breaches of security without undue delay. A similar obligation is included in the current ePrivacy Directive,[505] and (regarding personal data breaches) in the GDPR.[506] Article 40 of the draft European Electronic Communications Code is without prejudice to the GDPR and the ePrivacy Directive.

Other security-related provisions in the draft European Electronic Communications Code are Articles 3, 41, 95, and 96, and Recital 90, 91, 92, and 236.

Security obligations are also covered in the Radio Equipment Directive[507], the NIS Directive,[508] and to a lesser extent the eIDAS Regulation.[509] But the combined scope of these five regulatory instruments may not include all the services envisaged within the scope of the ePrivacy Regulation.

The EDPS calls for clarification, by stating that Article 40 of the draft European Electronic Communications Code applies *mutatis mutandis* to all services within the scope of the ePrivacy proposal, regardless of whether those services are also within the scope of that Code. [510] A recital accompanying such a provision could include more specific additional security measures such as:

(i) Security and privacy standards for networks and services,
(ii) Security requirements for software used in combination with communications services (e.g. smartphone operating systems),

---

[503] Explanatory memorandum to the ePrivacy proposal, section 1.3.
[504] Article 2(22) of the draft European Electronic Communications Code. See generally on the legal concept of security: Arnbak 2016.
[505] Article 4(3) of the 2009 ePrivacy Directive.
[506] Article 33 of the GDPR.
[507] Directive 2014/53/EU of the European Parliament and of the Council of 16 April 2014 on the harmonisation of the laws of the Member States relating to the making available on the market of radio equipment and repealing Directive 1999/5/EC (OJ L 153, 22.5.2014, p. 62–106).
[508] Directive (EU) 2016/1148 of the European Parliament and of the Council of 6 July 2016 concerning measures for a high common level of security of network and information systems across the Union (OJ L 194, 19.7.2016, p. 1–30).
[509] Regulation (EU) No 910/2014 of the European Parliament and of the Council of 23 July 2014 on electronic identification and trust services for electronic transactions in the internal market and repealing Directive 1999/93/EC (OJ L 257, 28.8.2014, p. 73–114).
[510] EDPS 2017/6, p. 34.

(iii) Inclusion of Internet of Things devices in the scope of security requirements,
(iv) Inclusion of all network components in the scope of security requirements (e.g. SIM cards and routers).[511]

The provisions in the ePrivacy proposal do not refer to the security requirements in the GDPR and the draft European Electronic Communications Code. The EU lawmaker could consider including a reference to those two texts in Article 17 of the ePrivacy Regulation. The inclusion of a reference to the definition of 'security' in the European Electronic Communications Code[512] in article 4(1)(b) of the ePrivacy Regulation could also be considered. It should also be considered whether 'security of networks' in Article 17 should be changed to 'security of electronic communications networks'.

**We recommend that the EU lawmaker examines whether EU legislation on security of devices should be improved**. **However, the ePrivacy Regulation may not be the right instrument for such rules.** Moreover, security of devices is a complicated topic.[513] A full analysis of that topic would take a long time, and might slow down the drafting of the ePrivacy Regulation too much.

### 4.7.2. Encryption

Nowadays, the primary way in which communications confidentiality is guaranteed in practice is through the application of encryption, and cryptographic protocols more generally.[514] Encryption protocols help to secure online banking and e-commerce. Encryption can secure business and government communications. Encryption can also help news websites to establish secure channels for receiving information from confidential sources.

Increasingly, online messaging services of the type that the new ePrivacy proposal aims to cover have deployed more advanced forms of encryption that protect communications confidentiality for their users. By implementing end-to-end encryption in their services, applications such as WhatsApp, Signal, and Telegram protect their users' private communications.

Considering the fundamental value of encryption for the protection of communications confidentiality, **the ePrivacy Regulation should recognise the value of encryption for the protection of privacy and confidentiality of communications.** The UN Special Rapporteur on the promotion and protection of the right to freedom of opinion and expression, David Kaye, concludes that encryption enables individuals to exercise their rights, and therefore deserves strong protection from a human rights perspective.[515] A recent UNESCO study on Encryption and Human Rights concludes:

'There needs to be recognition of cryptographic methods as an essential element of the media and communications landscape. What ultimately matters, from a human rights perspective, is that cryptographic methods empower individuals in their enjoyment of privacy and freedom of expression, as they allow for the protection of human-facing properties of information, communication and computing. These properties include the confidentiality, privacy, authenticity, availability, integrity and anonymity of information and communication.'[516]

Confidentiality cannot be ensured when backdoors might be contained in encryption software. The EDPS recommends prohibiting such backdoors.[517] The Article 29 Working Party and the

---

[511] EDPS 2017/6, p. 34.
[512] Article 2(22) of the draft European Electronic Communications Code.
[513] See Arnbak 2016; Asghari 2016.
[514] Felten 2017.
[515] Kaye 2015.
[516] Schulz and Van Hoboken 2016.
[517] EDPS 2017/6, p. 34.

EDPS also suggest that obligations to use encryption could enhance the protection of confidentiality of communications.[518]

---

# 5. INDEPENDENT SUPERVISORY AUTHORITIES AND ENFORCEMENT (PROPOSAL CHAPTER IV)

## 5.1. Article 18, independent supervisory authorities

Article 18 reads as follows:

'(1) The independent supervisory authority or authorities responsible for monitoring the application of [the GDPR] shall also be responsible for monitoring the application of this Regulation. Chapter VI and VII of [the GDPR] shall apply *mutatis mutandis*. The tasks and powers of the supervisory authorities shall be exercised with regard to end-users.
(2) The supervisory authority or authorities referred to in paragraph 1 shall cooperate whenever appropriate with national regulatory authorities established pursuant to the [Directive Establishing the European Electronic Communications Code].'

See also Recitals 38 and 39.

**Comment**
Article 18 establishes the duties of the supervisory authorities responsible for monitoring the application of the ePrivacy Regulation. The responsible supervisory authorities for the GDPR (Data Protection Authorities for short) are the same supervisory authorities responsible for the ePrivacy Regulation.

Both the EDPS and the Article 29 Working Party have advised that the new ePrivacy Regulation should be monitored and supervised by Data Protection Authorities.[519] Recital 38 states that the task of monitoring the application of the ePrivacy Regulation should not jeopardise the task of monitoring the application of the GDPR. Recital 38 adds that additional financial and human resources, premises, and infrastructure should be provided to the supervisory authority responsible for the ePrivacy Regulation.

Recital 38 also mentions that Member States should provide the supervisory authorities with additional assets necessary for the effective performance of its tasks. The text of the ePrivacy proposal is not clear on whether a hierarchy between the monitoring of the GDPR and ePrivacy Regulation exists. **The EU lawmaker could consider clarifying that the monitoring of the GDPR and the ePrivacy Regulation are of equal importance.**

Recital 38 mentions the monitoring of the application of the Regulation regarding electronic communications data for legal entities. We are of the opinion that the supervisory authorities should monitor the application of the Regulation to legal entities regardless of whether electronic communications data are involved. As mentioned in the comments on the scope of the Regulation,[520] the Regulation concerns more than the protection of electronic communications data. For example, Chapter III may not always concern electronic communications data. The provisions of that Chapter are applicable to end-users and their protection, thereby also protecting legal entities that are end-users. The monitoring of the application of the Regulation should also concern these provisions.

Further clarification regarding the application of key concepts of the consistency and cooperation mechanisms under the GDPR to the ePrivacy Regulation would be welcome. For example, it is unclear how the lead authority mechanism would apply in cases of cross-border processing[521] regarding interference with terminal equipment.[522] **In sum, we recommend that the EU lawmaker considers clarifying the provision and the relevant recitals.**

---

[519] Article 29 Working Party 2017 (WP247), p. 7; EDPS 2017/6, p. 8.
[520] See our comment on Articles 2 and 3 of the ePrivacy proposal.
[521] Art. 56(1) of the GDPR.
[522] Article 29 Working Party 2017 (WP247), p. 35.

## 5.2. Article 19, European Data Protection Board

Article 19 reads as follows:

'The European Data Protection Board, established under Article 68 of [the GDPR], shall have competence to ensure the consistent application of this Regulation. To that end, the European Data Protection Board shall exercise the tasks laid down in Article 70 of [the GDPR]. The Board shall also have the following tasks:
(a) advise the Commission on any proposed amendment of this Regulation;
(b) examine, on its own initiative, on request of one of its members or on request of the Commission, any question covering the application of this Regulation and issue guidelines, recommendations and best practices in order to encourage consistent application of this Regulation.'

**Comment**
Article 19 establishes the competences of the European Data Protection Board (EDPB) regarding the consistent application of the ePrivacy Regulation. The tasks laid down in Articles 19(a) and 19(b) of the ePrivacy proposal resemble the tasks in Articles 70(1)(b) and 70(1)(e) GDPR. **We recommend that the EU lawmaker considers clarifying that Article 70 of the GDPR applies *mutatis mutandis*.**[523] Article 18 of the proposal already adds that the specified articles of the GDPR apply *mutatis mutandis*.

## 5.3. Article 20, cooperation and consistency procedures

Article 20 reads as follows:

'Each supervisory authority shall contribute to the consistent application of this Regulation throughout the Union. For this purpose, the supervisory authorities shall cooperate with each other and the Commission in accordance with Chapter VII of [the GDPR] regarding the matters covered by this Regulation.'

See also Recital 39.

**Comment**
Article 20 establishes the duties of supervisory authorities regarding the consistent application of the ePrivacy Regulation. To this end, Chapter VII (on cooperation and consistency) of the GDPR applies. Recital 39 adds that 'this Regulation relies on the consistency mechanism of [the GDPR].'

Recitals 43 and 44 of the December 2016 draft mentioned cooperation and consistency procedures. These recitals have been removed in the ePrivacy proposal. It is unclear why.

As with other provisions of the proposal that refer to provisions of the GDPR, issues concerning interpretation may arise when dealing with parties which do not entirely fit the definitions of 'controller' and 'processor' as defined in the GDPR.[524] **The EU lawmaker should consider clarifying the ePrivacy proposal on this point.**

---

[523] See Article 29 Working Party 2017 (WP247), p. 35.
[524] See Article 29 Working Party 2017 (WP247), p. 35

# 6. REMEDIES, LIABILITY, AND PENALTIES (PROPOSAL CHAPTER V)

## 6.1. Article 21, remedies

Article 21 reads as follows:

'(1) Without prejudice to any other administrative or judicial remedy, every end-user of electronic communications services shall have the same remedies provided for in Articles 77, 78, and 79 of [the GDPR].
(2) Any natural or legal person other than end-users adversely affected by infringements of this Regulation and having a legitimate interest in the cessation or prohibition of alleged infringements, including a provider of electronic communications services protecting its legitimate business interests, shall have a right to bring legal proceedings in respect of such infringements.'

**Comment**
Article 21 of the ePrivacy proposal establishes the remedies that end-users of electronic communications services have a right to, in cases where their rights under the ePrivacy Regulation have been violated. Article 21 refers to the remedies provided in the GDPR.[525]

Article 21 creates uncertainty. For example, it is not clear what remedies Article 21(1) offers to those whose rights under Chapter III of the ePrivacy Regulation have been violated. For example, Article 21(1) states that the end-user has the *same remedies* as provided for in Article 79 GDPR. Article 79 of the GDPR concerns effective judicial remedies against a *controller or processor* who processes *personal data* in violation of the GDPR. Hence, it is uncertain whether Article 21(1) of the ePrivacy Regulation grants the end-user a right to an effective judicial remedy in the case of a violation of his or her rights under the ePrivacy Regulation, when the violating party does not meet the criteria of Article 79 GDPR.

The December 2016 draft of the ePrivacy Regulation offered more certainty, as it explicitly stated that the end-user should have the right to an effective judicial remedy when he or she considers that his or her rights under the ePrivacy Regulation have been violated. It would be advisable to carefully examine whether Articles 77-79 of the GDPR, to which Article 21(1) of the ePrivacy Regulation refers, provide remedies for the end-user in all cases of potential violation of the ePrivacy Regulation, especially the provisions of Chapter III.

**Furthermore, the ePrivacy proposal does not provide the opportunity for a not-for-profit organisation to represent an end-user when exercising the rights of Article 21(1) of the ePrivacy Regulation.** The December 2016 draft did provide the option for end-users to mandate not-for-profit organisations to represent them to exercise their rights to a remedy in Article 23(6) of the ePrivacy Regulation.

**The December 2016 draft also provided for a right for NGOs to lodge complaints with the supervisory authority independently of a mandate from end-users.[526]** In addition, the December 2016 draft provided for the right of NGOs to an effective judicial remedy against a binding decision of a supervisory authority or a violation of rights under the ePrivacy Regulation. This right was removed from the ePrivacy proposal. **We recommend that the EU lawmaker considers reinserting such provisions.**

**The ePrivacy regulation could also refer to Article 80 of the GDPR, which provides for similar collective redress mechanisms as those of Articles 23(5) and 23(6) of**

---

[525] In comparison to the December 2016 draft, Article 21 has been condensed significantly. Instead of explicitly stating every the end–user of electronic communication services has, the ePrivacy proposal now refers to the relevant articles of the GDPR.
[526] Article 23(6) of the December 2016 draft.

**the December 2016 draft. It is unclear why the collective redress mechanism is left out of the ePrivacy Regulation.** This new mechanism is important for upholding the rights of end-users of electronic communications services.[527] **In sum, we recommend that the EU lawmaker revises Article 21.**

## 6.2. Article 22, right to compensation and liability

Article 22 reads as follows:

'Any end-user of electronic communications services who has suffered material or non-material damage as a result of an infringement of this Regulation shall have the right to receive compensation from the infringer for the damage suffered, unless the infringer proves that it is not in any way responsible for the event giving rise to the damage in accordance with Article 82 of [the GDPR].'

### Comment

Article 22 establishes a right to compensation for end-users of electronic communications services who have suffered (non-)material damage as a result of an infringement of the ePrivacy Regulation.

**Article 22 should be carefully reviewed to ensure that the right to compensation and liability encompasses all foreseeable situations under the ePrivacy Regulation, especially violations of the provisions of Chapter III.** Article 22 of the ePrivacy Regulation refers to Article 82 of the GDPR. That latter article uses the definitions of 'processor' and 'controller' from the GDPR.[528] It is not clear whether the terms 'processor' and 'controller' that are used in Article 82 GDPR cover all situations in which liability for a violation of rights under the ePrivacy Regulation should exist**. Perhaps it should be explicitly stated that violations of the provisions of Chapter III of the ePrivacy Regulation fall within the scope of Article 22.**

## 6.3. Article 23, general conditions for imposing administrative fines

Article 23(1) states:

'1. For the purpose of this Article, Chapter VII of [the GDPR] shall apply to infringements of this Regulation.
2. Infringements of the following provisions of this Regulation shall, in accordance with paragraph 1, be subject to administrative fines up to EUR 10 000 000, or in the case of an undertaking, up to 2 % of the total worldwide annual turnover of the preceding financial year, whichever is higher:
(a) the obligations of any legal or natural person who process electronic communications data pursuant to Article 8;
(b) the obligations of the provider of software enabling electronic communications, pursuant to Article 10;
(c) the obligations of the providers of publicly available directories pursuant to Article 15;
(d) the obligations of any legal or natural person who uses electronic communications services pursuant to Article 16.
3. Infringements of the principle of confidentiality of communications, permitted processing of electronic communications data, time limits for erasure pursuant to Articles 5, 6, and 7 shall, in accordance with paragraph 1 of this Article, be subject to administrative fines up to 20 000 000 EUR, or in the case of an undertaking, up to 4 % of the total worldwide annual turnover of the preceding financial year, whichever is higher.
4. Member States shall lay down the rules on penalties for infringements of Articles 12, 13, 14, and 17.

---

[527] See EDPS 2017/6, p. 35.
[528] Article 4(7) and 4(8) of the GDPR.

5. Non-compliance with an order by a supervisory authority as referred to in Article 18, shall be subject to administrative fines up to 20 000 000 EUR, or in the case of an undertaking, up to 4 % of the total worldwide annual turnover of the preceding financial year, whichever is higher.

6. Without prejudice to the corrective powers of supervisory authorities pursuant to Article 18, each Member State may lay down rules on whether and to what extent administrative fines may be imposed on public authorities and bodies established in that Member State.

7. The exercise by the supervisory authority of its powers under this Article shall be subject to appropriate procedural safeguards in accordance with Union and Member State law, including effective judicial remedy and due process.

8. Where the legal system of the Member State does not provide for administrative fines, this Article may be applied in such a manner that the fine is initiated by the competent supervisory authority and imposed by competent national courts, while ensuring that those legal remedies are effective and have an equivalent effect to the administrative fines imposed by supervisory authorities. In any event, the fines imposed shall be effective, proportionate and dissuasive. Those Member States shall notify to the Commission the provisions of their laws which they adopt pursuant to this paragraph by [xxx] and, without delay, any subsequent amendment law or amendment affecting them.'

See also Recital 40.

**Comment**

Article 23 establishes the general conditions for imposing administrative fines. The provision is important, says the preamble, to 'strengthen the enforcement of the rules'.[529]

**We recommend that the EU lawmaker gives careful consideration to the relative weight of Articles 5, 6, and 7 of the ePrivacy Regulation, and Article 8 of the ePrivacy Regulation.** In the ePrivacy proposal, more weight seems to be attached to the rights of Articles 5, 6 and 7 compared to Article 8. A fine of € 20.000.000 or 4% of the annual worldwide turnover may be imposed for a violation of Articles 5 to 7; in contrast to € 10.000.000 or 2% of the annual worldwide turnover for a violation of Article 8.

Furthermore, Article 23 does not fully harmonise fines for all infringements of the ePrivacy Regulation.[530] Member States can lay down rules on infringement of Articles 12, 13, 14, 17 and 18. **For the purpose of harmonisation it should be considered whether it is beneficial to let Member States lay down rules on infringement of the aforementioned articles of the ePrivacy Regulation.** The Article 29 Working Party and the EDPS also suggest that fines for infringements of Articles 12, 13, 14, 17 and 18 should be included in the ePrivacy Regulation itself.[531]

## 6.4. Article 24, penalties

Article 24 reads as follows:

'(1) Member States shall lay down the rules on other penalties applicable to infringements of this Regulation in particular for infringements which are not subject to administrative fines pursuant to Article 23, and shall take all measures necessary to ensure that they are implemented. Such penalties shall be effective, proportionate and dissuasive.

(2) Each Member State shall notify to the Commission the provisions of its law which it adopts pursuant to paragraph 1, no later than 18 months after the date set forth under Article 29(2) and, without delay, any subsequent amendment affecting them.'

---

[529] Recital 40.
[530] Article 29 Working Party 2017 (WP247), p. 25. See also EDPS 2017/6, p. 35.
[531] Article 29 Working Party 2017 (WP247), p. 25; EDPS 2017/6, p. 35.

See also Recital 7.

## Comment

Article 24 concerns penalties. Article 24 grants Member States the right to lay down rules on other penalties applicable to infringements of the ePrivacy Regulation, in particular infringements that are not subject to fines under Article 23. The penalties must be effective, proportionate, and dissuasive. Such provisions should be notified to the Commission no later than 18 months after the ePrivacy Regulation enters into force.

The phrase 'other penalties applicable to infringements of this Regulation in particular for infringements which are not subject to administrative fines pursuant to Article 23' makes it unclear whether penalties can *only* be laid down for infringements that are not subject to Article 23. The Article can also be read in a way that makes it possible for Member States to impose other penalties for infringements which are regulated in Article 23, which might lead to differences in the application of the Regulation across Member States. **We recommend that this uncertainty be addressed.**

**As mentioned,[532] the EU lawmaker should consider whether it is desirable from the perspective of harmonisation that Member States can lay down rules on infringements that are not subject to Article 23 of the proposal.[533]**

---

[532] See our comment on Article 23 of the ePrivacy proposal.
[533] See also Article 29 Working Party 2017 (WP247), p. 25.

# 7. DELEGATED ACTS AND IMPLEMENTING ACTS (PROPOSAL CHAPTER VI)

## 7.1. Article 25, exercise of the delegation

Article 25 reads as follows:

'(1) The power to adopt delegated acts is conferred on the Commission subject to the conditions laid down in this Article.

(2) The power to adopt delegated acts referred to in Article 8(4) shall be conferred on the Commission for an indeterminate period of time from [the date of entering into force of this Regulation].

(3) The delegation of power referred to in Article 8(4) may be revoked at any time by the European Parliament or by the Council. A decision to revoke shall put an end to the delegation of the power specified in that decision. It shall take effect the day following the publication of the decision in the Official Journal of the European Union or at a later date specified therein. It shall not affect the validity of any delegated acts already in force.

(4) Before adopting a delegated act, the Commission shall consult experts designated by each Member State in accordance with the principles laid down in the Inter- institutional Agreement on Better Law-Making of 13 April 2016.

(5) As soon as it adopts a delegated act, the Commission shall notify it simultaneously to the European Parliament and to the Council.

(6) A delegated act adopted pursuant to Article 8(4) shall enter into force only if no objection has been expressed either by the European Parliament or the Council within a period of two months of notification of that act to the European Parliament and the Council or if, before the expiry of that period, the European Parliament and the Council have both informed the Commission that they will not object. That period shall be extended by two months at the initiative of the European Parliament or of the Council.'

See also Recital 41.

**Comment**

Article 25 establishes the conditions for the adoption of delegated acts by the Commission. Recital 41 states that one the objectives of such delegated acts is the protection of fundamental rights of 'natural persons'. However, the ePrivacy Regulation mostly uses the term 'end-user', which also includes legal entities. No reference is made to legal entities in Recital 41 of the ePrivacy proposal.

We also note that Recital 41 refers to 'the objectives of this Regulation, namely to protect the fundamental rights and freedoms of natural persons and in particular their right to the protection of personal data and to ensure the free movement of personal data within the Union'. Article 1(1) of the ePrivacy proposal however lists 'the rights to respect for private life and communications and the protection of natural persons with regard to the processing of personal data' as one of the objectives of the ePrivacy Regulation. Article 1(2) lists 'free movement of electronic communications data and electronic communications services within the Union' as one of the objectives of the ePrivacy Regulation. **Recital 41 seems to refer more to the objectives of the GDPR than to the objectives of the ePrivacy Regulation. We recommend resolving this discrepancy.**

Furthermore, Recital 41 refers to the necessity of delegated acts to specify a code to identify direct marketing calls, including those made through automated calling and communication systems. But Article 16(7) of the ePrivacy Regulation, regarding unsolicited communications, empowers the Commission to take *implementing measures* instead of *delegated acts* and refers to Article 26(2) of the ePrivacy Regulation. **We recommend addressing this ambiguity.**

## 7.2. Article 26, committee

Article 26 reads as follows:

'(1) The Commission shall be assisted by the Communications Committee established under Article 110 of the [Directive establishing the European Electronic Communications Code]. That committee shall be a committee within the meaning of Regulation (EU) No 182/2011.[534] (2) Where reference is made to this paragraph, Article 5 of Regulation (EU) No 182/2011 shall apply.'

**Comment**
**See our comment on Article 16(7).**

---

[534] Original footnote: Regulation (EU) No 182/2011 of the European Parliament and of the Council of 16 February 2011 laying down the rules and general principles concerning mechanisms for control by Member States of the Commission's exercise of implementing powers (OJ L 55, 28.2.2011, p. 13–18).

# 8. FINAL PROVISIONS (PROPOSAL CHAPTER VII)

## 8.1. Articles 27 and 29, repeal of the directive and entry into force of the ePrivacy Regulation

Article 27 reads as follows:

'(1) Directive 2002/58/EC is repealed with effect from 25 May 2018.
(2) References to the repealed Directive shall be construed as references to this Regulation.'

Article 29 reads as follows:

'(1) This Regulation shall enter into force on the twentieth day following that of its publication in the Official Journal of the European Union.
(2) It shall apply from 25 May 2018.'

See also Recital 43.

### Comment
The planned timeline is ambitious. On the one hand, there are good reasons to aim for a finalised text of the ePrivacy Regulation that enters into force at the same time as the GDPR. The current ePrivacy Directive is not fit for application in a GDPR era and questions regarding the interpretation of provisions of the current ePrivacy Directive could arise once the GDPR enters into force.

On the other hand, the ambitious planning should not lead to too much haste. Companies will already have to adopt their practices to comply with the GDPR. If companies are confronted with a new ePrivacy Regulation that imposes new responsibilities on them at short notice, compliance might not be optimal.

We believe that it should be possible to finalise the ePrivacy Regulation in time. Hence, **we recommend that the EU lawmaker does aim for a quick adoption of the ePrivacy Regulation. The EU lawmaker should, however, consider whether companies should be given a grace period for complying with the ePrivacy Regulation and adapting their practices.** The length of such a grace period could depend on the time between adoption and entry into force.

## 8.2. Article 28, monitoring and evaluation clause

Article 28 reads as follows:

'By 1 January 2018 at the latest, the Commission shall establish a detailed programme for monitoring the effectiveness of this Regulation.
No later than three years after the date of application of this Regulation, and every three years thereafter, the Commission shall carry out an evaluation of this Regulation and present the main findings to the European Parliament, the Council and the European Economic and Social Committee. The evaluation shall, where appropriate, inform a proposal for the amendment or repeal of this Regulation in light of legal, technical or economic developments.'

### Comment
**We recommend that Article 28 be retained. See also our comment on Article 6 ('no legitimate interests provision should be added').**

## Legal texts and drafts (a selection)

- Regulation (EU) No 182/2011 of the European Parliament and of the Council of 16 February 2011 laying down the rules and general principles concerning mechanisms for control by Member States of the Commission's exercise of implementing powers, OJ L 55, 28.2.2011, p. 13–18 <http://eur-lex.europa.eu/eli/reg/2011/182/oj>.

* * *

**DIRECTORATE-GENERAL FOR INTERNAL POLICIES**

# POLICY DEPARTMENT C
## CITIZENS' RIGHTS AND CONSTITUTIONAL AFFAIRS

## Role

Policy departments are research units that provide specialised advice
to committees, inter-parliamentary delegations and other parliamentary bodies.

## Policy Areas

- Constitutional Affairs
- Justice, Freedom and Security
- Gender Equality
- Legal and Parliamentary Affairs
- Petitions

## Documents

Visit the European Parliament website:
**http://www.europarl.europa.eu/supporting-analyses**

PHOTO CREDIT: iStock International Inc.

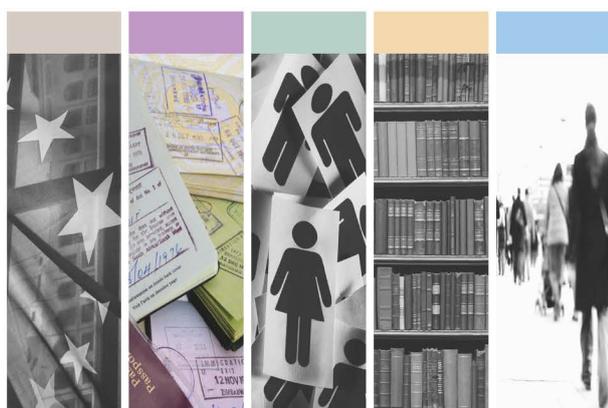

Publications Office